\documentclass[seceq,preprint]{ptptex}

\usepackage{graphicx}
\usepackage{color} 

\def\metap{m_{\eta'}}
\def\mpi{m_\pi}
\def\msea{m_\text{sea}}
\def\mval{m_\text{val}}
\def\mres{m_\text{res}}

\def\calO{{\cal O}}

\def\vev#1{\Big\langle #1 \Big\rangle}           

\title{
$\eta'$ meson from two flavor dynamical domain wall fermions%
}

\author{
Koichi \textsc{Hashimoto}$^{1,2}$ and %
Taku \textsc{Izubuchi}$^{1,3}$ \\
(for the RBC collaboration)%
}

\inst{
$^1$Institute for Theoretical Physics, Kanazawa University, Kakuma, Kanazawa 920-1192, Japan\\
$^2$Radiation Laboratory, RIKEN, Wako, Saitama 351-0198, Japan\\
$^3$ RIKEN-BNL Research Center, Brookhaven National Laboratory,Upton, NY 11973, USA
}

\abst{
We explore the spectrum of 
a flavor singlet pseudoscalar meson, 
$\eta'$, in two-flavor ($N_f=2$) lattice 
Quantum Chromo Dynamics (QCD).
The continuum-like relation between the topology of the QCD vacuum and
the $U(1)_A$ anomaly, which prevents the $\eta'$ 
meson from being a  Nambu-Goldstone boson, 
is expected to hold in the domain wall fermions (DWF) used as
a lattice quark field in this work.
Although our simulation is limited to  relatively heavy quark masses
and the statistical error is not small despite the
improvements in the measurements and fitting procedures for meson propagators, 
we obtained $\metap=819(127)$ MeV for the $N_f=2$ QCD, where the error is only statistical.
Several sources of systematic errors, which may be significant,
are discussed. Results for the other mesons  are also reported.
}

\begin{document}

\maketitle

\section{Introduction}

One of the most fascinating puzzles in the meson mass spectrum
is  $U(1)_A$ problem: why  the mass of the flavor singlet
pseudoscalar meson, $\eta'$, is  so heavier, $\metap=957.78(14)$ MeV, than 
that of its flavor nonsinglet counterparts in nature, 
$m_{\pi^{0}}=134.9766(6)$ MeV, $m_{K^{0}}=497.648(22)$ MeV, and 
$m_\eta=547.51(18)$ MeV\cite{Yao:2006px}. 

Nonsinglet mesons behave as Nambu-Goldstone (NG) bosons with the spontaneous 
breaking of $SU(3)_{A}$ symmetry in the quark massless limit $(m_{\rm quark}\to0)$,
ignoring the small QED effects 
whereas $\eta'$ is not an NG boson because 
$U(1)_{A}$ symmetry 
is broken by the quantum effect, the $U(1)_{A}$ anomaly.
The nonvanishing divergence of the flavor singlet axial current, 
${\cal A}_\mu^0(x)$, in the axial Ward-Takahashi identity (AWTI) occurs
for an operator ${\cal O}$
in  the case of degenerate quarks up to the contact term,
\begin{eqnarray}
\partial_{\mu} \vev{{\cal{A}}_{\mu}^{b}(x) {\cal O}} 
= 2m_{\rm quark}\vev{P^{a}(x) {\cal{O}}} 
+ \delta_{b,0} {2N_f}
\vev{\rho_\text{top} (x) {\cal O}},
\label{eq:flavor-singlet-AWTI} 
\end{eqnarray}
and expresses the anomalous breaking of chiral symmetry in
the last term, which is proportional to
the topological charge density, $\rho_\text{top}(x)$.
For a sufficiently smooth gauge field,
\begin{eqnarray}
\rho_\text{top}(x)=
 {1\over 32\pi^2} \epsilon_{\mu\nu\rho\sigma}
 \text{tr} F_{\mu\nu}F_{\rho\sigma}(x).
\end{eqnarray}

The difference in the pseudoscalar meson masses
between the flavor singlet sector, $m_{\eta'}$,
and the non-singlet sector,$m_\pi$,
was estimated by the Witten-Veneziano (WV) 
relation,\cite{Witten:1979vv, Veneziano:1979ec},
\begin{eqnarray}
m_{0}^{2}=m_{\eta'}^{2}-m_{\pi}^{2}={2N_{f}\over f_{\pi}^{2}} \chi_\text{top}
\label{eq:witten-veneziano}
\end{eqnarray}
at the limit of $N_{c}\to\infty$. Here 
$\chi_\text{top}$ is the susceptibility of the topological charge ($Q_\text{top}$) :
\begin{eqnarray}
\chi_\text{top}={\vev{Q_\text{top}^2}\over VT},~~
Q_\text{top} = \int \rho_\text{top}(x)\, d^4x,~~
\label{eq:Qtop}
\end{eqnarray}
in pure Yang-Mills (YM) theory in a four dimensional volume $VT$.
A recent result in $N_c=3$ YM theory with the overlap 
fermion\cite{DelDebbio:2004ns} shows that $\chi = (191(5) {\rm MeV})^{4}$. 
${\eta'}$ mass from this estimation for $N_f=3$ 
and in the chiral limit, $\mpi^{2}\to0$,
is  $\metap\approx970$ MeV, which is very close to experimental values. 

The direct numerical calculation of the $\eta'$ spectrum
is important for checking theoretical scenarios such as
the WV relation, and should result in its correction in finite $N_c$ and nonzero 
quark masses.
 
Simulations of ${\eta'}$ physics in 
pure YM theory with quenched Wilson
fermions were carried out in pioneer works\cite{Itoh:1987iy, Kuramashi:1994aj}. 
The relation between the topological charge and the mass of $\eta'$ 
was also explored\cite{Fukugita:1994iw}. 
Unquenched simulations\cite{Lesk:2002gd, McNeile:2000hf, Allton:2004qq, Struckmann:2000bt, Schilling:2004kg}
were performed for two-flavor and  for 2+1-flavor\cite{Aoki:2006xk} 
of Wilson fermions. Using staggered fermions, 
$m_{\eta'}$ has been calculated for $N_f=0,2$\cite{Venkataraman:1997xi} and
$N_f=2+1$\cite{Gregory:2007ev,Gregory:2007ce}.
Recently there are other interesting investigations, such as using 
twisted-mass quarks \cite{Michael:2007vn} or 
a local imaginary $\theta$-term \cite{Izubuchi:2008mu}.

In this paper, we discuss the mass of $\eta'$ in $N_{f}=2$ QCD with domain wall fermions (DWF). 
DWF \cite{Kaplan:1992bt, Shamir:1993zy, Furman:1994ky} 
is one of the lattice chiral fermions, which has both
flavor and  chiral symmetries even at finite lattice spacing $(a>0)$, 
and is thus suitable for investigation of nonperturbative physics of chiral anomalies. 
These features of DWF make their use preferable to the other alternative 
methods of discretization.
Wilson fermions break chiral symmetry at $a>0$ and 
discretization errors start at ${\cal O}(a\Lambda_\text{\rm QCD})$. 
The singlet flavor meson in staggered fermions is a very important 
subject as it may be related to the potential issue about the locality of 
the formalisms in the continuum limit\cite{Sharpe:2006re, Creutz:2007rk}.

Chiral and flavor symmetry are particularly important for $\eta'$ physics, 
and the DWF is the natural choice of lattice quark in investigations. 
Chiral symmetry in a DWF is not realized perfectly, it is broken 
due to its finite extent in the fifth direction, $L_s$.
The amount of breaking can be measured by a shift in quark mass:
$m_\text{quark}= m_f+m_\text{res}$, 
so that the nonsinglet axial current is conserved at $m_\text{quark}=0$.
$m_\text{res}$ is called the residual quark mass and vanishes at large $L_s$
for a sufficiently smooth gauge configuration\cite{Hernandez:1998et}.

Although it is  desirable to take $L_s\to\infty$ limit,
to reduce the computational cost, we restrict ourselves to finite $L_s=12$ 
with the combination of DBW2 improved gauge action 
\cite{Takaishi:1996xj, deForcrand:1999bi}, which smoothen gauge field at short
distance and reduces $\mres$ significantly \cite{Aoki:2002vt}. 

The RBC collaboration examined the first large scale dynamical DWF simulation 
\cite{Aoki:2004ht}. 
Pseudoscalar meson masses and decay constants were computed 
and fit to the chiral perturbation theory (ChPT) formula. 
$m_{\pi}$, $m_{K}$, $m_{\rho}$, $f_{\pi}$, and $f_{K}$  calculated 
in their work are reasonably consistent with values obtained in experiments.
The $J$ parameter is closer to the phenomenological value than the value obtained 
in the quenched simulation.
The nonsinglet scalar meson, $a_{0}$, mass and the decay constant have also been
examined both in dynamical QCD and partially quenched QCD using partially quenched 
ChPT \cite{Prelovsek:2004jp}.

We will mainly focus on the $\eta'$ meson in this paper, but
we will also report on the results of other mesons belong to
other Lorentz and flavor representations, and  also investigate the
signal of the mesons in their excited state and 
their decay constants. 

As the results are limited to the isospin symmetric case
and the number of dynamical quarks is two,
our focal interest in this paper is to provide a benchmark calculation
for the study of the general meson spectrum on a dynamical DWF 
ensemble with various (smeared) meson field using  a larger statistical sample 
than in the previous study.

In \S 2,  the theoretical expectations
of $\eta'$ meson physics are summarized.
We  explain the details of the simulation 
including  improvements in the
signal-to-noise ratio and  the fitting methods used to relate
the simulation data to physical quantities
in \S 3. The numerical results are presented 
in \S 4 with a list of their systematic uncertainties.
We will summarize in \S 5.

\section{Theoretical results on
 physics of flavor singlet meson}
\label{sec:theoretical_expectation}
In (continuum Euclidean) QCD with $N_{f}$ degenerated quarks,   
the operator of the flavor singlet pseudoscalar meson, $\eta'$: $I(J^{P})=0(0^{-})$, 
is defined by quark operators, $q_{f}$, as
\begin{equation}
\eta'(x)={1\over\sqrt {N_{f}}}\sum_{f=1}^{N_{f}}
\bar q_{f}(x)i\gamma_{5}q_{f}(x), 
\label{eq:eta_operator}
\end{equation}
where $f=1, ..., N_{f}$ is the flavor index. 
The $\eta'$ propagator  consists of two parts: 
\begin{eqnarray}
&&\int d^3x \langle \eta'(\vec{x},t) \eta'^{\dag}(\vec{0},0)\rangle
=C_{\gamma_5}(t) - N_{f}D_{\gamma_5}(t), \label{eq:eta_def}\\
&&C_{\gamma_5}(t)= - \int d^3x \left\langle{{1\over N_{f}}
\sum_{f}^{N_{f}}\overbrace{\bar
q_{f}(\vec{x},t)\gamma_{5}\underbrace{q_{f}(\vec{x},t) \bar
q_{f}(\vec{0},0)}\gamma_{5}q_{f}(\vec{0},0)}}\right\rangle, \label{eq:eta_def2}\nonumber\\
&&D_{\gamma_5}(t)=\int d^3x \left\langle{{1\over N_{f}}
\sum_{f}^{N_{f}}\overbrace{\bar
q_{f}(\vec{x},t)\gamma_{5}q_{f}(\vec{x},t)}{1\over N_{f}}
\sum_{g}^{N_{f}}\overbrace{\bar
q_{g}(\vec{0},0)\gamma_{5}q_{g}(\vec{0},0)}}\right\rangle\nonumber.
\end{eqnarray}
The braces represent the contraction of the quark propagators, $S_q(0,t)$.
Thus, for example,  $C_{\gamma_5}(t)$ is $\langle S_q(t,0)\gamma_5 S_q(0,t)\gamma_5\rangle$, 
the same as the nonsinglet meson (pion)
propagator, and $D_{\gamma_5}(t)$ is the correlation function between
disconnected quark loops, which exists  in the flavor singlet mesons.
When $D_{\gamma_5}(t)$ is suppressed by the OZI rule, it propagates and
acquires  $U(1)_{A}$ anomaly. 

In dynamical QCD, in which the mass of quark polarizing the gluon, $\msea$,
is equal to that of  the valence quark consisting the meson operator, $\mval$,
the $\eta'$ propagator is an exponential function of time 
with its damping factor being the mass of the meson, $m_\eta'$,
\begin{eqnarray}
\int d^3x {\langle \eta'(\vec{x},t) \eta'^{\dag}(\vec{0}, 0)\rangle} 
= C_{\gamma_5}(t) - N_{f}D_{\gamma_5}(t) = 
A_{\eta'}e^{-m_{\eta'}t}+\cdots,
\label{eq:eta_prop}
\end{eqnarray}
at large $t$.

\begin{figure}
\begin{center}
\includegraphics[angle=-00,scale=0.40,clip=true]{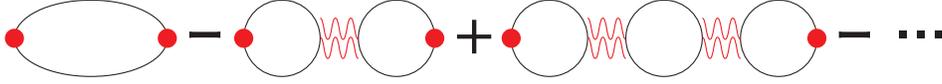}
\end{center}
\caption{Diagram of $\eta'$ propagator.}
\label{fig:eta_expand}
\end{figure}

A model of the $\eta'$ propagator is depicted in Fig. \ref{fig:eta_expand}.
The meson propagator is expressed as a series expansion in number of 
the quark loops  with signs reflecting the Grassmannian feature of the
quark, and the blobs at the ends are the meson operators
(\ref{eq:eta_operator}). The wavy lines connecting the quark loops represent the
coupling between disconnected loops attached to the pseudoscalar density,
which is related to the $U(1)_A$ anomaly.

The meson propagator in momentum space can be calculated from the model.
The first term in  Fig. $\ref{fig:eta_expand}$ is the same as the
nonsinglet pseudoscalar meson (pion), $1/(p^2+m^2_\pi)$, and
the second term is given by two pion propagators coupled to each other by 
gluons, $1/(p^2+m^2_\pi)\times m_0^2/N_f \times 1/(p^2+m^2_\pi)$,
whose coupling we parameterize as $m_0^2/N_f$. There are $N_f$ 
combinations of quark loops in the second term.
Repeating the similar identification of connected pion propagators to 
the $n$th order, the momentum space representation of the $\eta'$ propagator can
be written  as a geometrical series:
\begin{eqnarray}
&&\langle \eta'(p) \eta'^{\dag}(-p) \rangle
\nonumber\\
&&\propto {1\over p^{2}+m_{\pi}^{2}} 
- N_{f}{1\over p^{2}+m_{\pi}^{2}}{m_{0}^{2}\over N_{f}}{1\over p^{2}+m_{\pi}^{2}}
+N_{f}^{2}{1\over p^{2}+m_{\pi}^{2}}
{m_{0}^{2}\over N_{f}}{1\over p^{2}+m_{\pi}^{2}}{m_{0}^{2}\over N_{f}}{1\over p^{2}+m_{\pi}^{2}}-\cdots \nonumber\\
&& = {1\over p^{2}+m_{\pi}^{2}}\sum_{n=0}^{\infty}
\left({-m_{0}^{2}\over p^{2}+m_{\pi}^{2}}\right)^n=
{1\over p^{2}+(m_{\pi}^{2}+m_{0}^{2})}, 
\label{eq:eta_expand}
\end{eqnarray}
The $m_\pi$ pole in the connected diagram, $C_{\gamma_5}$, 
is exactly canceled by  part of the disconnected diagram, $D_{\gamma_5}$,
and thus the square of the $\eta'$ meson mass, $m_{\eta'}^{2}$, is identified by
$m_{\pi}^{2}+m_{0}^{2}$, which means $\eta'$ does not behave as 
an NG boson. In terms of this model, $\eta'$ spectroscopy calculated in
lattice simulation should reveal the magnitude of  $m^2_0$, 
and if it is consistent with the WV relation (\ref{eq:witten-veneziano}).

In a later section, we will also calculate the ratio between
$D_{\gamma_5}(t)$ and $C_{\gamma_5}(t)\sim A_\pi e^{-m_\pi t}$. 
From (\ref{eq:eta_prop}), the ratio at large $t$ should behave as
\begin{eqnarray}
&&{N_{f}D_{\gamma_5}(t)\over C_{\gamma_5}(t)}
=1- B {e^{m_{\eta'} t}+e^{-m_{\eta'} (T-t)} \over  
e^{m_\pi t}+e^{-m_\pi (T-t)}} +\cdots \label{eq:eta_ratio_1} \\
&&\ \ \ \ \ \ \ 
\stackrel{ (T-t) \gg 1 }{\longrightarrow}
1-Be^{-\Delta mt}+\cdots, 
\label{eq:eta_ratio} 
\\
&& \Delta m = m_{\eta'}-m_{\pi}, \ \ B={A_{\eta'}\over A_{\pi}}. 
\end{eqnarray}

The ratio at large $t$  exponentially approaches  unity with
the exponent  being the mass difference between  $\eta'$ and pion, 
which is a signature of the dynamical sea quark.
This is in contrast to the quenched QCD, in which $\msea$ is taken to 
be infinitely heavy while $\mval$ is kept finite.
In this nonunitary theory, 
the third and higher terms in the quark loop expansion (\ref{eq:eta_def2})
are missing due to the decoupling of the sea quark, and the resulting
meson propagator has an unphysical double pole,
\begin{eqnarray}
&&\langle \eta'(p) \eta'^{\dag}(-p) \rangle_\text{quenched}
\propto {1\over p^{2}+m_{\pi}^{2}} 
- N_{f}{1\over p^{2}+m_{\pi}^{2}}{m_{0}^{2}\over N_{f}}{1\over
p^{2}+m_{\pi}^{2}}.
\end{eqnarray}
The ratio $D_{\gamma_5}(t)/C_{\gamma_5}(t)$ in this case behaves as a 
linear function of time,
\begin{eqnarray}
&&{N_{f}D^\text{(quenched)}_{\gamma_5}(t)\over C_{\gamma_5}(t)}
={m_{0}^{2}\over 2m_{\pi}}t + {\rm const}+\cdots, 
\label{eq:eta_ratio_quenched}  
\end{eqnarray} 
which is clearly different from (\ref{eq:eta_ratio}). 
Thus, to obtain a physical $\eta'$ we must simulate  
the dynamical theory $(\msea=\mval)$.
We will examine  $m_{\eta'}$ in the domain wall QCD 
only at the dynamical points.  
 
\section{Simulation details}

\subsection{Domain wall fermion (DWF)}
The DWF action is defined as 
\begin{eqnarray}
&& S_{F}=\sum_{x,y,s,s'}\bar\psi(x,s)D_{\rm DWF}(x,s;y,s')\psi(y,s'), \\
&& D_{\rm DWF}(x,s;y,s')=\delta_{s,s'}D^\parallel_{x,y}+\delta_{x,y}D^\bot_{s,s'}, \label{eq:DWF-action}\\
&& D^\parallel_{x,y}={1\over2}\sum_{\mu=1}^{4}
\left[(1-\gamma_{\mu})U_{\mu}(x)\delta_{x+\hat\mu,y}+
(1+\gamma_{\mu})U_{\mu}^{\dag}(y)\delta_{x-\hat\mu,y}\right]+
(M_{5}-4)\delta_{x,y}, \\
&& D^\bot_{s,s'}=
{1\over2}\left[(1-\gamma_{5})\delta_{s+1,s'}+(1+\gamma_{5})\delta_{s-1,s'}\right]
\nonumber\\&&~~~~~~~+
{m_{f}\over2}\left[(1-\gamma_{5})\delta_{s,L_{s}-1}\delta_{0,s'}
+(1+\gamma_{5})\delta_{s,0}\delta_{L_{s}-1,s'}\right],\label{eq:DWF-action-bot}
\end{eqnarray}
where $\psi(x,s)$ is a DWF that is
located in five dimensional space, $(x,s)$, $L_s$ is the size of the fifth 
direction, and the parameter $M_5$ is the domain wall height.
By setting $M_5$ in a region around $[0,2]$, from (\ref{eq:DWF-action-bot}),
left-(right-)handed zero modes are localized around $s=0 (L_s-1)$ and the
zero modes undergo exponential damping as $s,~(L_s-1-s)$ increases. 
When a four-dimensional fermion  and antifermion, $q(x)$ and $\bar{q}(x)$, 
are defined as 
\begin{eqnarray}
&&q(x)={1-\gamma_5 \over 2}\psi(x,0)+{1+\gamma_5 \over 2}\psi(x,L_s-1), \\
&&\bar{q}(x)=\bar{\psi}(x,0){1+\gamma_5 \over 2}+\bar{\psi}(x,L_s-1){1-\gamma_5 \over 2},
\end{eqnarray}
chiral symmetry is fulfilled even with finite lattice spacing $(a>0)$ 
at the $L_s\to\infty$ limit.

However, in the simulation,  $L_s$ is restricted to  be finite,
and the AWTI is modified from its expression in the continuum theory to,
\begin{eqnarray}
\partial_{\mu} \vev{{\cal{A}}_{\mu}^{b}(x) {\cal O}} 
= 2(m_{f}+\mres)\vev{P^{b}(x) {\cal{O}}},
\end{eqnarray}
i.e., the physical quark mass is shifted to $m_{\rm quark}= m_{f}+\mres$. 
$\mres$ is a small lattice artifact called the residual quark mass, defined as 
\begin{eqnarray}
\mres=\lim_{t\to\infty}{\sum_{\vec{x}}\vev{J_{5q}^b(\vec{x},t)
P^b(\vec{0},0)}\over \sum_{\vec{x}}\vev{P^b(\vec{x},t) P^b(\vec{0},0)}},
\label{eq:mres_def}
\end{eqnarray}
where $J_{5q}^b(\vec{x},t)$ is an operator similar to the pseudoscalar operator
but made of fermions at the midpoint of the fifth direction\cite{Furman:1994ky},
$s\sim L_s/2$, thus the numerator of (\ref{eq:mres_def}) includes the
contractions between the surface fermions at $s=0$ or $s=L_s$ and 
the midpoint fermions at $s\sim L_s/2$. For the flavor non-singlet case, 
$b\ne0$,  $\mres$ is an exponential function of $L_s$ as a consequence of the
exponentially localized zero modes to the surface, and vanishes as $L_s\to\infty$. 

One could further argue\cite{Blum:2001sr} that the effective
Lagrangian contains the diverging, ${\cal O}(1/a)$,
discretization error, which can be corrected by the small shift
of the quark mass, $m_\text{quark}= m_f+m_\text{res}$.
The remaining error is ${\cal O}(a)$, similar to that of Wilson fermions,
however, it is an exponentially small number, $e^{\alpha L_s}$, or
${\cal O}(m_\text{res})$. Although $m_\text{res}$ is small
compared with the  statistical errors we will have in most of
observables, we will treat the shifted quark mass
$m_\text{quark}=m_f+m_\text{res}$ as the physical quark mass
so that our analysis is precise modulo ${\cal O}(m_\text{res} a, a2)$,
which is a few percent in our simulation.

On the other hand, for flavor singlet $(b=0)$ case, 
$J_{5q}^b(\vec{x},t)$ in (\ref{eq:mres_def}) can 
be attached to a quark loop that does not propagate in the entire $L_s$ 
in the fifth direction, and is free from suppression.
the counterparts of $\mres$ in the flavor singlet case 
remains finite even as $L_s\to\infty$, and 
reproduces the following anomalous term\cite{Shamir:1993yf}
\begin{eqnarray}
\sum_{\vec{x}}\vev{J_{5q}^b(\vec{x},t) \calO}\to\delta_{b,0}
\sum_{\vec{x}}\vev{\rho_\text{top}(\vec{x},t)\calO}~.
\end{eqnarray}
In summary, DWF even for finite $L_s$ correctly reproduces the quantum anomaly of
axial symmetry with  small error due to lattice discretization. 

\subsection{Ensemble: actions and parameters}
We employ the $N_{f}=2$ QCD ensemble\cite{Aoki:2004ht} with DWF actions described in
the previous subsection. Our gauge action contains  an improvement 
in the sense of the renormalization group invariance, 
DBW2\cite{deForcrand:1999bi}:
\begin{eqnarray}
&& S_{G}={\beta \over 3}\left[(1-8c_{1})\sum_{x,\mu>\nu}
{\text{ReTr}}[1-P_{\mu\nu}(x)]
+c_{1}\sum_{x,\mu\ne\nu}
{\text{ReTr}}[1-R_{\mu\nu}(x)]\right], \\ 
&& P_{\mu\nu}(x)=U_\mu(x)U_\nu(x+\hat\mu)
U^{\dag}_\mu(x+\hat\nu)U^{\dag}_\nu(x), \\
&& R_{\mu\nu}(x)=U_\mu(x)U_\mu(x+\hat\mu)U_\nu(x+2\hat\mu)
U^{\dag}_\mu(x+\hat\mu+\hat\nu)U^{\dag}_\mu(x+\hat\nu)
U^{\dag}_\nu(x), 
\end{eqnarray} 
with $\beta=0.80$ and $c_{1}=-1.4069$.
The parameters of the DWF action (\ref{eq:DWF-action-bot}) are set as
$L_{s}=12$ and $M_{5}=1.8$.
We measure observables on  a 470-940 lattice configuration samples for three different masses, 
$m_{f}$=0.02, 0.03, and 0.04, which correspond to $m_{\pi}/m_{\rho}\approx$
0.51-0.64. The lattice size is $16^{3}\times32$, 
the lattice scale is $a^{-1}\approx 1.5$ GeV ($a\approx 0.13$ fm), 
and the residual chiral breaking $m_{\rm res}=0.00137(4)$ which is about 
an order of magnitude smaller than the input quark masses. Throughout this paper
we estimate the statistical error using the blocked jackknife method.
The size of the block is determined to be 50 trajectories by monitoring
the autocorrelation of the hadron propagators.
A summary of lattice ensembles and parameters is given  in
Table \ref{tab:parameter}. Other results on these ensembles 
can be found in \citen{Aoki:2004ht,  Prelovsek:2004jp,Gadiyak:2005ea,Dawson:2006qc,Blum:2007cy}.

\begin{table}[t]
\caption{%
Lattice ensembles and simulation parameters.
}
\label{tab:parameter}
\begin{center}
\begin{tabular}{ccccccc}\hline \hline
$\beta$ & $c_{1}$ & $V\times T$ & $a^{-1}$ [GeV] &
$a$ [fm] &  $Va^3$ [fm$^3$] & $\mres$ \cr \hline
0.80 & $-1.4069$ & $16^{3}\times32$ & 
1.537(26) & 0.1284(22) & $(2.054)^3$ & 
0.00137(4) \cr \hline 
$m_{f}$ & $m_{\pi}/m_{\rho}$ & 
\multicolumn{3}{c}{begin-end(step) traj.} & \#config.  & $N_{\text{noise}}$ \cr
\hline
 0.02 & 0.5121(36) & \multicolumn{3}{c}{656-5351(5)} & 940 & 1 \cr 
 0.03 & 0.5984(31) & \multicolumn{3}{c}{615-6205(10)} & 560 & 3 \cr
 0.04 & 0.6415(33) & \multicolumn{3}{c}{625-1765(10), 2075-5615(10) $^{\rm a}$}
& 470 & 2 \cr
\hline 
\end{tabular}
\end{center}
\footnotesize{$^{\rm (a)}$ For the $m_{f}=0.04$ ensemble, we do not use 
trajectories 1775-2065 due to a hardware error on trajectory 
1772 that was not detected until 
lattice generation was finished. } 
\end{table}

\subsection{Improvements: smearing and sources}

Before constructing the meson propagators, we describe an improvement
for the quark propagators in this section.
It is known to be difficult to reduce the
statistical error of the flavor singlet meson spectrum. 
As we have seen in the previous section, 
the meson propagator includes the correlation function between 
disconnected loops, $D_\Gamma$
$(\Gamma=\gamma_5,\gamma_i,{\bf 1},\gamma_5\gamma_i,\gamma_i\gamma_j)$, 
whose statistical fluctuation is very large, particularly for large $t$ as we
will see. We have implemented  smearing for a quark operator in a 
gauge-covariant manner called Wuppertal smearing \cite{Gusken:1989qx}.
The smeared quark operator $q_S$ is a gauge-covariant superposition 
of the local quark operator $q_L$:
\begin{eqnarray}
&& q^c_{L}(\vec{x},t)\to q^{c}_{S}(\vec{x},t)=\sum_{\vec{y},c'}
F^{c,c'}(\vec{x},\vec{y}) q^{c'}_{L}(\vec{y},t), \\\
&& F^{c,c'}(\vec{x},\vec{y})=\left[\left\{{\bf 1}+{\omega^2\over 4N}\sum_{i=1}^3
\left(\nabla_i+\nabla_{i}^{\dag}\right)
\right\}^N\right]_{\vec{x},c;\vec{y},c'}, \label{eq:gauss-smear}\\
&& [{\bf 1}]_{\vec{x},c;\vec{y},c'}=\delta^{c,c'}\delta_{\vec{x},\vec{y}}, \\
&& [\nabla_{i}]_{\vec{x},c;\vec{y},c'} = U_{i}(\vec{x},t)^{c,c'}\delta_{\vec{x}+\hat i, \vec{y}}-\delta^{c,c'}\delta_{\vec{x},\vec{y}}, \\
&& [\nabla^{\dag}_{i}]_{\vec{x},c;\vec{y},c'} = U_{i}^{\dag}(\vec{y},t)^{c,c'}\delta_{\vec{x}-\hat i, \vec{y}}-\delta^{c,c'}\delta_{\vec{x},\vec{y}}. 
\end{eqnarray}
The shape of $q_S$ in terms of $q_L$ is Gaussian with width $\omega$ 
as $N\to\infty$. We set  $\omega=4.35$ and $N=40$.
The overlap between the ground state and the meson operator made of
smeared quarks is expected to be larger the meson made of unsmeared quarks, 
and the excited state contamination is suppressed for small $t$, 
where the statistical error is smaller.

Both the  quark correlation functions, $C_\Gamma(t)$ and $D_\Gamma(t)$,
are calculated for a complex ${\boldsymbol{Z}}_2$ noise source, $\xi$, defined by
\begin{eqnarray}
&& \xi^{(n)}(\vec{x},t) = {1\over\sqrt2} [\xi_{1}^{(n)}(\vec{x},t)+i\xi_{2}^{(n)}(\vec{x},t)],  
\end{eqnarray}
where $n=1, 2, \dots, N_{\rm noise}$ 
are random noise ensembles and $\xi_{1}$ and $\xi_2$ take 
values of $\pm 1$ randomly. 
$\xi(\vec{x},t)$ is statistically independent of space-time: thus, it
satisfies
\begin{eqnarray}
&& \lim_{N_{\rm noise}\to\infty}{1\over N_{\rm noise}}
\sum_{n=1}^{N_{\rm noise}}\xi^{(n)}(\vec{x},t) \xi^{(n)}(\vec{y},t') = 0,   
\label{eq:noise_zero}
\\
&& \lim_{N_{\rm noise}\to\infty}{1\over N_{\rm noise}}
\sum_{n=1}^{N_{\rm noise}}\xi^{(n)}(\vec{x},t) \xi^{(n)*}(\vec{y},t') 
= \delta_{\vec{x},\vec{y}}\delta_{t,t'},   
\label{eq:noise_one}
\end{eqnarray}
which is useful for calculating the disconnected loops as we will see in the
next subsection.
We use the source restricted to a time slice (wall source) for $C_\Gamma(t)$
and a space-time volume source for $D_\Gamma(t)$, and 
$N_{\text{noise}}=1$, 3, and 2 for $m_f$=0.02, 0.03, and 0.04, respectively. 

\subsection{Meson operators and correlation functions}
Our naming convention for meson fields is
similar to that used by the particle data group \cite{Yao:2006px},
but our simulation is limited to having only up and down quarks 
($N_f=2$) with degenerate masses and zero electric charges; thus,
the meson spectra are inevitably different from those 
in the real world.  The systematic error from these omission 
may be comparable or smaller to our target precision of $\sim 10$ \%. 
This point certainly needs further investigation.

The Hermitian interpolation fields for flavor nonsinglet meson in our
simulation, $\pi$, $\rho$, $a_0$, $a_1$, and $b_1$, and singlet fields,
 $\eta'$, $\omega$, $f_0$, $f_1$, and $h_1$ are defined in terms of  
quark operators, $q_{I,f}$ and $\bar{q}_{J,f}$ as follows: 
\begin{eqnarray}
&&\pi_I(\vec{x},t)={1\over\sqrt {2}}\sum_{f,g=1}^{2}
\bar q_{I,f}(\vec{x},t)\tau^b_{f,g} i \gamma_{5}q_{I,g}(\vec{x},t), \\
&&\rho_I(\vec{x},t)={1\over\sqrt {6}}
\sum_{i=1}^3 \sum_{f,g=1}^{2}
\bar q_{I,f}(\vec{x},t)\tau^b_{f,g} i \gamma_{i}q_{I,g}(\vec{x},t)  \\
&&a_{0I}(\vec{x},t)={1\over\sqrt {2}}\sum_{f,g=1}^{2}
\bar q_{I,f}(\vec{x},t)\tau^b_{f,g}q_{I,g}(\vec{x},t), \\
&&a_{1I}(\vec{x},t)={1\over\sqrt {6}}
\sum_{i=1}^3 \sum_{f,g=1}^{2}
\bar q_{I,f}(\vec{x},t)\tau^b_{f,g}i\gamma_{5}\gamma_{i}q_{I,g}(\vec{x},t) \\
&&b_{1I}(\vec{x},t)={1\over\sqrt {6}}
\sum_{\substack{1\le i\le 3\\i<j\le 3}} \sum_{f,g=1}^{2}
\bar q_{I,f}(\vec{x},t)\tau^b_{f,g}i\gamma_{i}\gamma_{j}q_{I,g}(\vec{x},t) \\ 
&&\eta'_I(\vec{x},t)={1\over\sqrt {2}}\sum_{f=1}^{2}
\bar q_{I,f}(\vec{x},t) i \gamma_{5}q_{I,f}(\vec{x},t), \\
&&\omega_I(\vec{x},t)={1\over\sqrt {6}}
\sum_{i=1}^3\sum_{f=1}^{2}
\bar q_{I,f}(\vec{x},t) i \gamma_{i}q_{I,f}(\vec{x},t)\\
&&f_{0I}(\vec{x},t)={1\over\sqrt {2}}\sum_{f=1}^{2}
\bar q_{I,f}(\vec{x},t)q_{I,f}(\vec{x},t), \\
&&f_{1I}(\vec{x},t)={1\over\sqrt {6}}
\sum_{i=1}^3\sum_{f=1}^{2}
\bar q_{I,f}(\vec{x},t)i\gamma_{5}\gamma_{i}q_{I,f}(\vec{x},t)\\
 &&h_{1I}(\vec{x},t)={1\over\sqrt {6}}
\sum_{\substack{1\le i\le 3\\i<j\le 3}}\sum_{f=1}^{2}
\bar q_{I,f}(\vec{x},t) i \gamma_{i}\gamma_{j}q_{I,f}(\vec{x},t)
\end{eqnarray}
where $\tau^b$ $(b=1, 2, 3)$ 
are the Pauli matrices for the flavor indices $f$ and $g$, and
$I$ and $J$ denotes whether we use the local quark field ($L$) or
the smeared field ($S$) to control the ground-state overlap.
In Table \ref{tab:meson-operator}, we summarize the 
quantum numbers of each meson field.

\begin{table}[t]
\caption{%
Meson operators in the simulation and their quantum numbers.
}
\label{tab:meson-operator}
\begin{center}
\begin{tabular}{ccccc}\hline \hline
Meson type  & $J^{PC}$& $\Gamma$ & nonsinglet & singlet \cr \hline 
pseudoscalar & $0^{-+}$ & $i \gamma_5$ & $\pi$ & $\eta'$ \cr
vector & $1^{--}$ & $i {\gamma_i}$ $^{\rm a}$ & $\rho$ & $\omega$ \cr
scalar & $0^{++}$ & {\bf{1}} & $a_0$ & $f_0$ \cr
pseudovector & $1^{++}$ & $i{\gamma_5\gamma_i}$ $^{\rm a}$ & $a_1$ & $f_1$ \cr
pseudovector & $1^{+-}$ & $i \gamma_i\gamma_j$ $^{\rm a}$
& $b_1$ & $h_1$ \cr
\hline 
\end{tabular}
\end{center}
\footnotesize{$^{\rm (a)}$ average over $i,j=1,2,3$ is taken.} 
\end{table}

The two-point correlation functions between the interpolation fields
are calculated as
\begin{eqnarray}
&&\sum_{\vec{x},\vec{y}}\langle \pi_I(\vec{x},t) \pi^{\dag}_J(\vec{y},0)\rangle
=C_{IJ,\gamma_5}(t), \\
&&\sum_{\vec{x},\vec{y}}\langle \rho_I(\vec{x},t) \rho^{\dag}_J(\vec{y},0)\rangle
={1 \over 3}\sum_{i=1}^3 C_{IJ,\gamma_i}(t), \\
&&\sum_{\vec{x},\vec{y}}\langle a_{0I}(\vec{x},t) a_{0J}^{\dag}(\vec{y},0)\rangle
=C_{IJ,\bf 1}(t), \\
&&\sum_{\vec{x},\vec{y}}\langle a_{1I}(\vec{x},t) a_{1J}^{\dag}(\vec{y},0)\rangle
={1 \over 3}\sum_{i=1}^3 C_{IJ,\gamma_5\gamma_i}(t), \\
&&\sum_{\vec{x},\vec{y}}\langle b_{I1}(\vec{x},t) b_{1J}^{\dag}(\vec{y},0)\rangle
={1 \over 3}\sum_{i<j} C_{IJ,\gamma_i\gamma_j}(t), \\
&&\sum_{\vec{x},\vec{y}}\langle \eta'_I(\vec{x},t) \eta'^{\dag}_J(\vec{y},0)\rangle
=C_{IJ,\gamma_5}(t) - 2 D_{IJ,\gamma_5}(t), \label{eq:eta_op_def}\\
&&\sum_{\vec{x},\vec{y}}\langle \omega_I(\vec{x},t) \omega^{\dag}_J(\vec{y},0)\rangle
={1 \over 3}\sum_{i=1}^3 \left[C_{IJ,\gamma_i}(t)-2 D_{IJ,\gamma_i}(t)\right], \\
&&\sum_{\vec{x},\vec{y}}\langle f_{0I}(\vec{x},t) f_{0J}^{\dag}(\vec{y},0)\rangle
=C_{IJ,\bf 1}(t) - 2 D_{IJ,\bf 1}(t), \\
&&\sum_{\vec{x},\vec{y}}\langle f_{1I}(\vec{x},t) f_{1J}^{\dag}(\vec{y},0)\rangle
={1 \over 3}\sum_{i=1}^3 
\left[C_{IJ,\gamma_5\gamma_i}(t)-2 D_{IJ,\gamma_5\gamma_i}(t)\right], \\
&&\sum_{\vec{x},\vec{y}}\langle h_{1I}(\vec{x},t) h_{1J}^{\dag}(\vec{y},0)\rangle
={1 \over 3}\sum_{i<j} 
\left[C_{IJ,\gamma_i\gamma_j}(t)-2 D_{IJ,\gamma_i\gamma_j}(t)\right],
\end{eqnarray}
in terms of the connected and disconnected quark loop contributions
(${\rm Tr}$ is for the trace over color and spinor indices only):
\begin{eqnarray}
&&C_{IJ,\Gamma}(t)=\sum_{\vec{x},\vec{y}}\left\langle{
\overbrace{\bar q_I(\vec{x},t)\Gamma \underbrace{q_I(\vec{x},t)
\bar q_J}(\vec{y},0)\Gamma q_J}(\vec{y},0)}\right\rangle \nonumber\\ 
&&\ \ \ \ \ \ \ \ 
=-\sum_{\vec{x},\vec{y}}\left\langle{\rm Tr} \left[
G_{IJ}(\vec{x},t;\vec{y},0)\Gamma G_{JI}(\vec{y},0;\vec{x},t)\Gamma
\right]\right\rangle\ \ (\Gamma=i\gamma_5, i\gamma_i, {\bf 1}, 
i\gamma_5\gamma_i, i\gamma_i\gamma_j), 
\label{eq:conn}
\\
&&D_{IJ,\Gamma}(t)=-\sum_{\vec{x},\vec{y}}\left\langle{
\overbrace{\bar q_I(\vec{x},t)\Gamma q_I}(\vec{x},t)
\overbrace{\bar q_J(\vec{y},0)\Gamma q_J}(\vec{y},0)}\right\rangle\nonumber\\ 
&&\ \ \ \ \ \ \ \ 
=-\sum_{\vec{x},\vec{y}}\left\langle\left\{{\rm Tr} \left[
G_{II}(\vec{x},t;\vec{x},t)\Gamma\right] -\sum_{\vec{x'}}\left\langle{\rm Tr} \left[ G_{II}(\vec{x'},t;\vec{x'},t)\Gamma\right]\right\rangle \right\} \right. \nonumber\\
&& \ \ \ \ \ \ \ \ \ \ \ \ \ \times
\left.\left\{{\rm Tr}\left[G_{JJ}(\vec{y},0;\vec{y},0)\Gamma\right]
-\sum_{\vec{y'}}\left\langle{\rm Tr} \left[ G_{JJ}(\vec{y'},0;\vec{y'},0)\Gamma\right] \right\rangle\right\}
\right\rangle\nonumber\\
&&\ \ \ \ \ \ \ \ \ \ \ \ \ \ 
(\Gamma=i\gamma_5, i\gamma_i, {\bf 1}, 
i\gamma_5\gamma_i, i\gamma_i\gamma_j)
\label{eq:disc}. 
\end{eqnarray}
Here  $G_{IJ}(\vec{x},t;\vec{y},t')$ is the propagator of the four dimensional quark field 
\begin{eqnarray}
&&G_{LL}^{c,\alpha;c',\alpha'}(\vec{x},t;\vec{y},t') = \left[D^{-1}(\vec{x},t;\vec{y},t')\right]^{c,\alpha;c',\alpha'}, \\
&&G_{LS}^{c,\alpha;c',\alpha'}(\vec{x},t;\vec{y},t') = \sum_{c''}\sum_{\vec{x'}}\left[D^{-1}(\vec{x},t;\vec{x'},t')\right]^{c,\alpha;c'',\alpha'}F^{c'',c'}(\vec{x'},\vec{y}), \\
&&G_{SL}^{c,\alpha;c',\alpha'}(\vec{x},t;\vec{y},t') = \sum_{c''}\sum_{\vec{x'}}F^{c,c''}(\vec{x},\vec{x'})\left[D^{-1}(\vec{x'},t;\vec{y},t')\right]^{c'',\alpha;c',\alpha'}, \\
&&G_{SS}^{c,\alpha;c',\alpha'}(\vec{x},t;\vec{y},t') = \sum_{c'',c'''}\sum_{\vec{x'},\vec{y'}}F^{c,c''}(\vec{x},\vec{x'})
\left[D^{-1}(\vec{x'},t;\vec{y'},t')\right]^{c'',\alpha;c''',\alpha'}F^{c''',c'}(\vec{y'},\vec{y}), 
\end{eqnarray}
where $D^{-1}$ is written in terms of the inverse of the five dimensional matrix $D_{\rm DWF}^{-1}$ (Eq. (\ref{eq:DWF-action})): 
\begin{eqnarray}
&&D^{-1}(x,y)=\vev{q(x)\bar{q}(y)} \nonumber\\
&=&\sum_{s,s'}\left({1-\gamma_5\over2}\delta_{s,0}+{1+\gamma_5\over2}\delta_{s,L_s-1}\right)\vev{\psi(x,s)\bar{\psi}(y,s')}
\left({1+\gamma_5\over2}\delta_{s,0}+{1-\gamma_5\over2}\delta_{s,L_s-1}\right)\nonumber\\
&=&\sum_{s,s'}\left({1-\gamma_5\over2}\delta_{s,0}+{1+\gamma_5\over2}\delta_{s,L_s-1}\right)D_{\rm DWF}^{-1}(x,s;y,s')
\left({1+\gamma_5\over2}\delta_{s,0}+{1-\gamma_5\over2}\delta_{s,L_s-1}\right).
\label{eq:DWF_q_propagator}
\end{eqnarray}
$F(\vec{x},\vec{y})$ is the smearing function which is defined in Eq.(\ref{eq:gauss-smear}). 

$c,c',c'',c'''$ are the color indices and $\alpha, \alpha'$ are 
the spin indices. We  apply the zero-momentum projection 
to obtain the meson mass from meson energy:
$E_{\vec{p}}=\sqrt{m_{\text{meson}}^2+\vec{p}^2}\to m_{\text{meson}}$,
by summing over spatial volume $\vec{x},\vec{x'},\vec{y},\vec{y'}$.  
In eq. (\ref{eq:conn}), the sum over $\vec{y}$ 
is stochastically evaluated by the ${\bf Z}_2$ noise source at
$t=0$, 
while the sums over $\vec{x}$ and $\vec{y}$ in (\ref{eq:disc})  
are evaluated by ${\boldsymbol Z}_2$ source spreads over the space-time volume,
{\it c.f.\/} (\ref{eq:noise_zero}) and (\ref{eq:noise_one}):
\begin{eqnarray}
&&~~~{1\over N_{\rm noise}}\sum_{n=1}^{N_{\rm noise}}\sum_{\vec{x},\vec{y},\vec{z}}\langle {\rm Tr}[\{G_{IJ}(\vec{x},t;\vec{y},0)\xi^{(n)}(\vec{y},0)\}\Gamma
\gamma_5
\{G_{IJ}(\vec{x},t;\vec{z},0)\xi^{(n)}(\vec{z},0)\}^{\dag}\gamma_5\Gamma]\rangle\nonumber\\
&&={1\over N_{\rm noise}}\sum_{n=1}^{N_{\rm noise}}\sum_{\vec{x},\vec{y},\vec{z}}\langle {\rm Tr}[G_{IJ}(\vec{x},t;\vec{y},0)\Gamma
\gamma_5
G^{\dag}_{IJ}(\vec{x},t;\vec{z},0)\gamma_5\Gamma]\rangle\xi^{(n)}(\vec{y},0)\xi^{(n)*}(\vec{z},0)\nonumber\\
&&\to \sum_{\vec{x},\vec{y}}\langle {\rm Tr}[G_{IJ}(\vec{x},t;\vec{y},0)\Gamma
\gamma_5
G^{\dag}_{IJ}(\vec{x},t;\vec{y},0)\gamma_5\Gamma]\rangle~~(N_{\rm noise}\to\infty),\nonumber\\
&&=\sum_{\vec{x},\vec{y}}\langle {\rm Tr}[G_{IJ}(\vec{x},t;\vec{y},0)\Gamma
G_{JI}(\vec{y},0;\vec{x},t)\Gamma]\rangle,\label{eq:conn_contract_noise}\\
&& ~~~{1\over N_{\rm noise}}\sum_{n=1}^{N_{\rm noise}}\sum_{\vec{x},\vec{y},{t'}}\langle {\rm Tr}[\xi^{(n)*}(\vec{x},t)\{G_{II}(\vec{x},t;\vec{y},t')\xi^{(n)}(\vec{y},t')\}\Gamma
]\rangle\nonumber\\
&&\to \sum_{\vec{x}}\langle{\rm Tr}[G_{II}(\vec{x},t;\vec{x},t)\Gamma
]\rangle~~(N_{\rm noise}\to\infty),
\end{eqnarray}
The dagger ($\dagger$) is taken only for color and spinor  (and not for space-time) indices, and
we use the $\gamma_5$ hermiticity, $\gamma_5 D^{-1}\gamma_5 =[ D^{-1} ]^\dagger$, 
of the propagator (\ref{eq:DWF_q_propagator}) in  (\ref{eq:conn_contract_noise}).
The trace over color and spinor indices is exactly carried out  by solving
the quark propagator $3\times 4$ times each for 
a random source. 

\subsection{Meson mass fit}

Throughout this paper, we assume that the one particle state is the ground state
for quantum numbers $I$ and $J^{PC}$, for compatibility with to the interpolation operator
in Table \ref{tab:meson-operator}.
This assumption is not entirely true for some cases.
For example, a $\rho$ meson may decay into pions.
In our simulation, quarks are heavy with 
the lightest quark mass about half the strange quark mass, 
and confined in a relatively small ($\sim 2$ fm)$^3$ box.
Many of the decay processes would not occur in this setting since
the decaying particles have energies above the threshold.
Also we restrict ourselves to degenerate up and down quarks, $N_f=2$, 
so that a meson such as $a_0$ can not decay due to exact symmetry.

To extract the meson masses, 
the following two analyses are carried out.
\begin{enumerate}
\item[(A)] 
Standard method: \\
Only the ground state of mass $m_O$ is assumed to exist in
the correlation function $\langle O_{S} O_{S}\rangle$, which is 
fitted by the hyperbolic cosine function reflecting the periodic
boundary condition for a meson at $t=T$;
\begin{equation}
\sum_{\vec{x},\vec{y}}\langle O_I(\vec{x},t) O^{\dag}_I(\vec{y},0)\rangle
= {V\over 2m_O}|\langle 0|O_I|O(\vec{p}=\vec{0})\rangle|^2
\left[ e^{-m_{O}t} + e^{-m_{O}(T-t)}\right],~~~~
(I=L, S)
\label{eq:method-a}
\end{equation}
for sufficiently large  $t$ and $T-t$.
Although our main results will be obtained from the smeared-quark case, $I=S$,
we also analyze  local quark case to monitor the excited-state contamination. 
The fitting range of $t$ is determined so that the effective meson mass
becomes independent of the time. We also avoid a too large $t$ for which
the statistical error becomes large and the results become unreliable.
\item[(B)] 
Variational method \cite{Michael:1985ne, Luscher:1990ck}: \\
In this case, we also assume the first excited state of mass $m_{O^*}$.
Both the local ($I,J=L$) and the smeared ($I,J=S$) interpolation fields
are used to construct the correlation function $\langle O_{I} O_{J} \rangle$.
The $2\times2$ matrix, 
\begin{eqnarray}
&&X(t)=\left(
\begin{array}{cc}
\sum_{\vec{x},\vec{y}}\langle O_{L}(\vec{x},t) O^{\dag}_{L}(\vec{y},0) \rangle & 
\sum_{\vec{x},\vec{y}}\langle O_{L}(\vec{x},t) O^{\dag}_{S}(\vec{y},0) \rangle \cr
\sum_{\vec{x},\vec{y}}\langle O_{S}(\vec{x},t) O^{\dag}_{L}(\vec{y},0) \rangle & 
\sum_{\vec{x},\vec{y}}\langle O_{S}(\vec{x},t) O^{\dag}_{S}(\vec{y},0) \rangle 
\end{array}
\right), 
\end{eqnarray}
is normalized at a reference time $t_0$ to reduce the statistical error,
then is diagonalized as
\begin{eqnarray}
&& X^{-1/2}(t_{0})X(t)X^{-1/2}(t_{0})
\stackrel{\rm diag.}{\longrightarrow}
\left(
\begin{array}{cc}
\lambda_O(t,t_{0}) & 0 \cr
0 & \lambda_{O^{*}}(t,t_{0})
\end{array}
\right)~~.
\label{eq:variational_diag}
\end{eqnarray}
The eigenvalues are fit as a function of $t$,
\begin{eqnarray}
&& \lambda_O(t,t_{0})={e^{-m_{O}t} + e^{-m_{O}(T-t)} \over 
e^{-m_{O}t_{0}} + e^{-m_{O}(T-t_{0})}} 
\left(\stackrel{t,t_{0}\ll T/2}{\to}e^{-m_{O}(t-t_{0})}\right), \label{eq:method-c}\\
&& \lambda_{O^{*}}(t,t_{0})={e^{-m_{O^{*}}t} + e^{-m_{O^{*}}(T-t)} \over 
e^{-m_{O^{*}}t_{0}} + e^{-m_{O^{*}}(T-t_{0})}} 
\left(\stackrel{t,t_{0}\ll T/2}{\to}e^{-m_{O^{*}}(t-t_{0})}\right)~
\label{eq:variational_excited_eig}
\end{eqnarray}
to obtain  the masses of the states.

The second method, called the variational method, is 
employed to extract the ground-state energy precisely 
and to determine the amount of excited state contamination.

To fit $\lambda(t,t_0)$ using eq. (\ref{eq:method-c}), without unknown
amplitudes in front of the exponentials, $t_0$ should be sufficiently large
to ignore the higher excited states.
By monitoring $\lambda(t,t_0)$, we verify, for our choice of $t_0$,
that such contamination is not apparent within the current statistics. 
As an example $a_0$ case is shown in Fig. \ref{fig:a0-bt}. 
$t_0=2$ (squares) is chosen for the final results as
$\lambda(t,t_0)$ for $t_0=1$ (circles) can't be fit to a linear function of
$t-t_0$ meaning the meson propagator is not a single exponential, 
while those of $t_0>2$ (diamonds, triangles) have much larger error bars.
If the number of available configurations were larger, we would have
observed the effect from the second excited state and should have
calculated for more variations of interpolation field. This point may
be important for future investigations with larger statistical sample.

\begin{figure}[t]
\begin{center}
\includegraphics[angle=-00,scale=0.40,clip=true]{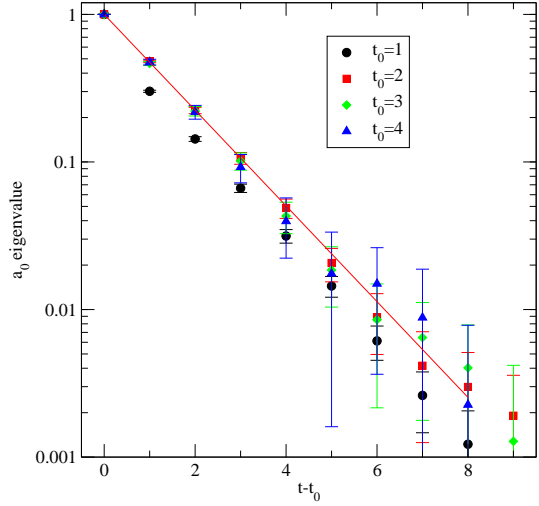}
\end{center}
\caption{
$t_0$ dependence of $a_0$ eigenvalue for $m_f=0.02$. 
We chose $t_0=2$ (squares) by determining the contamination from the higher 
excited states.}
\label{fig:a0-bt}
\end{figure}

\end{enumerate}

\begin{figure}[t]
\begin{center}
\includegraphics[angle=-00,scale=0.40,clip=true]{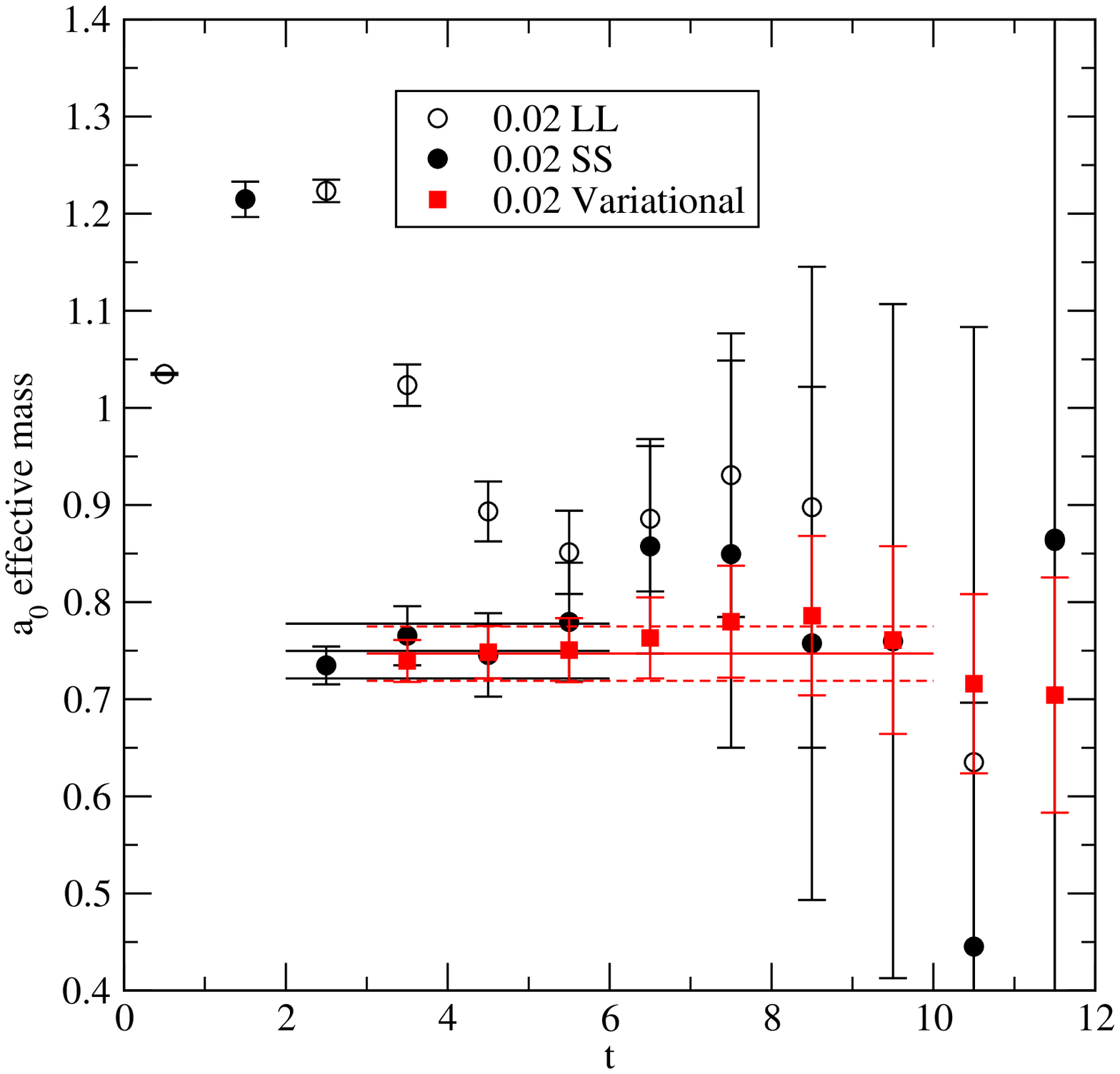}
\end{center}
\caption{
Comparison of the effective mass of $a_0$ obtained by the two methods.
Open (filled) circles show the results from the local-local
(smeared-smeared) interpolation field in the standard method
while squares show the effective mass obtained in the variational method. }
\label{fig:eta-comp}
\end{figure}

For another example, the effective mass of $a_0$, which we will define in
(\ref{eq:def-effmass}), obtained from the two methods is plotted in
Fig.~\ref{fig:eta-comp}. The effective mass obtained from the variational method (squares) has
the smallest statistical error, which is consistent with that of standard method
using a smeared-smeared interpolation field (filled circles). 
For the variational method, the plateau appears after a smaller time distance
when the excited-state is separated from the ground-state.
The global fits to the plateaux are almost identical to each other.
The clear signal of contamination from larger excited states 
for the local-local interpolation field (open circles) is observed.
The identical central values and error bars from the standard single exponential fit
and the variational method indicate that the effect from excited states is small for both
methods with these settings.

We analyze all masses by both methods
and compare the results to estimate the systematic uncertainty due to higher excited-states.
We also explore the first excited state for pseudoscalar and vector
mesons, $\pi^*$ and $\rho^*$, using the variational method.

\subsection{Decay constant}

The leptonic decay constant can be obtained from the
amplitude of the two-point correlation function of a meson.
We analyze decay constants for  a pion, $\pi^*$ and $\rho$ mesons.
Their respective decay constants, $f_\pi$, $f_{\pi^*}$ and $f_\rho$ can be defined 
through the  conserved axial and vector currents, 
${\cal A}_\mu^b(x)$ and ${\cal V}_i^b(x)$,
\begin{eqnarray}
&& f_O m_O = \langle 0 | {\cal A}_4^b(x) | O(\vec{p}=\vec{0}) \rangle 
=  Z_A\langle 0 | {A}_4^b(x) | O(\vec{p}=\vec{0}) \rangle
\ \ (O=\pi, \pi^*)\  
\label{eq:fps_def}
\\  
&& f_\rho m_\rho \epsilon_i = \langle 0 | {\cal V}_i^b(x) |
\rho(\vec{p}=\vec{0}) \rangle = Z_V\langle 0 | {V}_i^b(x) | \rho(\vec{p}=\vec{0}) \rangle
\ \ (i=1,2,3) 
\label{eq:fvec_def}
\end{eqnarray}
where $\epsilon_i$ is the polarization vector of the vector meson state,
and $Z_A$ and $Z_V$ are the matching factors between the lattice local currents,
\begin{eqnarray}
A^b_\mu(x) &=& \bar q(x) \tau^b \gamma_\mu \gamma_5 q(x),\\
V^b_\mu(x) &=& \bar q(x) \tau^b \gamma_\mu  q(x),
\end{eqnarray}
and an appropriate renormalization is used scheme  in the continuum QCD,
which, in our case, is $\overline{MS}$ at $\mu=2$ GeV.

For $f_\pi$ and $f_{\pi^*}$, the first matrix element in
 (\ref{eq:fps_def}) can be related to  pseudoscalar density
$P^b(\vec{x},t) = \bar q(x) \tau^b \gamma_5 q(x)$
using the (flavor nonsinglet) AWTI,
\begin{eqnarray}
\partial_\mu \langle 0 | {\cal A}_\mu^b(x) O(0) | 0 \rangle = 2(m_f+\mres) 
\langle 0 | P^b(x) O(0)| 0 \rangle, 
\label{eq:singlet AWT}
\end{eqnarray}
which leads to 
\begin{eqnarray}
f_O m_O^2 = 2(m_f+\mres)\langle 0 | P^b | O(\vec{p}=\vec{0}) \rangle \ \ (O=\pi,\pi^*). 
\label{eq:fpi}
\end{eqnarray}

The actual determination of the decay constants is performed by
the standard method (C) for a pion and $\rho$ meson,
and the variational method (D) for a pion and $\pi^*$ meson:
\begin{enumerate}

\item[(C)]
Standard method \\
In this case we assume the $\langle \pi_{L} \pi_{L}\rangle$ and $\langle \rho_{L} \rho_{L}\rangle$ correlation functions 
contain only  propagation of the ground-state.
$\langle \pi_{L} \pi_{L}\rangle$ and $\langle \rho_{L} \rho_{L}\rangle$ are fitted 
by a standard hyperbolic cosine function: 
\begin{eqnarray}
\sum_{\vec{x},\vec{y}}\langle \pi_L(\vec{x},t) \pi^{\dag}_L(\vec{y},0)\rangle
&=& {V\over 2m_{\pi}}|\langle 0 | P^a_L | \pi(\vec{p}=\vec{0}) \rangle|^2\left[ e^{-m_{\pi}t} + e^{-m_{\pi}(T-t)}\right] \nonumber\\
&=&{V f_{\pi}^2 m_{\pi}^3 \over 8(m_f+\mres)^2}\left[ e^{-m_{\pi}t} + e^{-m_{\pi}(T-t)}\right], \label{eq:method-C_fpi}\\
\sum_{\vec{x},\vec{y}}\langle \rho_L(\vec{x},t) \rho^{\dag}_L(\vec{y},0)\rangle
&=& {V\over 2m_{\rho}}|\langle 0 | V^a_L | \pi(\vec{p}=\vec{0}) \rangle|^2\left[ e^{-m_{\rho}t} + e^{-m_{\rho}(T-t)}\right] \nonumber\\
&=&{V f_{\rho}^2 m_{\rho} \over 2Z_V^2}\left[ e^{-m_{\rho}t} + e^{-m_{\rho}(T-t)}\right],  
\label{eq:method-C_frho}
\end{eqnarray}
to extract the quantities $m_\pi$, $f_\pi$, $m_\rho$, and
$f_\rho/Z_V$.

\item[(D)]
Variational method \\
In this case, the second excited state, $\pi^*$, in the correlation function
of the local meson operator, $\langle \pi_{L} \pi_{L}\rangle$, is 
also taken into account. $\langle \pi_{L} \pi_{L}\rangle$ is fitted by a
double hyperbolic cosine function:
\begin{eqnarray}
&&\sum_{\vec{x},\vec{y}}\langle \pi_L(\vec{x},t) \pi^{\dag}_L(\vec{y},0)\rangle
= {V\over 2m_{\pi}}|\langle 0 | P^a_L | \pi(\vec{p}=\vec{0}) \rangle|^2\left[ e^{-m_{\pi}t} + e^{-m_{\pi}(T-t)}\right] \nonumber \\
&& \ \ \ \ \ \ \ \ \ \ \ \ \ \ \ \ \ \ \ \ \ \ \ \ \ \ \ \ 
+{V\over 2m_{\pi^*}}|\langle 0 | P^a_L | \pi^*(\vec{p}=\vec{0}) \rangle|^2\left[ e^{-m_{\pi^*}t} + e^{-m_{\pi^*}(T-t)}\right] \nonumber\\
&&\!\!\!\!\!\!\!\!={V f_{\pi}^2 m_{\pi}^3 \over 8(m_f+\mres)^2}\left[ e^{-m_{\pi}t} + e^{-m_{\pi}(T-t)}\right] 
+{V f_{\pi^*}^2 m_{\pi^*}^3 \over 8(m_f+\mres)^2}\left[ e^{-m_{\pi^*}t} + e^{-m_{\pi^*}(T-t)}\right].
\label{eq:method-b2}
\end{eqnarray}
In this fitting procedure, we first determine $m_\pi$ and $m_{\pi^*}$ by
the variational method, (B) in the previous subsection and
then  fit the two-point function data to (\ref{eq:method-b2}) 
to determine  $f_{\pi}$ and $f_{\pi^*}$ using the
results from the first fitting.

\end{enumerate}

\subsection{Chiral Extrapolation}

To obtain  the masses and decay constants of various mesons
at the physical quark mass point, $m_f=m_{u,d}$ \cite{Aoki:2004ht}, 
we need to extrapolate the numerical value calculated
at heavier quark mass points.
As the number of simulation points is limited
and the statistical error is too large, we do not
use the fitting formula of chiral perturbation theory
at the next leading order or higher in this work.

As a crude estimation of the mass of $\eta'$  at the physical
point, we examine the formula valid in the lowest-order
approximation from the flavor singlet AWTI given by
eq. (\ref{eq:singlet AWT}):
\begin{eqnarray}
&& m_{\eta'}^2 = C_0+C_1(m_f+\mres) \ \ {\text{(AWTI\ type)}}, 
\label{eq:chiral_sqrt_formula}
\end{eqnarray}
We also examine the simplest linear extrapolation for 
all meson masses as well as the decay constants,
\begin{eqnarray}
&& O  = C_0+C_1(m_f+\mres) \ \ {\text{(linear\ type)}}~
\label{eq:chiral_lin_formula}
\end{eqnarray}
where $O$ is either a meson mass or a decay constant.

\section{Numerical results}

\subsection{Mass of $\rho$ meson and lattice scale}

\begin{figure}[t]
\begin{center}
\includegraphics[angle=-00,scale=0.37,clip=true]{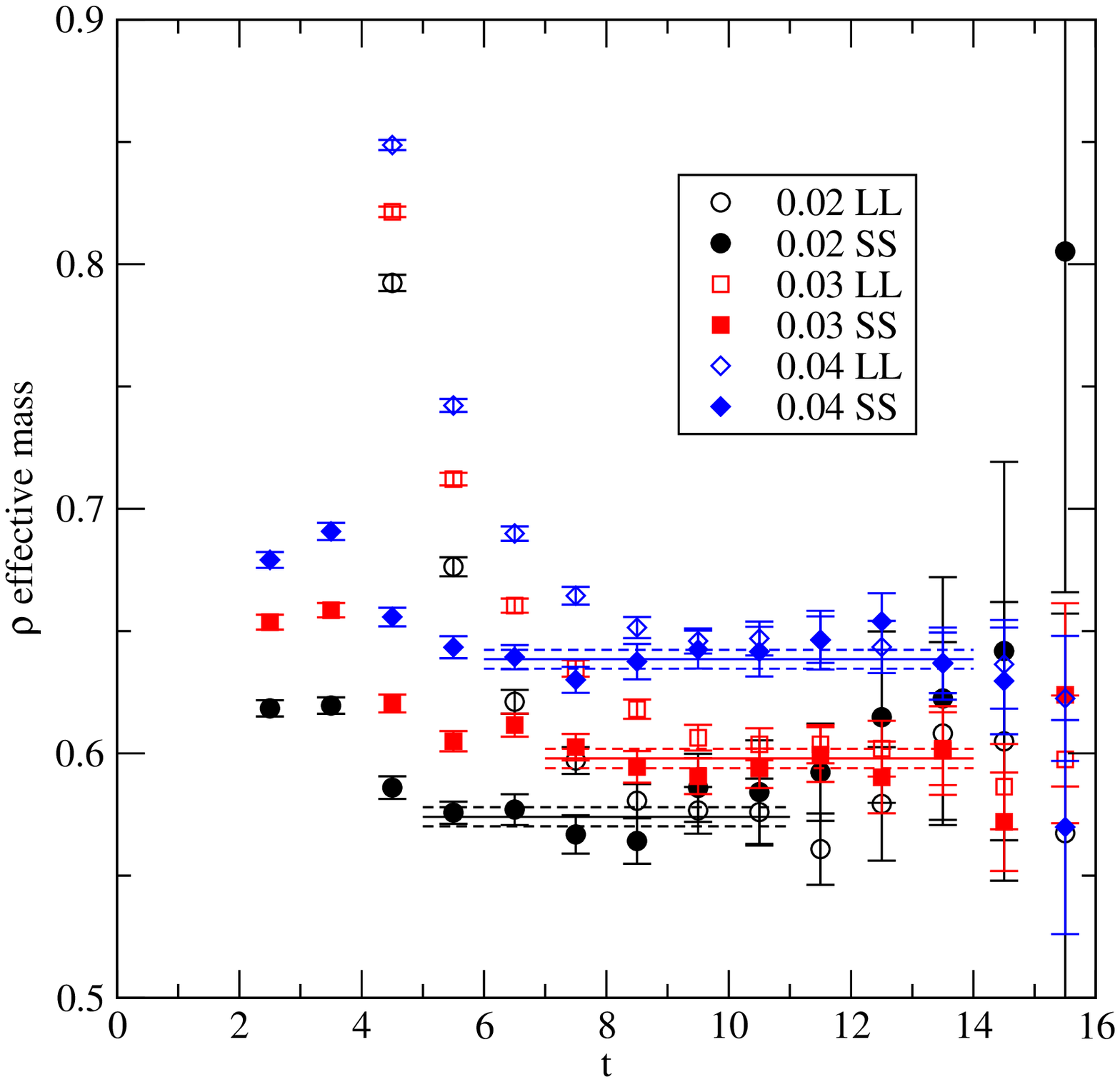}
\includegraphics[angle=-00,scale=0.37,clip=true]{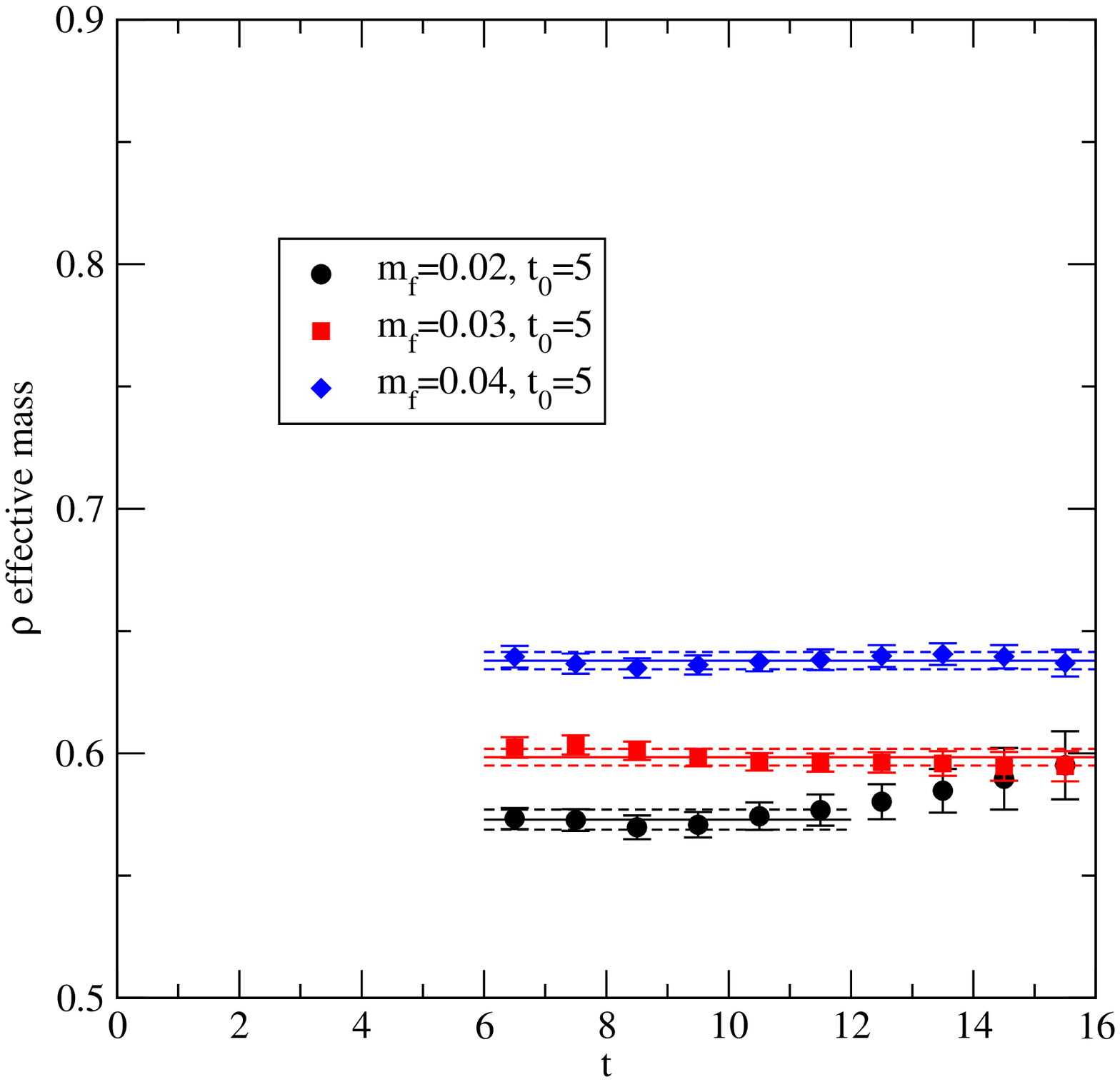}
\includegraphics[angle=-00,scale=0.37,clip=true]{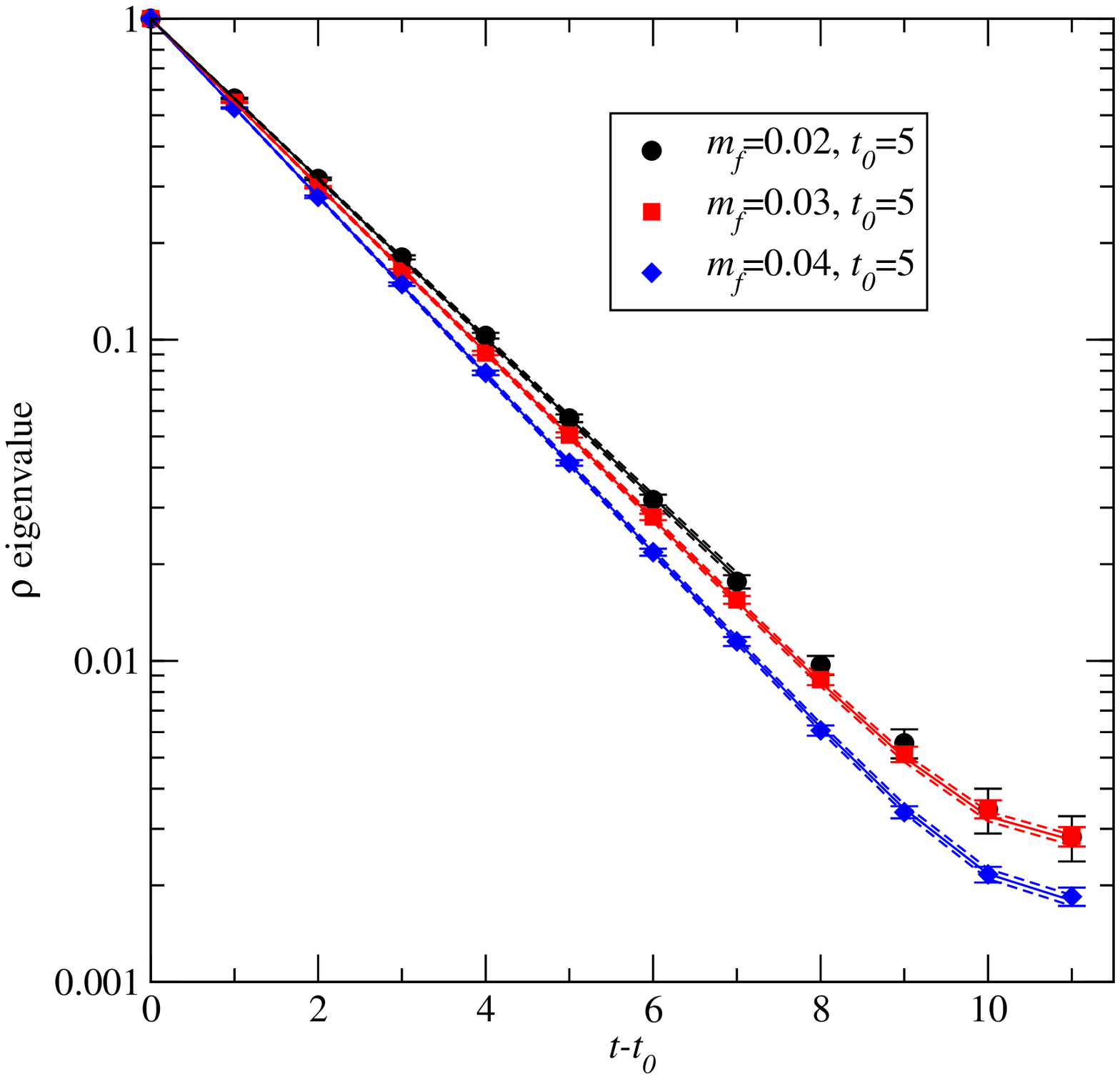}
\end{center}
\caption{
Effective mass of $\rho$ vs $t$ using method (A) (left) and method (B) (right), and $\rho$ eigenvalue vs $t-t_0$ (bottom). 
Lines show the globally fitted result with errors and the ranges of $t$. 
}
\label{fig:rho-a}
\end{figure}

First we analyze the mass of a $\rho$ meson using the methods (A) and (B) and 
determine the lattice scale from $m_\rho$ assuming that it is a stable particle, which
is true for the relatively heavy quark in the small box used in our simulation. 
In Fig. \ref{fig:rho-a}, the effective mas of a $rho$ meson,
taken from the damping rate between meson propagators at 
two neighboring times, $m_{\rho,IJ}^{\rm eff}(t+1/2)$,
which is defined as 
\begin{eqnarray}
{\sum_{\vec{x},\vec{y}}\vev{O_I(\vec{x},t) O_J^{\dag}(\vec{y},0)}\over\sum_{\vec{x},\vec{y}}\vev{O_I(\vec{x},t+1) O_J^{\dag}(\vec{y},0)}}
= {e^{-m_{O, IJ}^{\rm eff}(t+{1\over2}) \  t} 
+  e^{-m_{O, IJ}^{\rm eff}(t+{1\over2}) \  (T-t)} \over
    e^{-m_{O, IJ}^{\rm eff}(t+{1\over2}) \  (t+1)} 
+  e^{-m_{O, IJ}^{\rm eff}(t+{1\over2}) \  (T-t-1)} }~, \label{eq:def-effmass}
\end{eqnarray}
is plotted in the top panels (method (A)  and method (B) are shown 
in the left and right panels, respectively).
The bottom panel shows an eigenvalue of the ground-state obtained from the
variational method. The results of $m_\rho$ obtained 
from the standard hyperbolic cosine fit (method (A))
and the variational method (method (B)) are listed in Table \ref{tab:rho-fit}. 
The masses obtained from both methods are consistent with each other within statistical
error for all $m_f$, and the ground-state mass can be successfully extracted
using the smeared operator.

\begin{table}[t] 
\caption{%
$m_{\rho}$
\label{tab:rho-fit}
}
\begin{center}
\begin{tabular}{cccccc}\hline \hline
$m_{f}$ & $m_{\rho}$  & $t_{0}$ 
& $t_{\rm min}$ & $t_{\rm max}$ & method \cr
\hline
0.02 & 0.5741(39) & & 5 & 11 & (A) \cr
& 0.5729(41) & 5 & $t_0+1$ & 12 & (B) \cr
& 0.5425(64) $^{\rm a}$ & & 5 & 16 & (A) \cr
\hline  
0.03 & 0.5979(40) & & 7 & 14 & (A) \cr
& 0.5984(34) & 5 & $t_0+1$ & 16 & (B) \cr
& 0.5946(58) $^{\rm a}$ & & 6 & 16 & (A) \cr
\hline
0.04 & 0.6385(39) & & 6 & 14 & (A) \cr
& 0.6379(35) & 5 & $t_0+1$ & 16 & (B) \cr
& 0.6323(70) $^{\rm a}$
& & 7 & 16 & (A) \cr
\hline 
\end{tabular}
\end{center}
\footnotesize{$^{(a)}$ These values are obtained from 
$(I,J)=(L,W)$ correlation functions and quoted by \cite{Aoki:2004ht}.} 
\end{table}

\begin{figure}[t]
\begin{center}
\includegraphics[angle=-00,scale=0.40,clip=true]{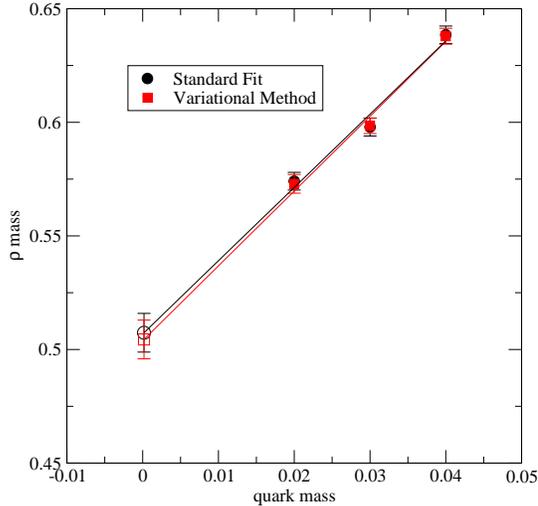}
\end{center}
\caption{
$m_{\rho}$ vs $m_f$}
\label{fig:rho-chiral}
\end{figure}

We perform linear extrapolation for both results and obtain $m_\rho$ at 
the physical quark mass point $(m_f=m_ {u,d})$. 
The result of the chiral extrapolation is shown in Fig.~\ref{fig:rho-chiral}
and Table~\ref{tab:rho-chiral}. 
The values obtained from both methods at the physical quark mass point are consistent 
within statistical error; we choose the value from method (B) as our main
value. The lattice scale determined from $m_\rho$=775.49 MeV \cite{Yao:2006px} is
\begin{eqnarray}
a_{m_\rho}^{-1}=1.537(26) \ \text{GeV}~.
\label{eq:a_value}
\end{eqnarray}
We have measured the potential energy between static quarks 
and extracted the Sommer scale $r_0$ from the potential
$r_0/a=4.278(54)$ \cite{Aoki:2004ht}. Using $a_{m_\rho}$ we obtained
\begin{eqnarray}
r_0^{\text{phys}}=0.5491(93)\ \text{fm}~~,
\end{eqnarray}
which is somewhat larger  than previously estimated values 
by $\sim$ 10\%. Although $r_0$ is one of the most precisely 
determined dimensionful quantities in the lattice QCD, its experimental
value is not known; thus, we could not judge whether our larger value is 
close to the physical value in QCD or whether it reflects some 
systematic errors, which we discuss in a later section.

By increasing the statistical sample size, the lattice scale changed from that
we reported in our previous paper\cite{Aoki:2004ht}.
Accordingly, the physical quark mass point, $m_f=m_{u,d}$, may change.
However, we use the old value of $m_{u,d}$ as the physical quark mass
point in this paper. This is because the number of quark mass points newly
obtained in this work is not sufficient to  repeat the same analysis as before
in which we used the formula of ChPT up to the next to the leading order.
We will discuss the decay constant and the excited-state meson, $\rho^*$, in later
sections.

\begin{table}[t]
\caption{%
$m_{\rho}$ at the physical quark mass point $(m_f= m_{u,d})$.
}
\label{tab:rho-chiral}
\begin{center}
\begin{tabular}{cccc}\hline \hline
$m_{\rho}$ 
& $a_{m_\rho}^{-1}$ [GeV] & $a_{m_\rho}$ [fm] 
& method \cr 
\hline
0.5073(85) & 1.528(26) & 0.1291(22) & (A) \cr
0.5044(85) & 1.537(26) & 0.1284(22) & (B) \cr
\hline 
\end{tabular}
\end{center}
\end{table}

\subsection{Pion mass}

\begin{figure}[t]
\begin{center}
\includegraphics[angle=-00,scale=0.37,clip=true]{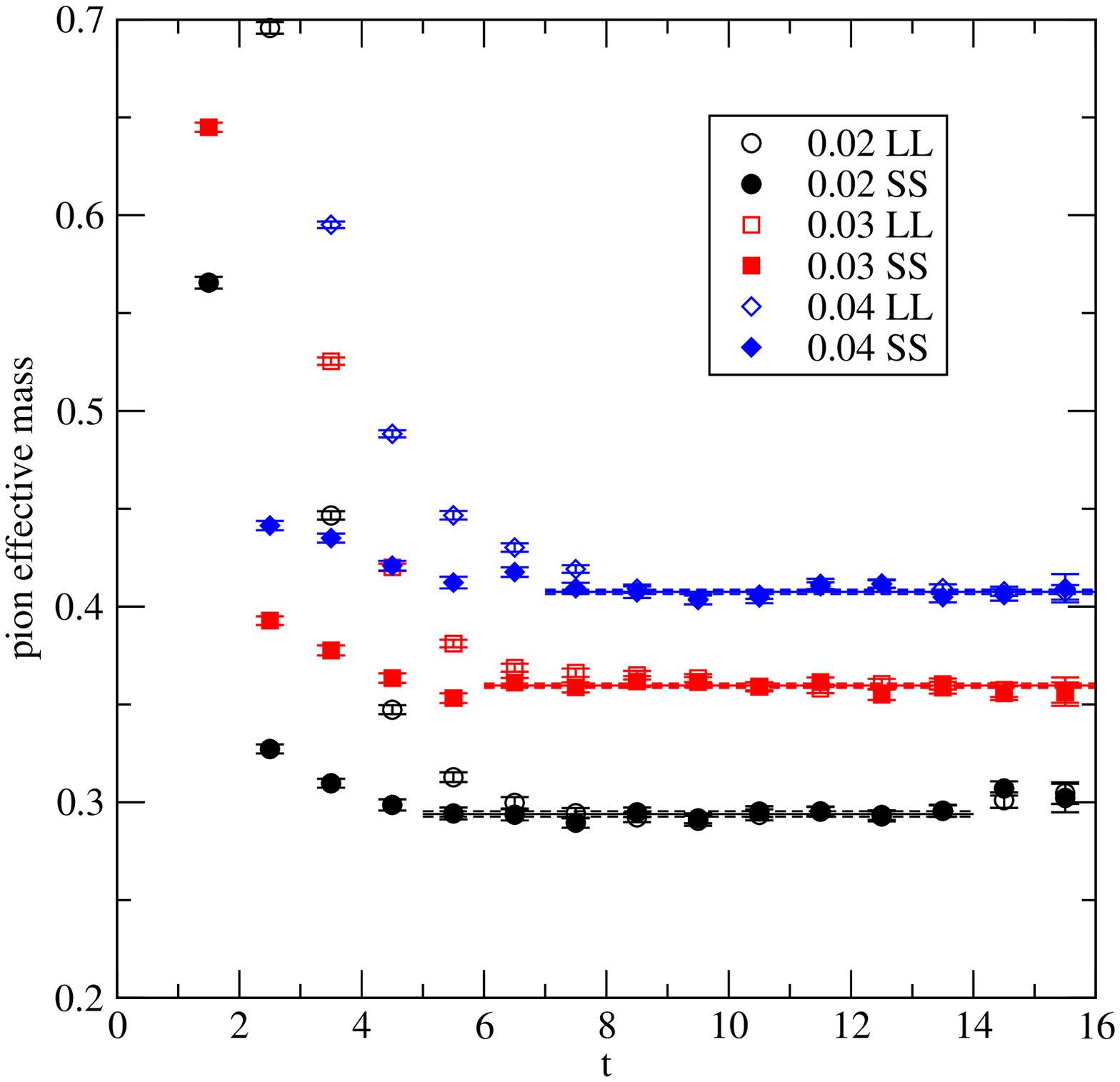}
\includegraphics[angle=-00,scale=0.37,clip=true]{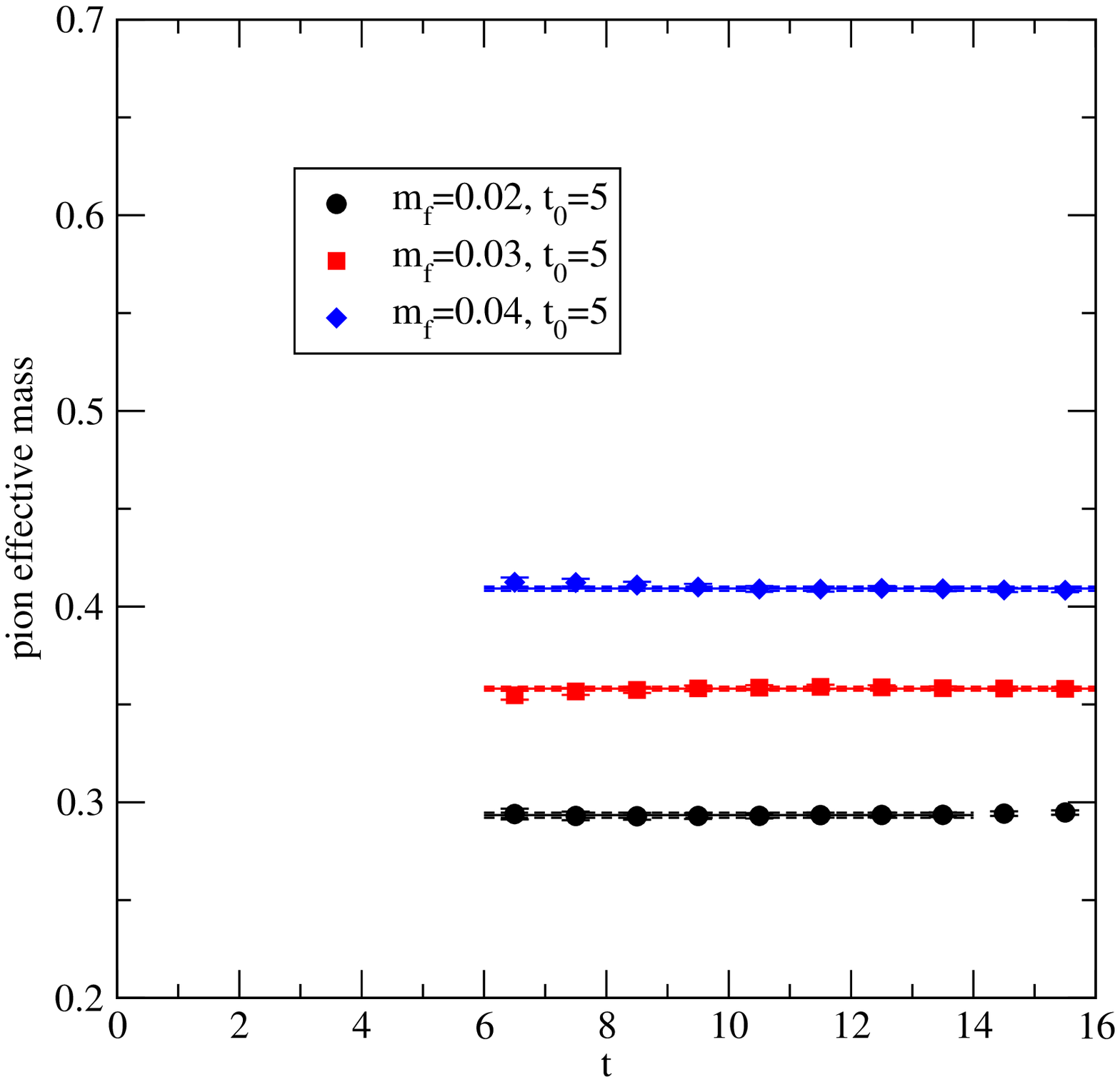}
\includegraphics[angle=-00,scale=0.37,clip=true]{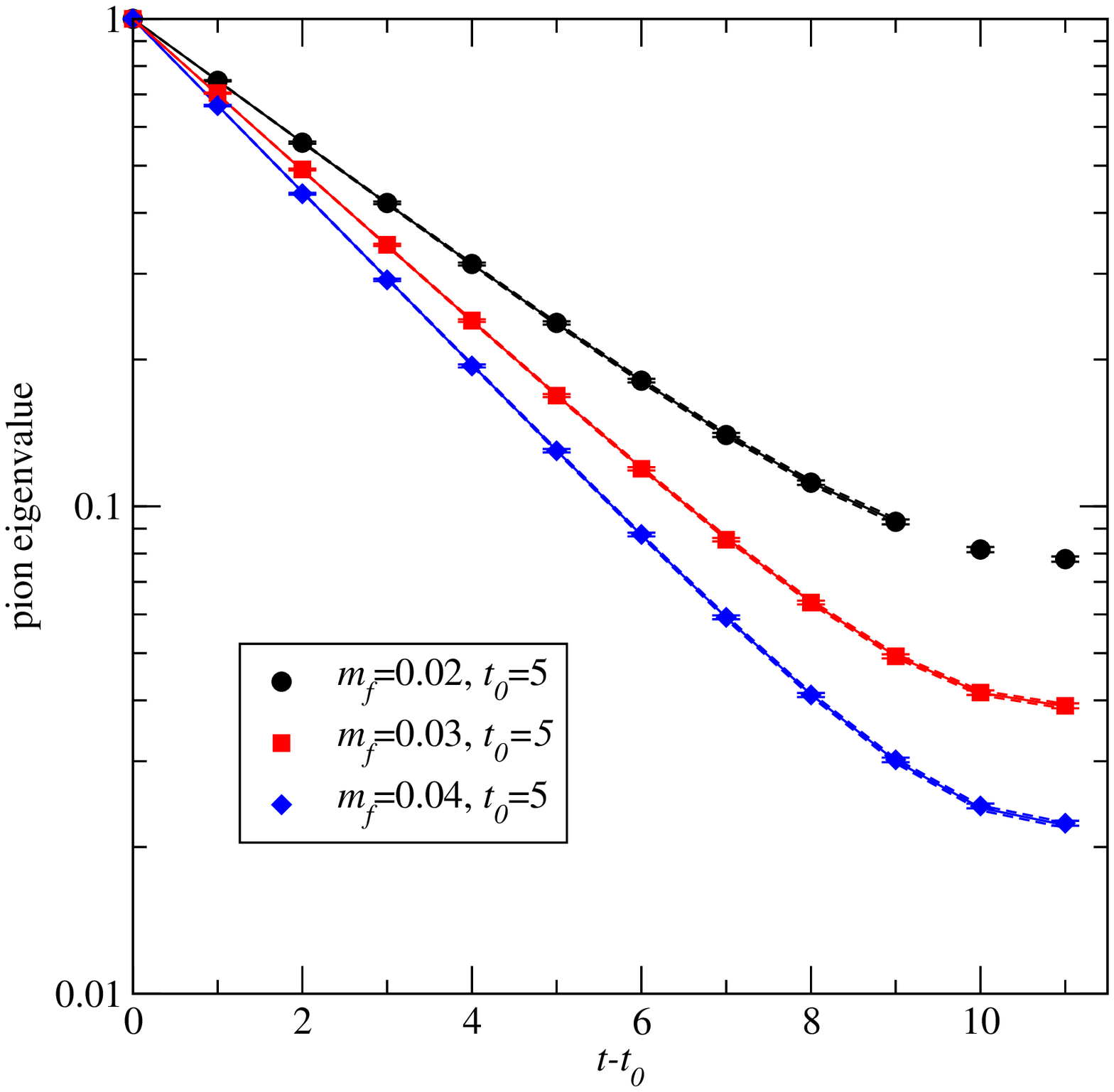}
\end{center}
\caption{
Pion effective mass vs. $t$  using method (A) (left) and method (B) (right), and pion eigenvalue vs. $t-t_0$ (bottom). 
Lines show fit values, errors and ranges. 
}
\label{fig:pion-a}
\end{figure}

In Fig. \ref{fig:pion-a}, we plot the effective mass of a pseudoscalar meson 
obtained by  method (A) on the left, that obtained by method (B) on the right, and
the ground-state eigenvalue obtained using method (B) on the bottom panel. 
Table \ref{tab:pion-fit} summarizes the values of the pion mass 
obtained by both methods.
By using 5-10 times more statical samples than in the previous analysis
and extracting the ground-state information from the meson propagator 
over shorter time distance, which becomes possible  using smeared operators, 
the statistical errors decrease to approximately half of those in 
the  previous results.
The fact that the reduction of the error size is closer to or even larger than
that expected from  the increase in the number of statistical samples,
$1/2 > 1/\sqrt{\text{5-10}}$,  suggests that the smearing itself does not 
necessarily cause the smaller statistical error for a pseudoscalar meson.
Rather, we  could determine the extent of the excited-state contamination
using smeared operators with different overlaps with the states.
In fact, our new results are consistent within statistical error with
the previous results. We will discuss the decay constant and 
the excited-state meson, $\pi^*$, later.

\begin{table}[t]
\caption{%
$m_{\pi}$
}
\label{tab:pion-fit}
\begin{center}
\begin{tabular}{cccccc}\hline \hline
$m_{f}$ & $m_{\pi}$ & $t_{0}$ &
$t_{\rm min}$ & $t_{\rm max}$ & method \cr
  \hline
0.02 & 0.2940(14) & & 5 & 14 & (A) \cr
& 0.2934(13) & 5 & $t_0+1$ & 14 & (B) \cr
& 0.2902(28) $^{\rm a}$ & & 9 & 16 & (A) \cr
\hline  
0.03 & 0.3596(11) & & 6 & 16 & (A) \cr
& 0.3581(10) & 5 & $t_0+1$ & 16 & (B) \cr
& 0.3575(19) $^{\rm a}$ & & 9 & 16 & (A) \cr
\hline
0.04 & 0.4075(11) & & 7 & 16 & (A) \cr
& 0.4092(11) & 5 & $t_0+1$ & 16 & (B) \cr
& 0.4094(25) $^{\rm a}$
& & 9 & 16 & (A) \cr
\hline 
\end{tabular}
\end{center}
\footnotesize{$^{(a)}$ These values are obtained from 
$(I,J)=(L,W)$ correlators and quoted by \cite{Aoki:2004ht}.}
\end{table}

\subsection{Mass of $a_0$}
From experiments, there are two flavor-non-singlet scalar mesons,
$a_0(980) $ and $a_0(1450)$, in nature. 
Although these are unstable particles in the more realistic $N_f=2+1$ case,
we assume a stable one-particle state to be the ground state 
in the scalar meson sector in our $N_f=2$ case with a relatively heavy 
quark and small space-time.
$a_0$ meson spectrum results previously obtained by lattice QCD calculation 
seem to fall roughly into two categories\cite{Liu:2007hma}, 
studies reporting lighter 
masses of $\sim 1$ GeV\cite{Hart:2002sp,McNeile:2006nv} and those reporting 
heavier masses $\sim 1.5$ GeV \cite{Bardeen:2001jm,Prelovsek:2004jp, Burch:2006dg,Liu:2007hma}.
Previous RBC results\cite{Prelovsek:2004jp} are 
$m_{a_0}=1.58(34)$ GeV by the analysis of unitary points and 
1.51 (19) GeV by partially quenched analysis.

\begin{figure}[t]
\begin{center}
\includegraphics[angle=-00,scale=0.37,clip=true]{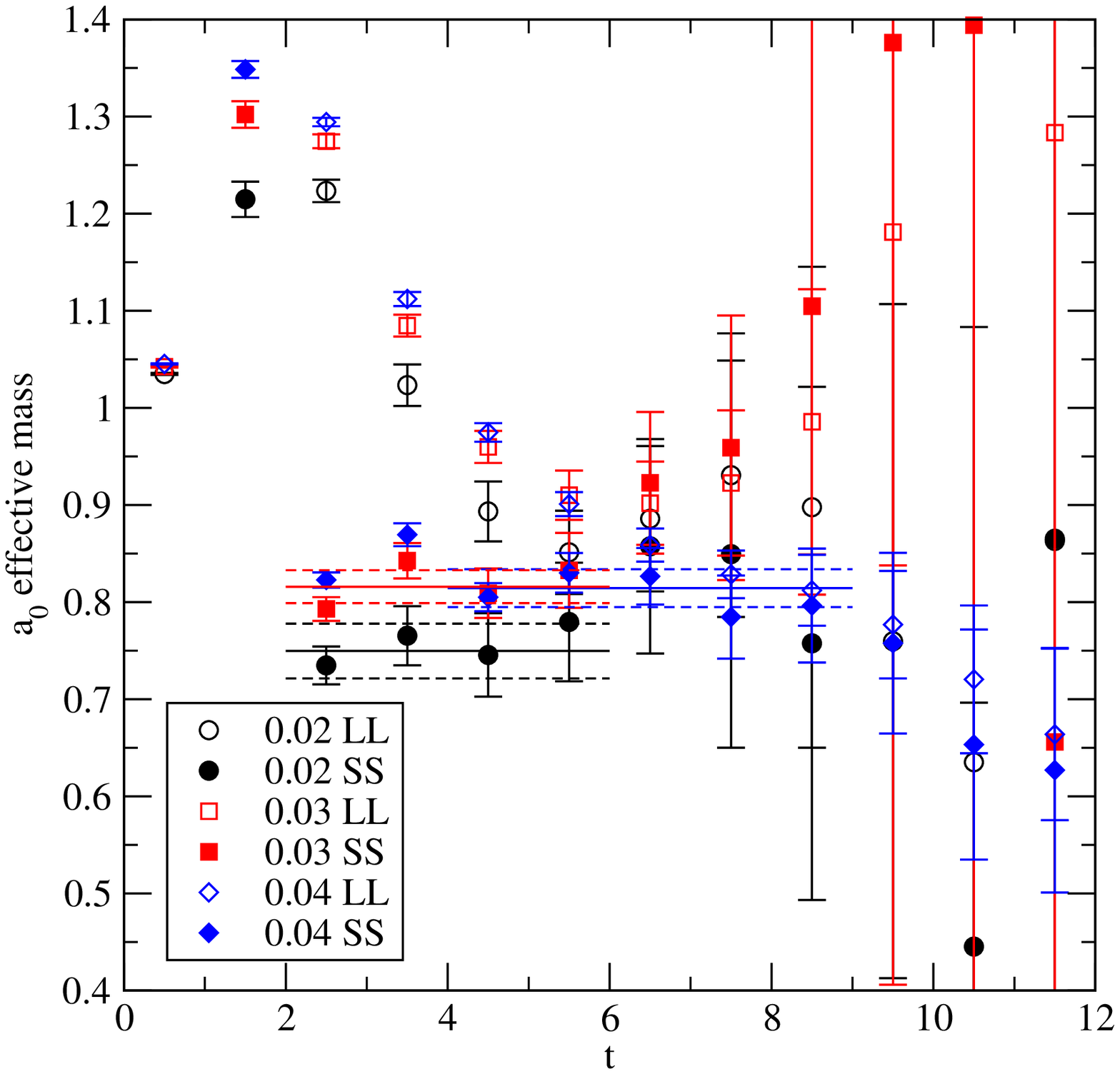}
\includegraphics[angle=-00,scale=0.37,clip=true]{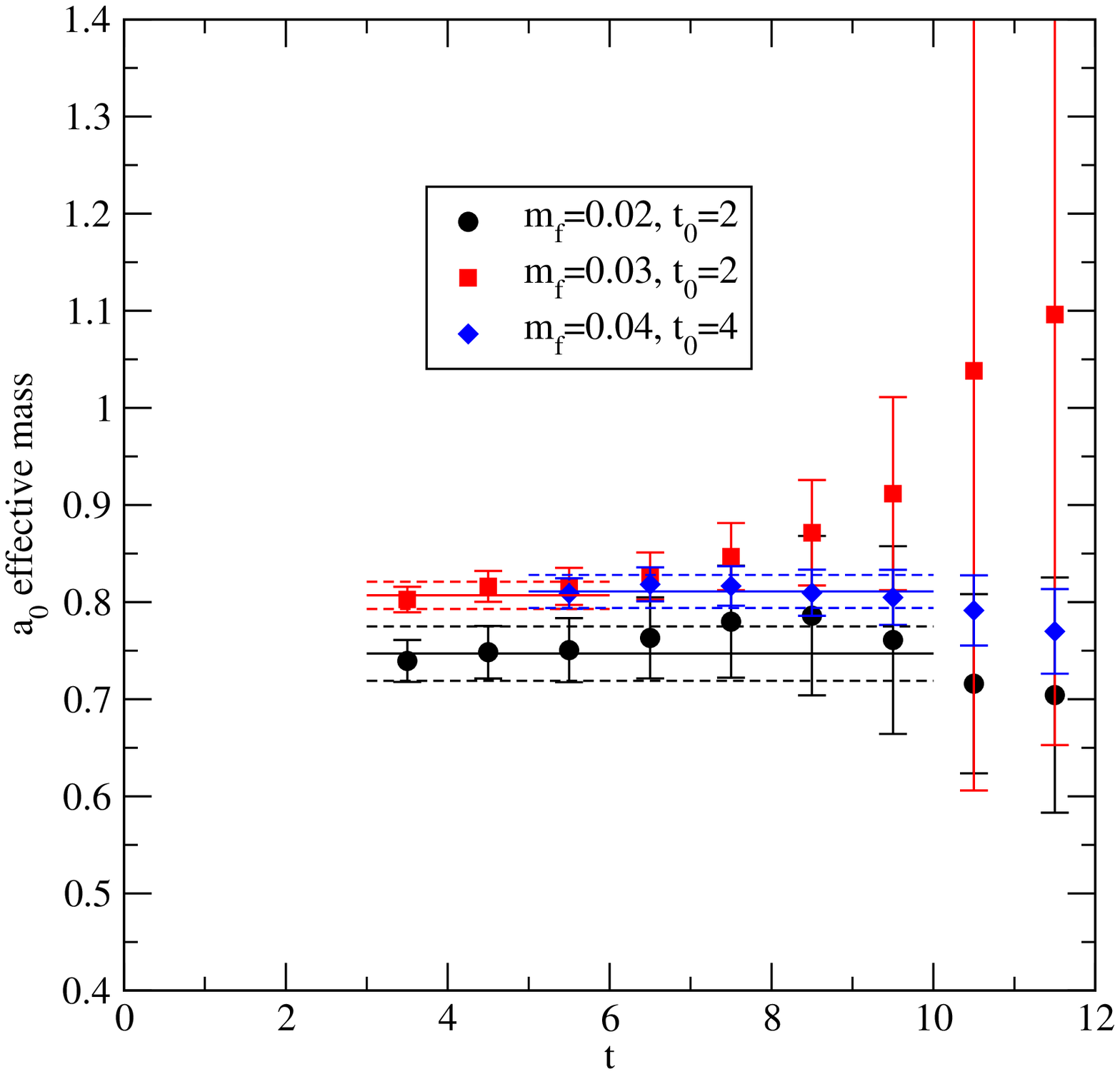}
\end{center}
\caption{
Effective mass of $a_0$ vs $t$ using method (A) (left) and method (B) (right).
Lines show fitted values, errors, and ranges. 
}
\label{fig:a0-a}
\end{figure}

Fig.~\ref{fig:a0-a} (left) shows the effective mass of $a_0$, 
Fig.~\ref{fig:a0-a} (right) shows the eigenvalue of the ground-state 
using the variational method (B). The numerical values are listed in 
Table \ref{tab:a0-fit}, in which we also quote the previous RBC
values\cite{Prelovsek:2004jp}.

Our new results for the mass of $a_0$ are significantly lighter than those of
previous results, as shown in  Table \ref{tab:a0-fit}.
Since the QCD ensemble used in both investigations is the same,
this discrepancy must originate from the difference in measuring the
meson operator. In the previous calculation the meson interpolation field
was constructed from quark fields at a point. 
Although the point operator was convenient for theoretical investigation
in the previous study, it is not necessarily optimal for extracting the ground state.
In fact, as shwon in the left panel of Fig. \ref{fig:a0-a}, the effective 
mass of the point operator (open symbols) is very large at a short distance, which 
implies a large amount of excited-state contamination in the point operator.
On the other hand, the effective mass obtained using the smeared operator  (filled symbols) reaches plateaux earlier in time and coincides to that obtained from
the ground-state eigenvalue by the variational method, shown in right panel.
Note that  the size of the statistical sample is increased by a factor
of five or more in this work compared with that in the previous report.

\begin{table}[t]
\caption{%
$m_{a_0}$
}
\label{tab:a0-fit}
\begin{center}
\begin{tabular}{cccccc}\hline \hline
$m_{f}$ & $m_{a_0}$  & $t_{0}$ 
& $t_{\rm min}$ & $t_{\rm max}$  
& method \cr
\hline
0.02 & 0.750(28) $^{\rm a}$ & & 2 & 6 & (A) \cr
& 0.747(28) & 2 & $t_0+1$ & 10 & (B) \cr
& 0.92(9) $^{\rm b}$ & & 4 & 10 & exponential fit \cr
\hline  
0.03 & 0.816(17) $^{\rm a}$ & & 2 & 6 & (A) \cr
& 0.807(14) & 2 & $t_0+1$ & 6 & (B) \cr
& 0.99(10) $^{\rm b}$ & & 5 & 10 & exponential fit \cr
\hline
0.04 & 0.814(19) $^{\rm a}$
& & 4 & 9 & (A) \cr
& 0.811(17) & 4 & $t_0+1$ & 10 & (B) \cr
& 0.94(5) $^{\rm b}$
& & 5 & 12 & exponential fit \cr
\hline 
\end{tabular}
\end{center}
{\footnotesize$^{(a)}$ These values are obtained by uncorrelated fitting.}
{\footnotesize$^{(b)}$ These values are obtained from 
$(I,J)=(L,L)$ correlators and quoted in \citen{Prelovsek:2004jp}.} 
\end{table}

\begin{figure}[t]
\begin{center}
\includegraphics[angle=-00,scale=0.40,clip=true]{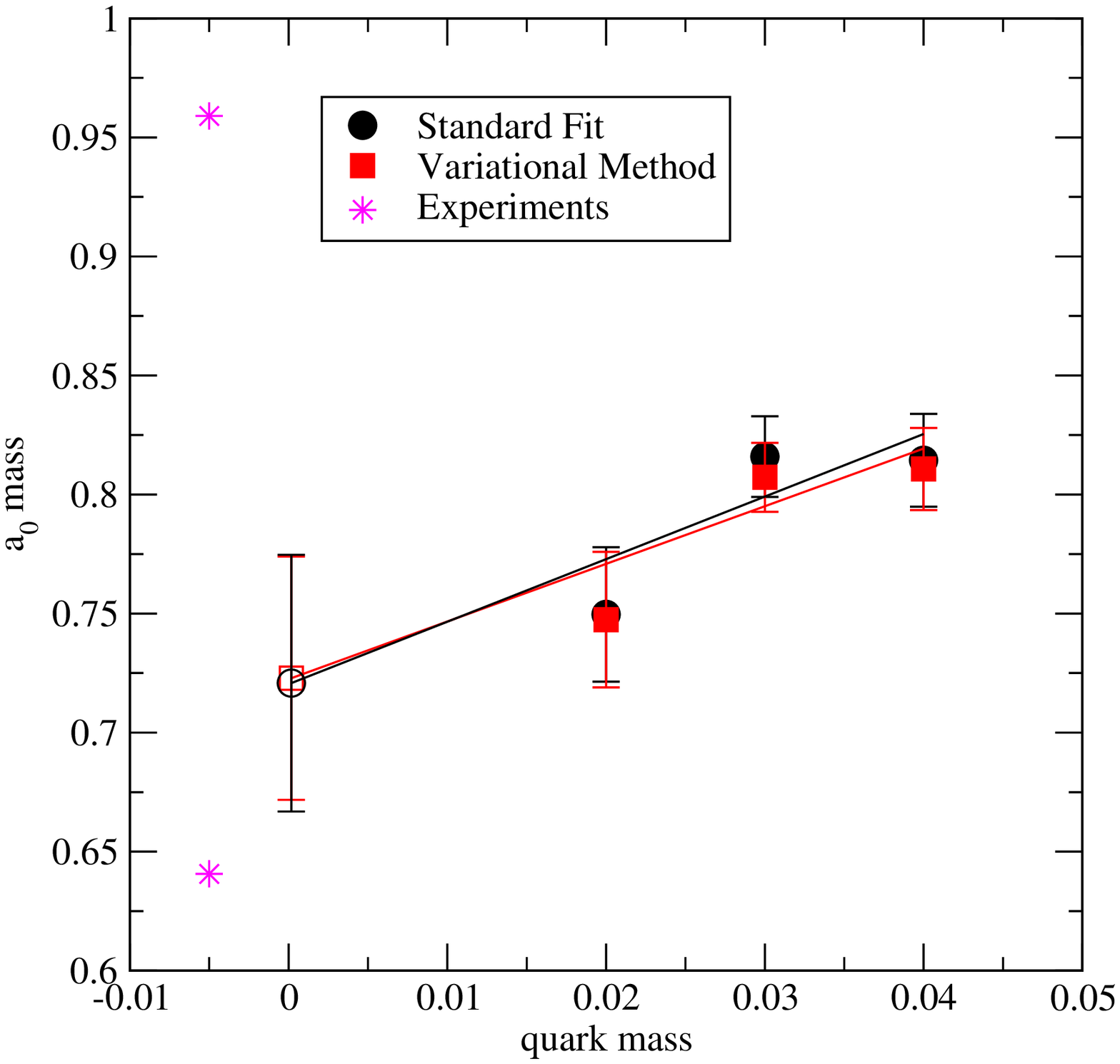}
\end{center}
\caption{$m_{a_0}$ vs $m_f$. 
The asterisk on the left shows the experimental values\cite{Yao:2006px}.}
\label{fig:a0-chiral}
\end{figure}

Fig. \ref{fig:a0-chiral} shows the results of extrapolation by linear fitting 
and Table \ref{tab:a0-chiral} shows $m_{a_0}$ 
at the physical quark mass point. 
Since both methods (A) and (B) are consistent with each other 
we choose 
\begin{eqnarray} 
m_{a_{0}}^{\text{phys}}=1.111(81)  \ \text{GeV} 
\end{eqnarray}
from method (B) as our final value in this work.

\begin{table}[t]
\caption{%
$m_{a_0}$ at the physical quark mass point $(m_f= m_{u,d})$.
}
\label{tab:a0-chiral}
\begin{center}
\begin{tabular}{cccc}\hline \hline
$m_{a_0}$ & $m_{a_0}^{\text{phys}}$ [MeV] 
& $m_{a_0} r_0$
& method \cr
\hline
0.721(54) & 1,108(85) & 3.08(23) & (A) \cr
0.723(51) & 1,111(81) & 3.10(22) & (B) \cr
\hline 
\end{tabular}
\end{center}
\end{table}

To clarify the discrepancies among results obtained from lattice calculations and 
experiments, further investigations including the calculation
multiparticle scattering states and 
the strange sea quark effects ($N_f=2+1$) are
needed.

\subsection{Mass of $\eta'$}
\begin{figure}[ht]
\begin{center}
\includegraphics[angle=-00,scale=0.40,clip=true]{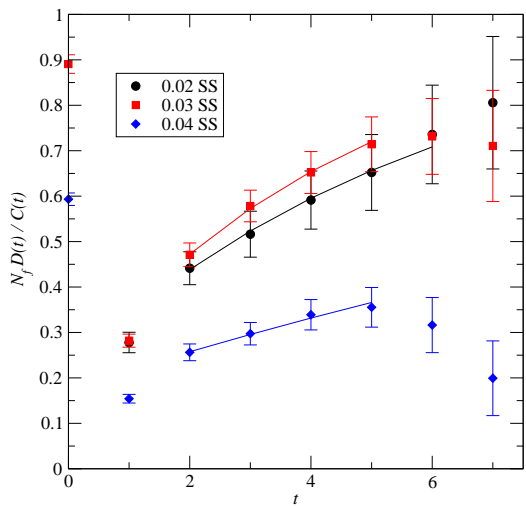}
\end{center}
\caption{$N_f D_{\gamma_5}(t)/C_{\gamma_5}(t)$ vs $t$.}
\label{fig:eta-ratio}
\end{figure}

Before presenting the mass spectrum results of  the flavor singlet pseudoscalar meson, $\eta'$, we check whether the theoretical expectation discussed in 
\S\ref{sec:theoretical_expectation} is realized.
The ratio of the correlation function between disconnected quark loops, $D_{\gamma_5}(t)$,
to the connected correlation function, $C_{\gamma_5}(t)$, was shown to 
approach  unity for a large time separation, which is  clearly different
from the expectation of the linear growth in the quenched QCD case
(\ref{eq:eta_ratio_quenched}).
In the discussion, only the pion and $\eta'$ states were considered when coupling to
the $I(J^P)=0(0^-)$ operator, leading  to
\begin{eqnarray}
{N_f D_{\gamma_5}(t) \over C_{\gamma_5}(t)} = 1 - B{e^{-m_{\eta'}t} + e^{-m_{\eta'}(T-t)} \over e^{-m_{\pi}t} + e^{-m_{\pi}(T-t)} } 
\stackrel{ (T-t) \gg 1 }{\longrightarrow}
1-B~e^{-(m_{\eta'}-m_\pi)t}. 
\label{eq:eta_ratio_again}
\end{eqnarray}
In Fig.~\ref{fig:eta-ratio}, the ratio extracted using the smeared operator,
$\eta'_S$, is plotted. Indeed the ratio asymptotically approaches  one
for the two lighter quark masses (circles, squares), 
although it is statistically uncertain at a large time distance.
However the heaviest quark mass point (diamonds) seems to approach  to a value lower
than one.
The mass difference between $\eta'$ and $\pi$ is smaller
for the heavier quark mass, and is close to zero for the heaviest 
quark mass (as we will discuss later); thus, the ratio only approaches unity
at a very large $t$ from (\ref{eq:eta_ratio}). 
Moreover, this deviation from the simplest theoretical explanation might be 
due to the omission of the excited  states
such as the $\pi^*$ or $0^{-+}$ glueball state, which may play a more significant
role in the heavier-quark-mass region. 
From the current results, we can not conclude whether the deviation from 
unity for the heaviest quark can be explained by the above-mentioned arguments
or is due to other reasons, for example, insufficient sampling of the different topological
sectors.

\begin{figure}[t]
\begin{center}
\includegraphics[angle=-00,scale=0.37,clip=true]{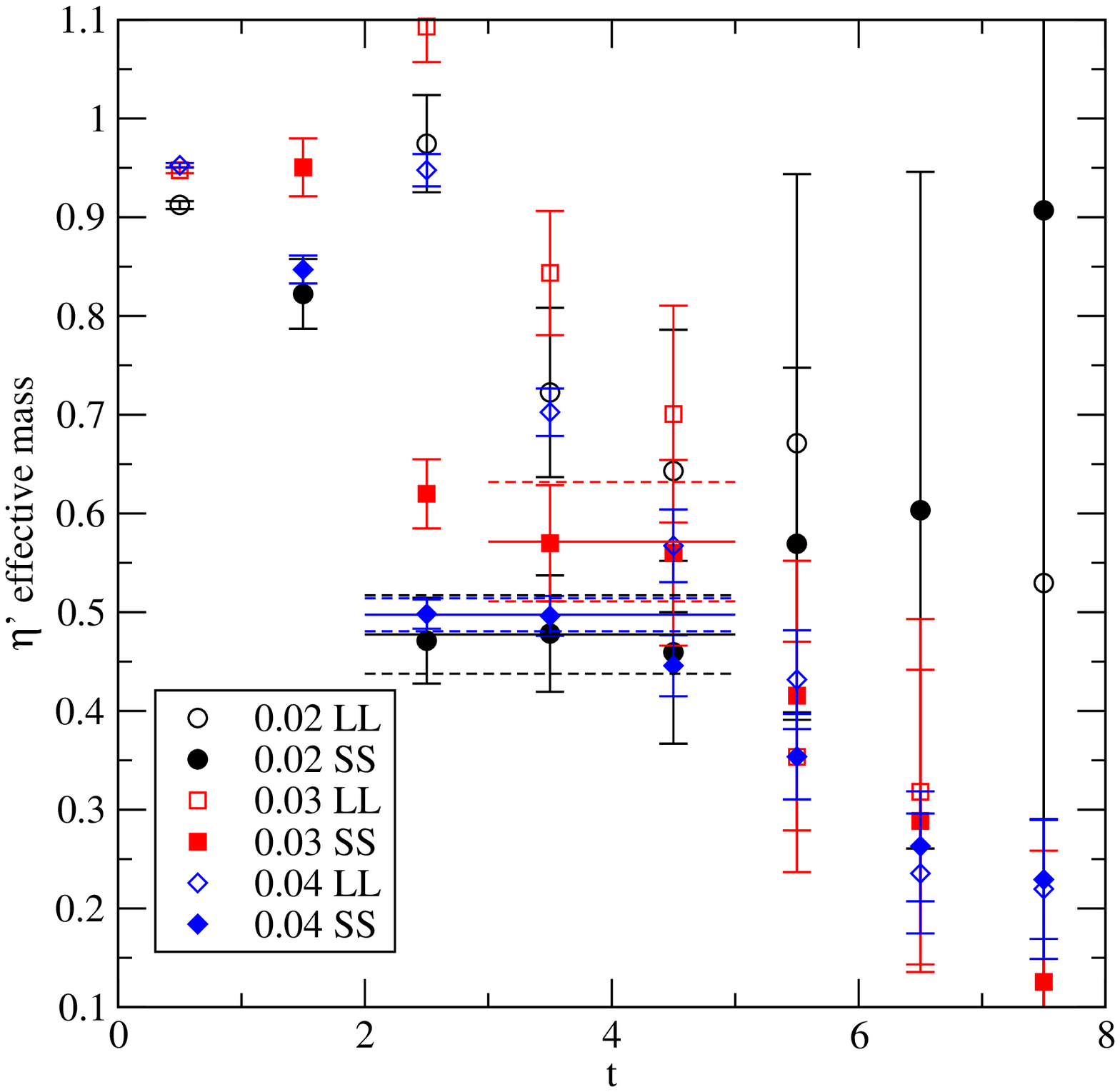}
\includegraphics[angle=-00,scale=0.37,clip=true]{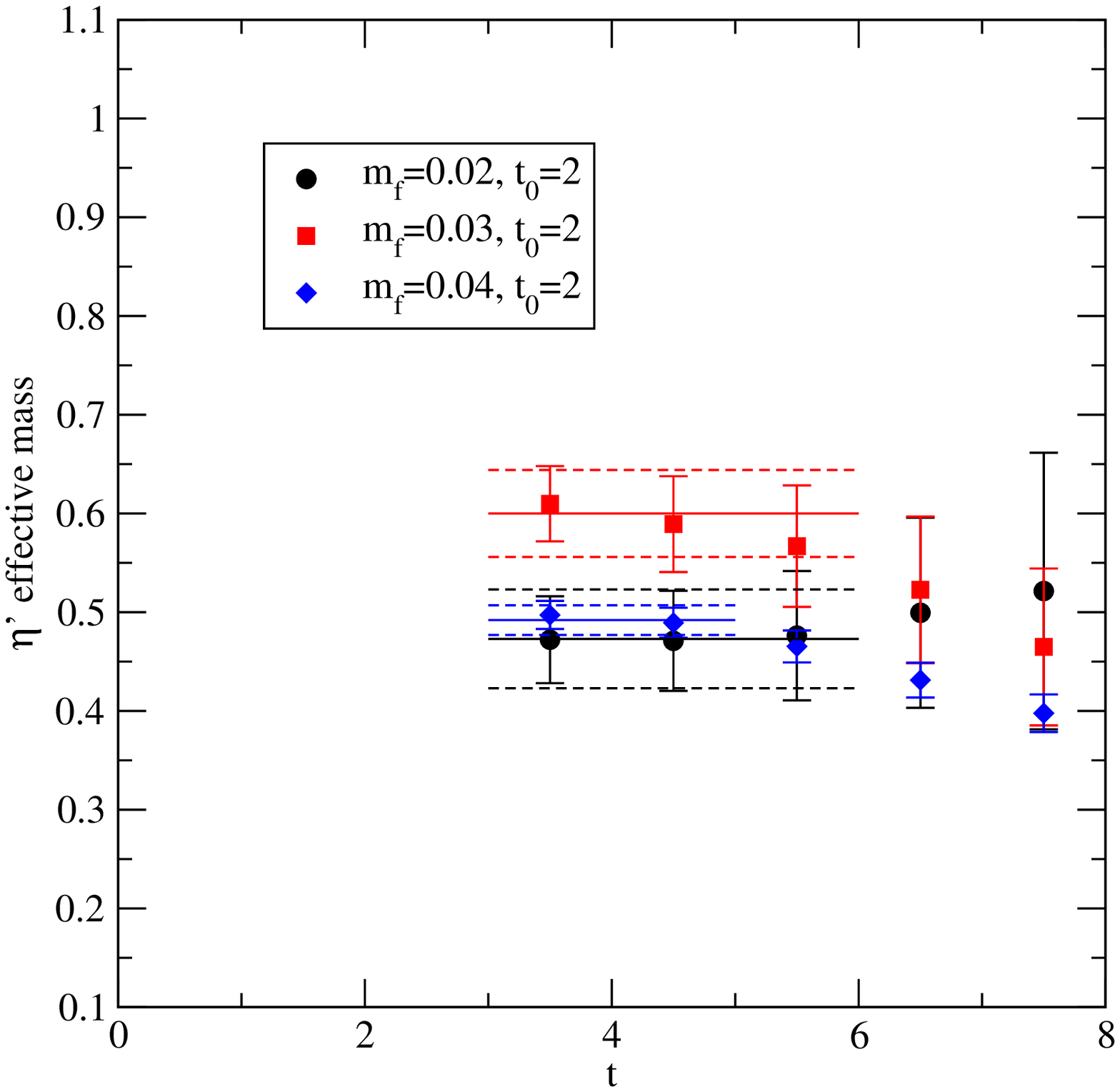}
\includegraphics[angle=-00,scale=0.37,clip=true]{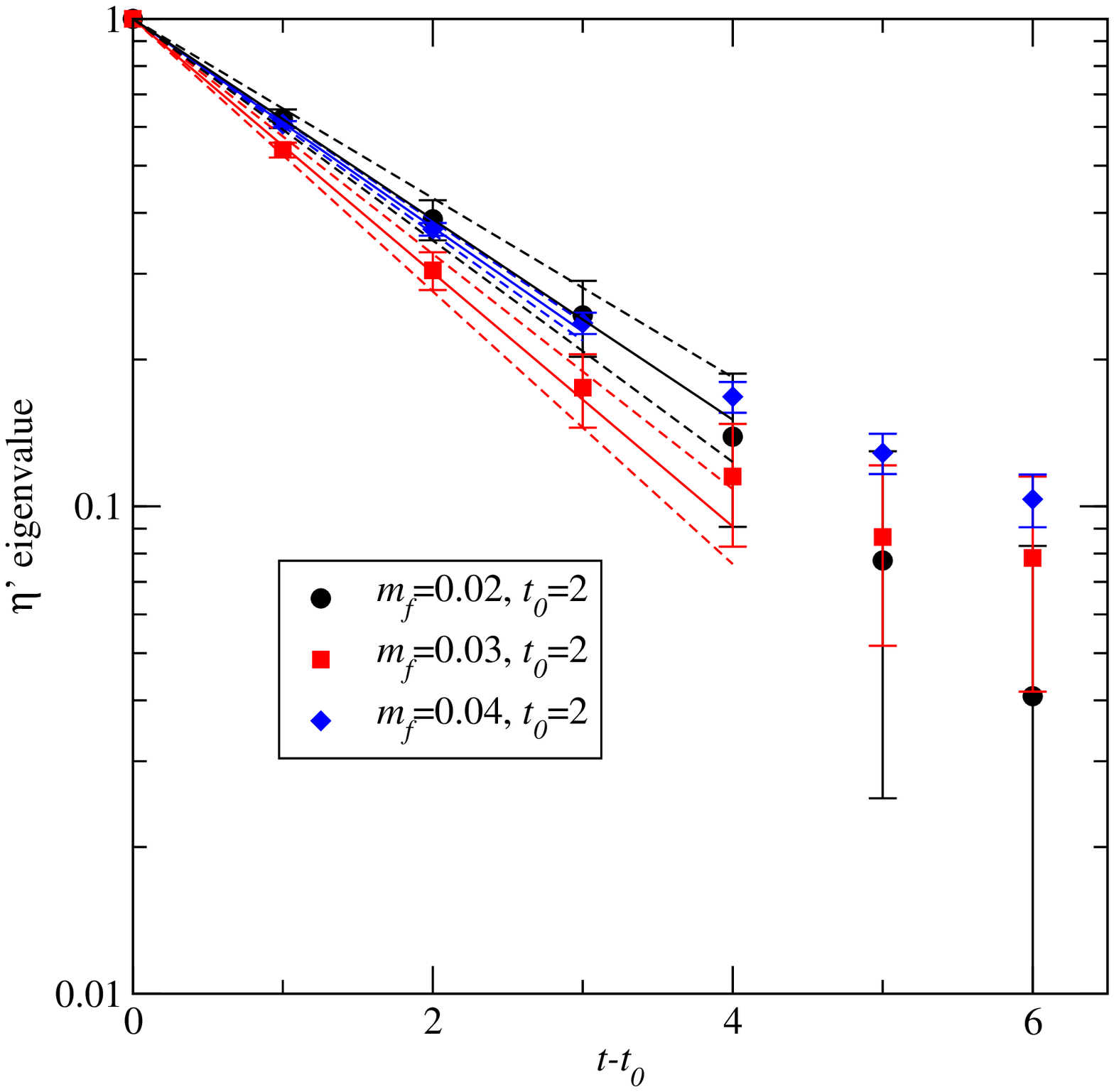}
\end{center}
\caption{
Effective mass of $\eta'$ vs $t$  using method (A) (left) and method (B) (right), 
and $\eta'$ eigenvalue vs $t-t_0$ (bottom). 
}
\label{fig:eta-a}
\end{figure}

We now describe the $\eta'$ spectrum obtained using methods (A) and (B).
Figure \ref{fig:eta-a} shows the effective mass (left: method (A), right: method (B)), and the ground-state eigenvalue, 
and their numerical values and
their fitted ranges are given in Table \ref{tab:eta-fit}.
We did not use a propagator from longer distance, where
the statistics are too poor and the standard error analysis 
would not be reliable, although the inclusion of a few more data points
does not change the fitted results for most of the masses.
Method (B) produces flatter plateaux than  method (A) 
for this meson.

As a consistency check, we also examined the temporal 
exponent of the ratio (\ref {eq:eta_ratio}) to extract the mass of $\eta'$. 
We have evaluated the effect of the finiteness of the lattice in the temporal direction
by using the fitting formula (\ref{eq:eta_ratio_1}), and found the results
to be unchanged. 
Combining the measured pion mass 
in Table \ref{tab:pion-fit},  the values obtained are $m_{\eta'}=0.458(58),$ 0.571(48), and 0.461(15) for 
$m_f=0.02$, 0.03, and 0.04, respectively. 
These estimations are slightly smaller than 
the results in Table \ref{tab:eta-fit}. 
One reason for this may be that the time range 
used in fitting the ratio is too short and the pion mass is overestimated,
which causes the estimation for the mass of $\eta'$ to be smaller than its
actual value. 
Because of this possibility,  we won't use the results obtained from the ratio fitting
in our main results.

\begin{figure}[t]
\begin{center}
\includegraphics[angle=-00,scale=0.37,clip=true]{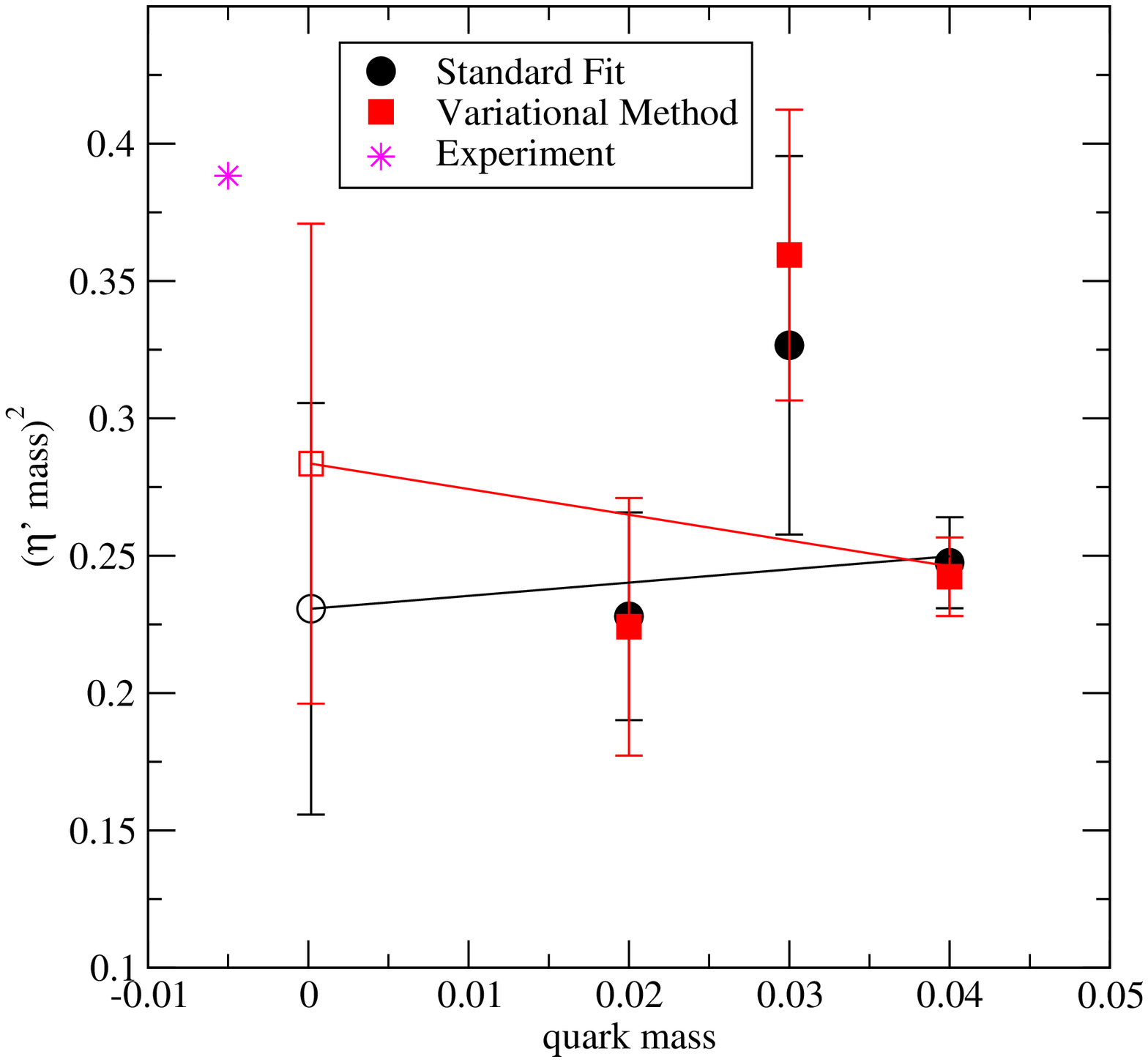}
\includegraphics[angle=-00,scale=0.37,clip=true]{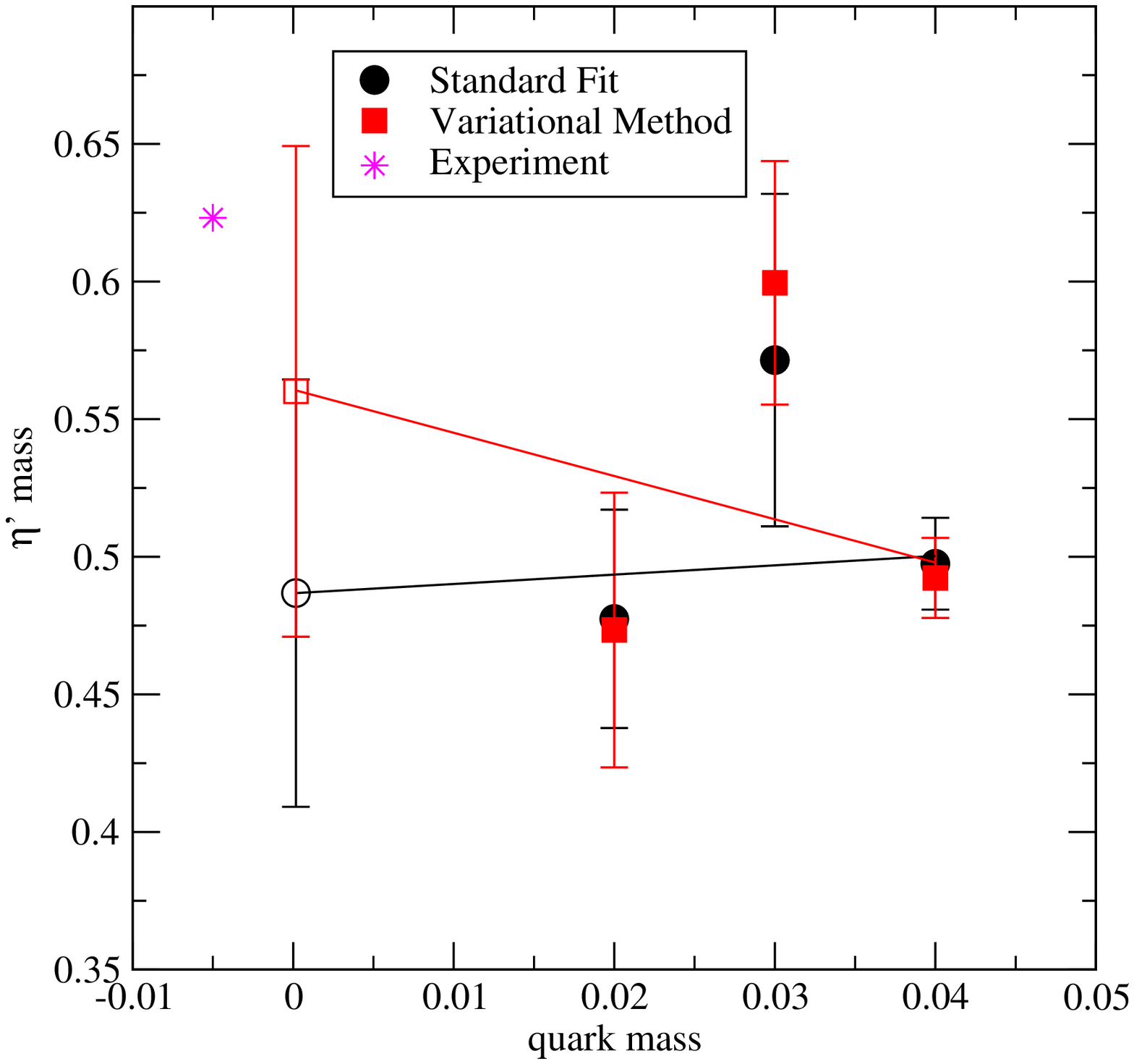}
\end{center}
\caption{
$m_{\eta'}^2$ vs $m_f$ (left), and $m_{\eta'}$ vs $m_f$ (right).
The open circles and squares are extrapolated values at the physical
quark mass point using (\ref{eq:chiral_sqrt_formula}) and (\ref{eq:chiral_lin_formula}),
respectively.
The asterisk on the left shows the experimental values\cite{Yao:2006px}.
}
\label{fig:eta-chiral}
\end{figure}

The mass of $\eta'$ has only slight dependence on the quark mass, as shown in 
Fig.~\ref{fig:eta-chiral}: all three masses are consistent within 
two to three standard deviations of statistical error. Their central value fluctuates 
nonmonotonically in quark mass order. Before being convinced of this 
nonmonotonicity, we should question the reliability of the error estimation 
and other systematic uncertainties such as  insufficient sampling over the topological charge since 
$\eta'$ is likely to depend strongly on the topological charge strongly.
In our simulation we use DBW2 gauge action to reduce the size of the 
residual chiral symmetry breaking, $\mres$, sacrificing the configuration
mobility among different topological sectors to some extent.

\begin{table}[t]
\caption{%
$m_{\eta'}$.
}
\label{tab:eta-fit}
\begin{center}
\begin{tabular}{cccccc}\hline \hline
$m_{f}$ & $m_{\eta'}$  & $t_{0}$ 
& $t_{\rm min}$ & $t_{\rm max}$  
& method \cr
\hline
0.02 & 0.477(40) & & 2 & 5 & (A) \cr
& 0.473(50) & 2 & $t_0+1$ & 6 & (B) \cr
\hline  
0.03 & 0.571(60) & & 3 & 5 & (A) \cr
& 0.600(44) & 2 & $t_0+1$ & 6 & (B) \cr
\hline
0.04 & 0.497(17) & & 2 & 5 & (A) \cr
& 0.492(15) & 2 & $t_0+1$ & 5 & (B) \cr
\hline 
\end{tabular}
\end{center}
\end{table}

Although the quark mass dependence has not been resolved sufficiently clearly,
we extrapolate the measured masses by the 
eqs. (\ref{eq:chiral_sqrt_formula}) 
and (\ref{eq:chiral_lin_formula}) to estimate 
the mass of $\eta'$ at the physical quark mass point.
The results are shown 
in Fig. \ref{fig:eta-chiral} and Table \ref{tab:eta-chiral}.
The central values of the estimation differ from each other by 15\%
but are within statistical error.
Our main estimation for the mass of $\eta'$ at the physical quark mass point
is obtained from the variational method (B) and chiral extrapolation using the
lowest order of ChPT (\ref{eq:chiral_sqrt_formula}), and is given by
\begin{eqnarray} 
m_{\eta'}^{\text{phys}}=819(127) \ \text{MeV}~.
\end {eqnarray}
This is the first estimation of the mass of $\eta'$
performed with the two flavors of  a dynamical (approximately) chiral fermion,
which is certainly heavier than a pion, which is thought to be related to 
the  chiral $U(1)_A$ anomaly.
Apart from the large statistical error and the various systematic errors
discussed above, the main results are close to the experimentally obtained mass of $\eta'$,
which suggests that further improvements can be made by especially calculation 
using an $N_f=2+1$ ensemble.

\begin{table}[t]
\caption{%
$m_{\eta'}$ at the physical quark mass point $(m_f= m_{u,d})$.
}
\label{tab:eta-chiral}
\begin{center}
\begin{tabular}{cccc}\hline \hline
$m_{\eta'}$ & $m_{\eta'}^{\text{phys}}$ [MeV] 
& $\metap r_0$ 
& method and chiral extrapolation \cr
\hline
0.480(78) & 738(121) & 2.05(33) & (A) AWTI type (\ref{eq:chiral_sqrt_formula}) \cr 
0.487(78) & 748(120) & 2.08(33) & (A) linear type (\ref{eq:chiral_lin_formula})\cr
0.532(82) & 819(127) & 2.28(35) & (B) AWTI type (\ref{eq:chiral_sqrt_formula})\cr
0.560(89) & 862(130) & 2.40(36) & (B) linear type (\ref{eq:chiral_lin_formula})\cr
\hline 
\end{tabular}
\end{center}
\end{table}

\subsection{Mass of $\omega$}

\begin{figure}[t]
\begin{center}
\includegraphics[angle=-00,scale=0.37,clip=true]{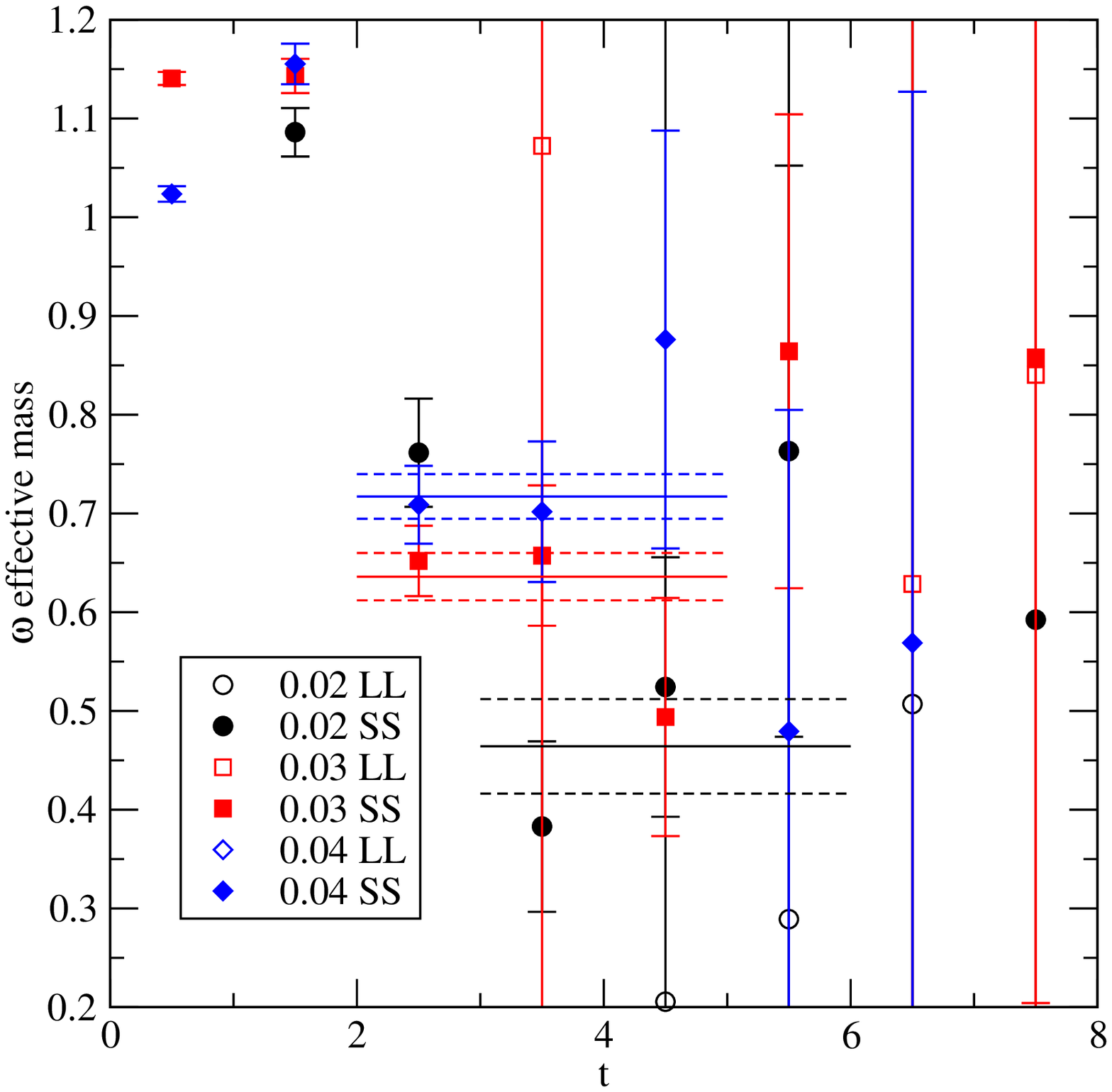}
\includegraphics[angle=-00,scale=0.37,clip=true]{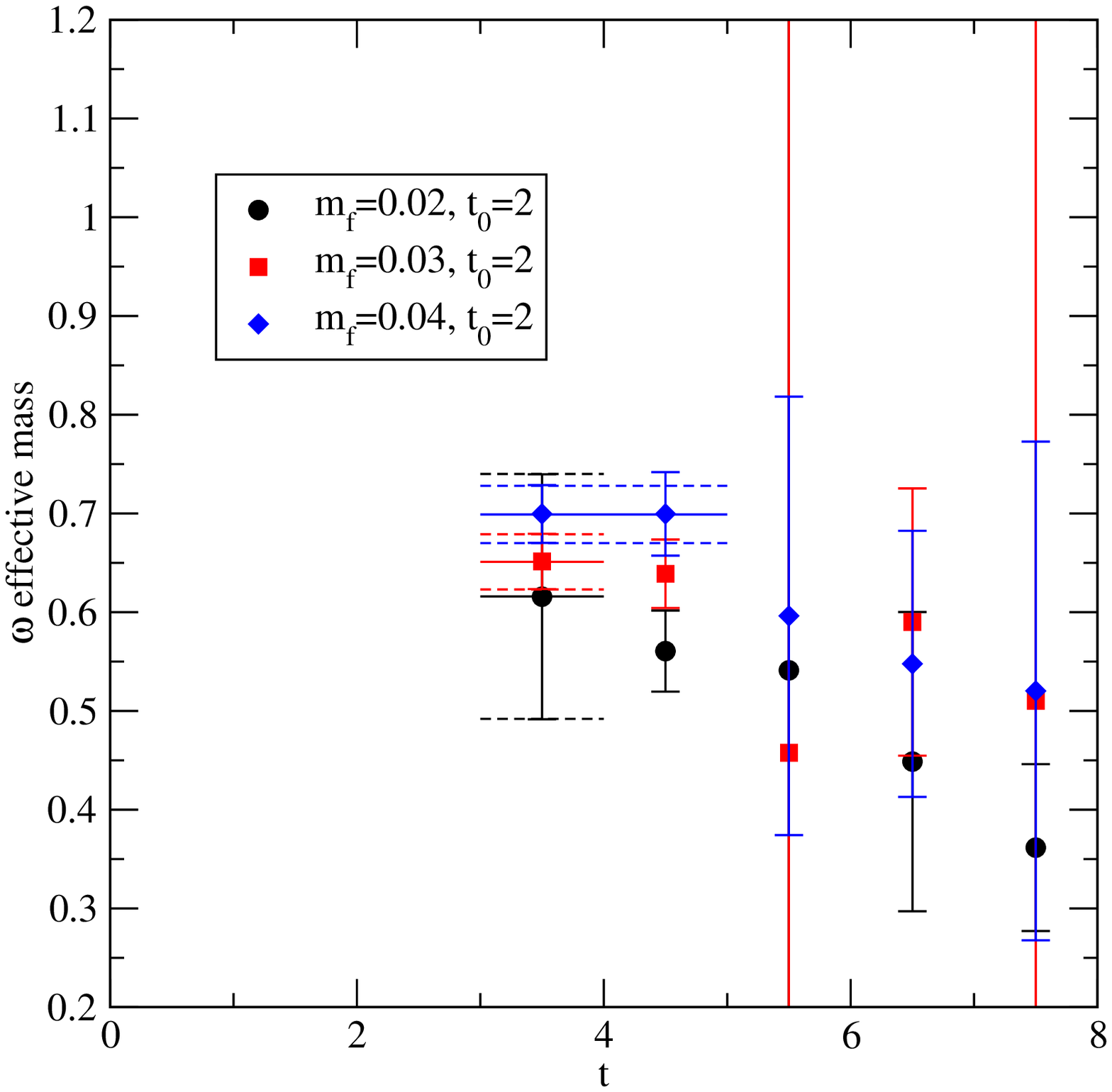}
\end{center}
\caption{
Effective mass of $\omega$ vs $t$ using method (A) (left) and method (B) (right).
}
\label{fig:omega-a}
\end{figure}

We also examine the flavor singlet vector meson, $\omega$,
using a similar procedure to that for $\eta'$. 
Fig.~\ref{fig:omega-a} shows the effective mass of $\omega$ (left: method (A), right: method (B)), 
which is also listed in Table \ref{tab:omega-fit}.
We are able to extract a non-zero signal, but from a shorter time distance; thus there
may be a significant distortion from the excited states. The results for the lightest 
point, $m_f=0.02$, has a particularly poor signal.

\begin{table}[t]
\caption{%
$m_{\omega}$.
}
\label{tab:omega-fit}
\begin{center}
\begin{tabular}{cccccc}\hline \hline
$m_{f}$ & $m_{\omega}$  & $t_{0}$ 
& $t_{\rm min}$ & $t_{\rm max}$  
& method \cr
\hline
0.02 & 0.464(48) & & 3 & 6 & (A) \cr
& 0.616(124) & 2 & $t_0+1$ & 4 & (B) \cr
\hline  
0.03 & 0.636(24) & & 2 & 5 & (A) \cr
& 0.651(28) & 2 & $t_0+1$ & 4 & (B) \cr
\hline
0.04 & 0.717(23) & & 2 & 5 & (A) \cr
& 0.699(29) & 2 & $t_0+1$ & 5 & (B) \cr
\hline 
\end{tabular}
\end{center}
\end{table}

\begin{figure}[t]
\begin{center}
\includegraphics[angle=-00,scale=0.40,clip=true]{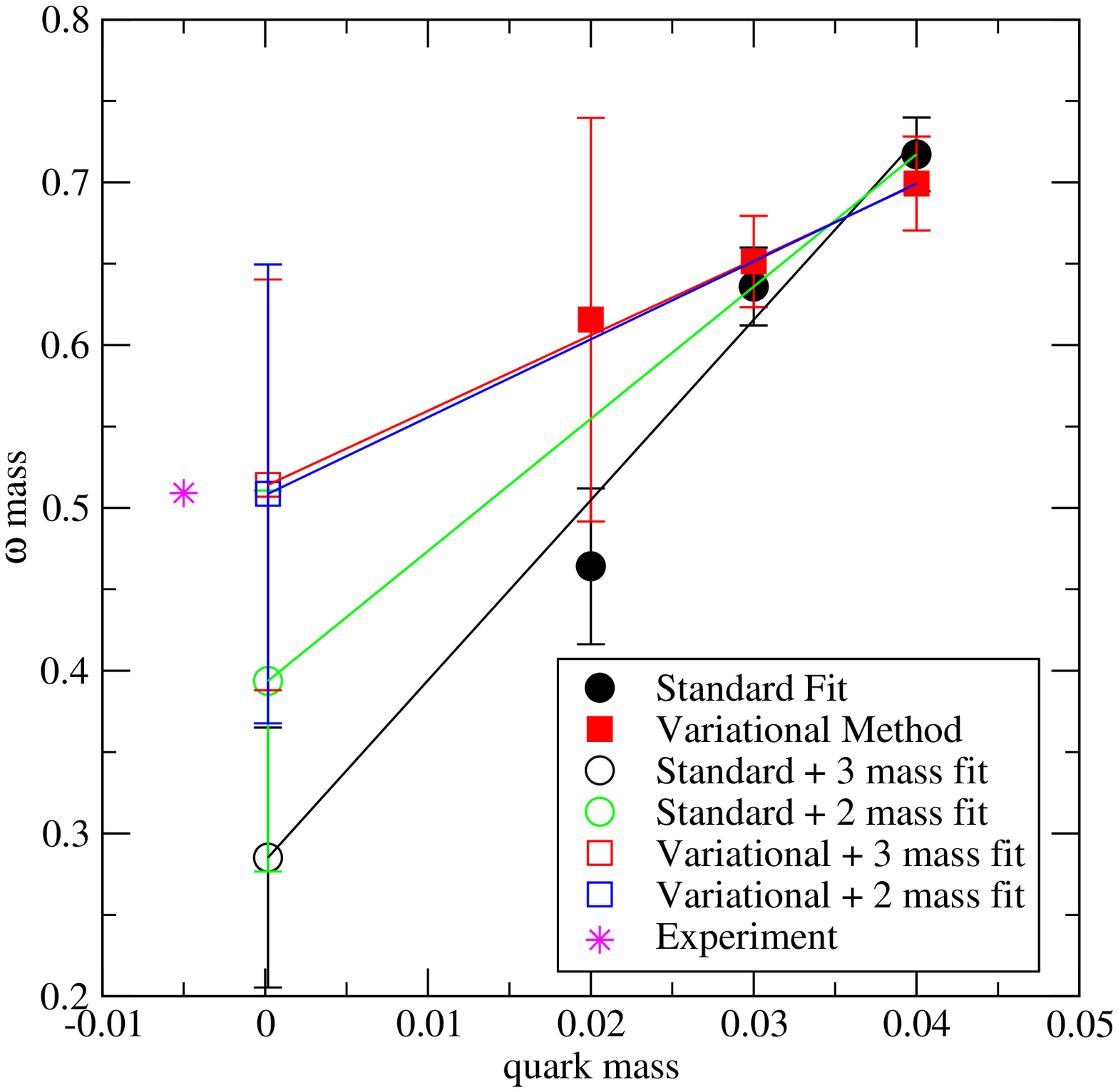}
\end{center}
\caption{$m_{\omega}$ vs $m_f$.
The asterisk on the left  shows the experimental value\cite{Yao:2006px}.
}
\label{fig:omega-chiral}
\end{figure}

We estimated the extrapolated mass at the physical quark mass point, as shown
in Fig.~\ref{fig:omega-chiral} and Table \ref{tab:omega-chiral}.
The fitting formula used is the linear extrapolation (\ref{eq:chiral_lin_formula}).
Since the statistical error for the lightest point $m_f=0.02$ is large,
as mentioned above, we examine two ways of the chiral extrapolation:
using all three masses or the heaviest two points. 
In Table \ref{tab:omega-chiral}, one can see that the results obtained from the 
``method (A) 3-masses fitting'' are significantly different from those of the 
``method (A) 2-masses fitting''.
At the physical quark mass point, $m_\omega$ is obtained from the "method (B) 3 masses fitting",
\begin{eqnarray}
m_{\omega}^{\text{phys}} = 790 (194) \ \text{MeV}~.
\end{eqnarray}
Our estimation for $\omega$ is consistent with the
experimental value, but with a large  statistical error $\sim$ 25\%.

\begin{table}[ht]
\caption{%
Estimation of $m_{\omega}$ at the physical quark mass point $(m_f= m_{u,d})$.
}
\label{tab:omega-chiral}
\begin{center}
\begin{tabular}{cccc}\hline \hline
$m_{\omega}$ & $m_{\omega}^{\text{phys}}$ [MeV] 
& $m_{\omega} r_0$
& fit method\cr
\hline
0.285(80) & 439(123) & 1.22(34) & (A) 3 masses fit \cr 
0.394(117) & 605(180) & 1.68(50) & (A) 2 masses fit \cr
0.514(126) & 790(194) & 2.20(54) & (B) 3 masses fit \cr
0.509(141) & 782(217) & 2.18(60) & (B) 2 masses fit \cr
\hline 
\end{tabular}
\end{center}
\end{table}

We also calculated the  propagators of the flavor singlet meson scalar, $f_0$, using
the same quark propagator for $\eta'$ and $\omega$, 
and found that they are too noisy to extract the spectrum for all values of $m_f$.

\subsection{Pseudovector meson ($a_1$, $b_1$, $f_1$, $h_1$) spectra}

\begin{figure}[t]
\begin{center}
\includegraphics[angle=-00,scale=0.37,clip=true]{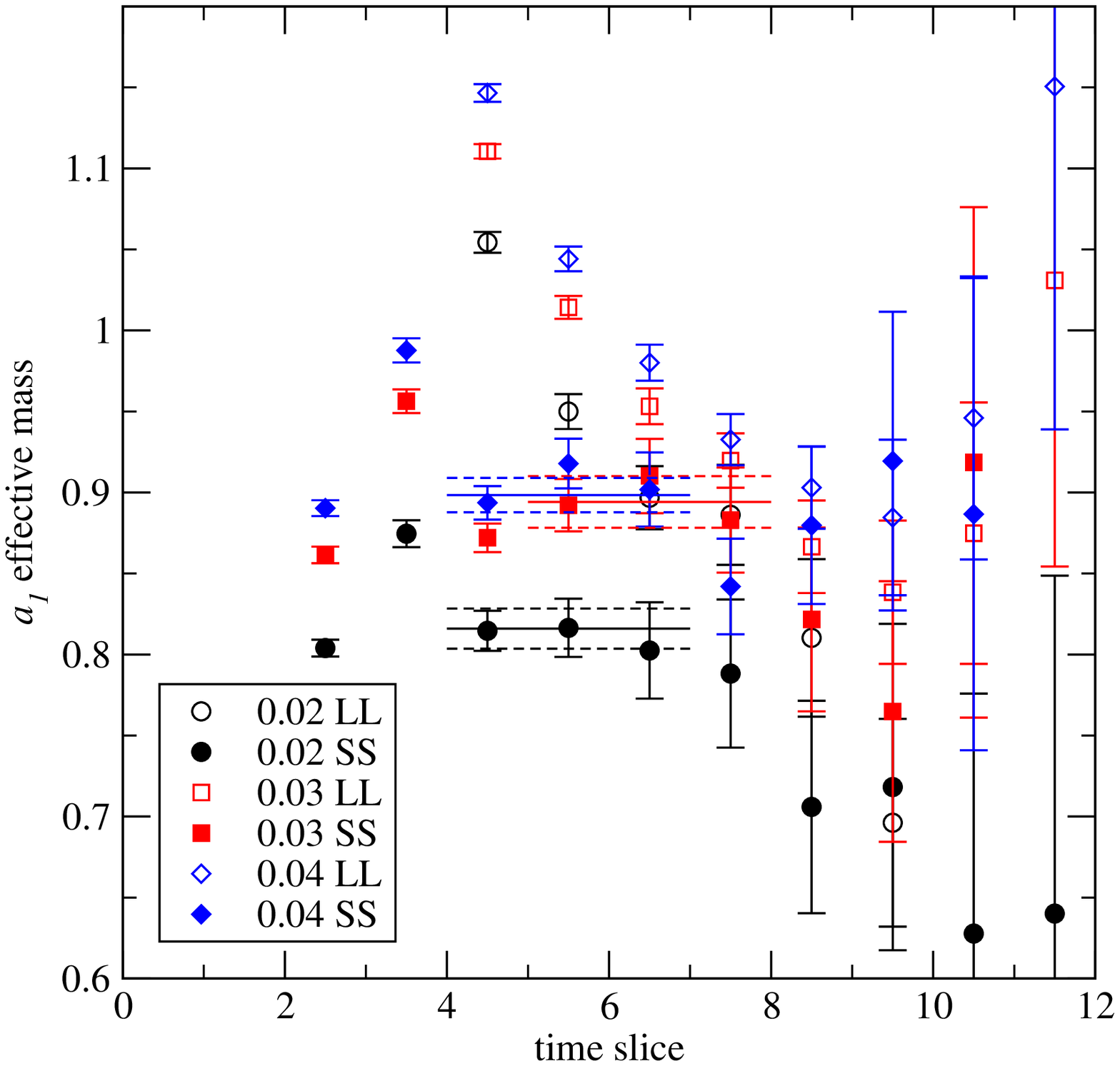}
\includegraphics[angle=-00,scale=0.37,clip=true]{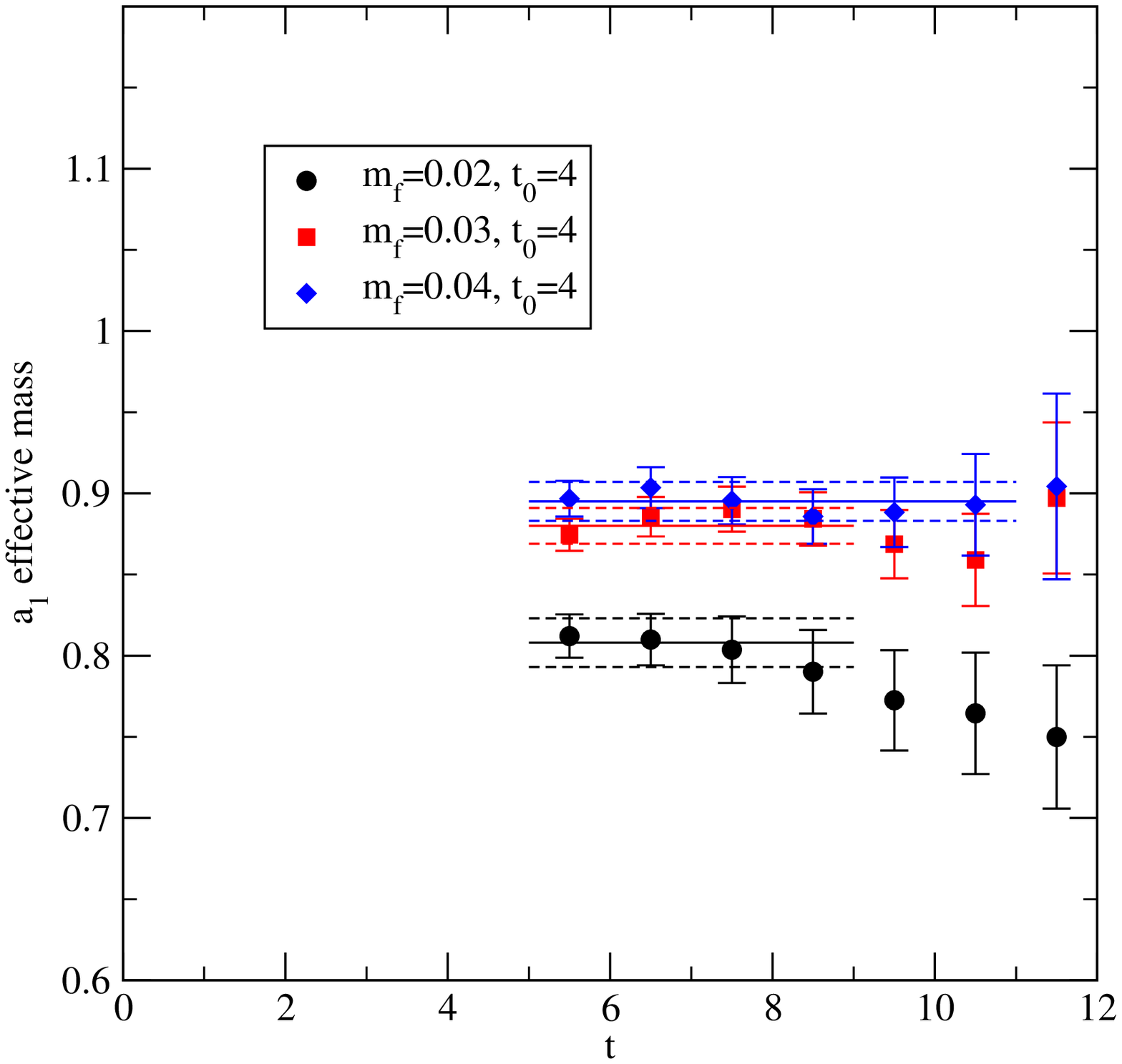}
\end{center}
\caption{
Effective mass of $a_1$ vs $t$ 
using method (A) (left) and method (B) (right).
}
\label{fig:a1-a}
\end{figure}

\begin{figure}[t]
\begin{center}
\includegraphics[angle=-00,scale=0.37,clip=true]{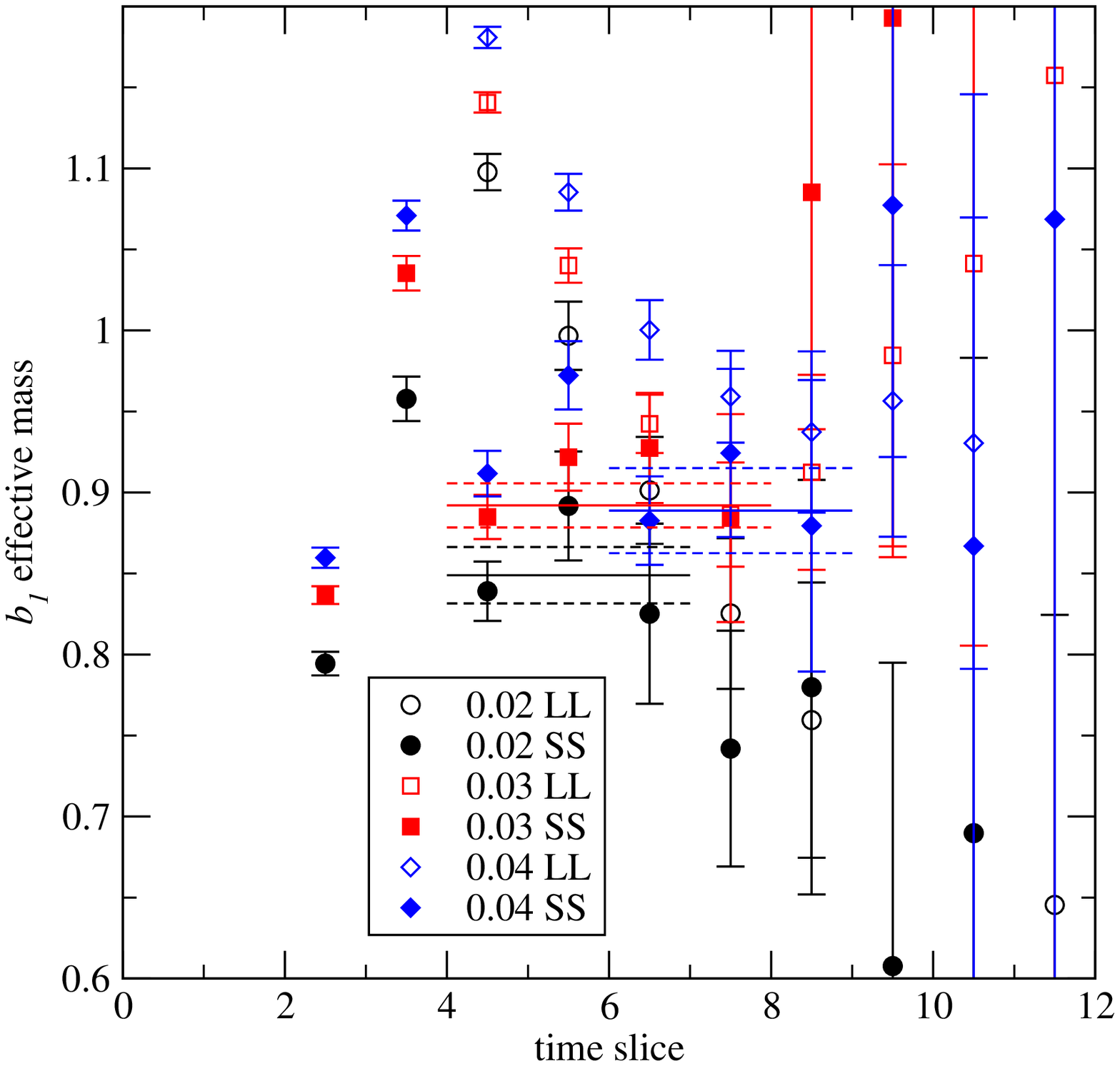}
\includegraphics[angle=-00,scale=0.37,clip=true]{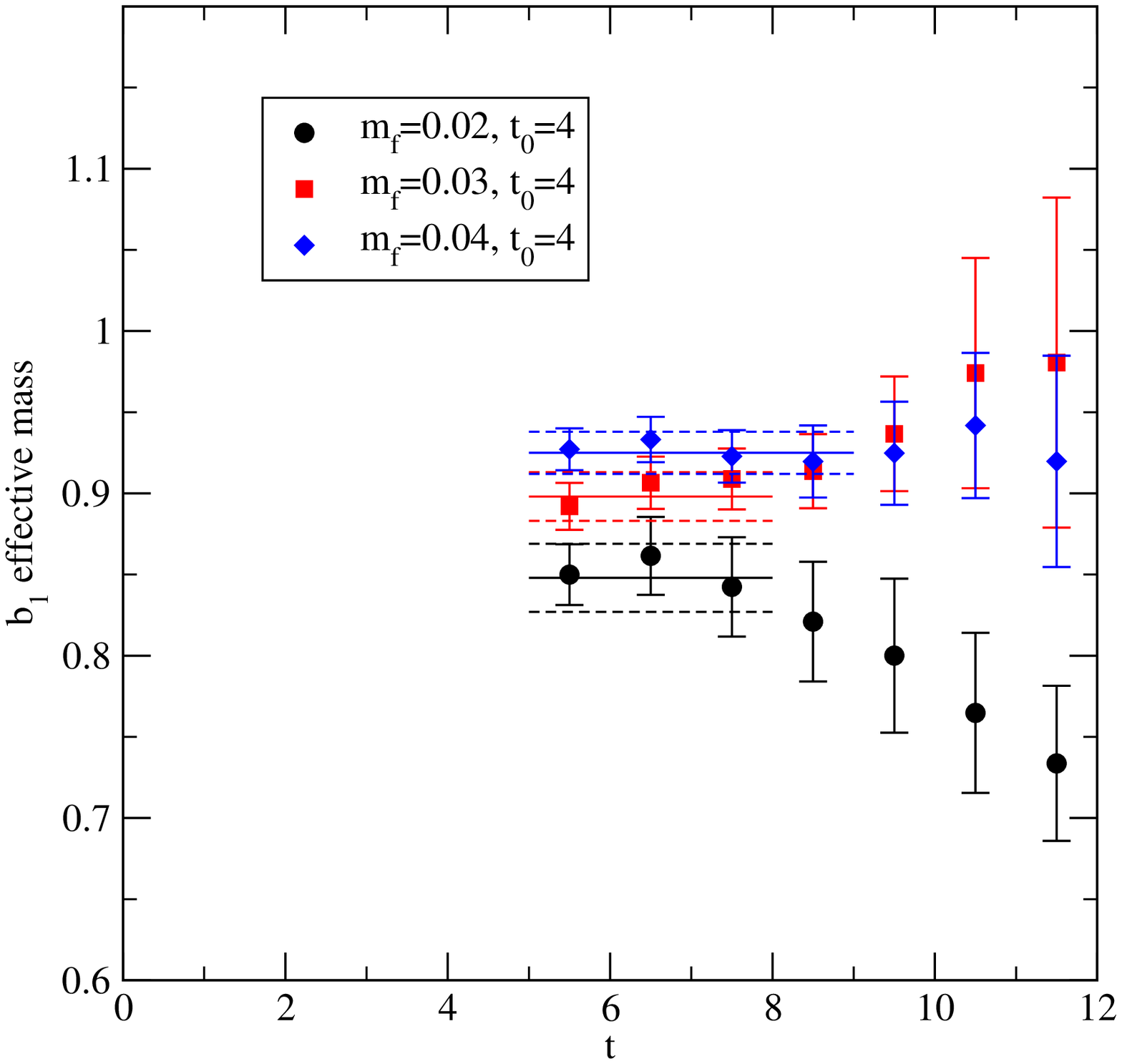}
\end{center}
\caption{
Effective mass of $b_1$ vs $t$ 
using method (A) (left) and method (B) (right).
}
\label{fig:b1-a}
\end{figure}

\begin{figure}[t]
\begin{center}
\includegraphics[angle=-00,scale=0.37,clip=true]{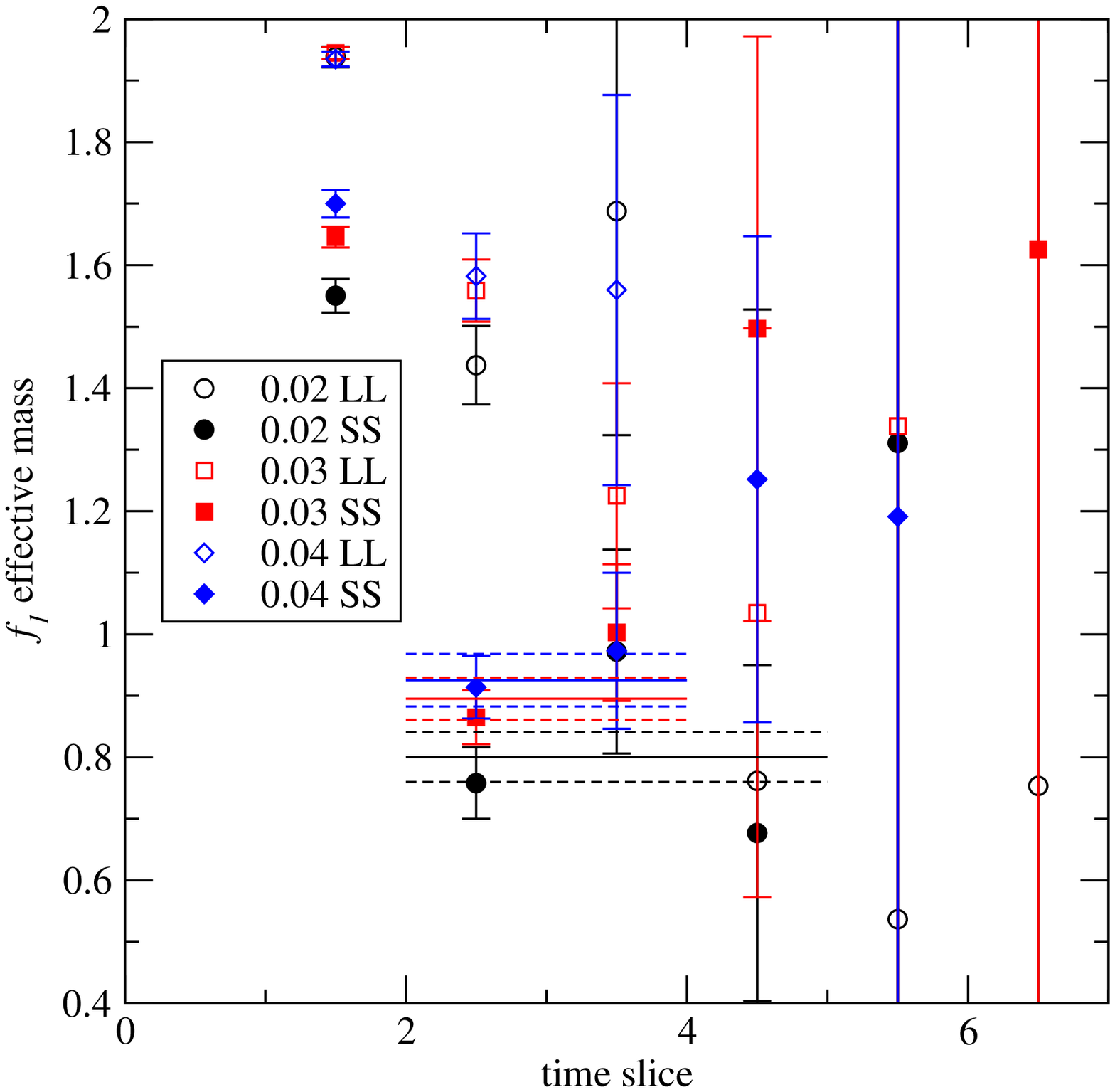}
\includegraphics[angle=-00,scale=0.37,clip=true]{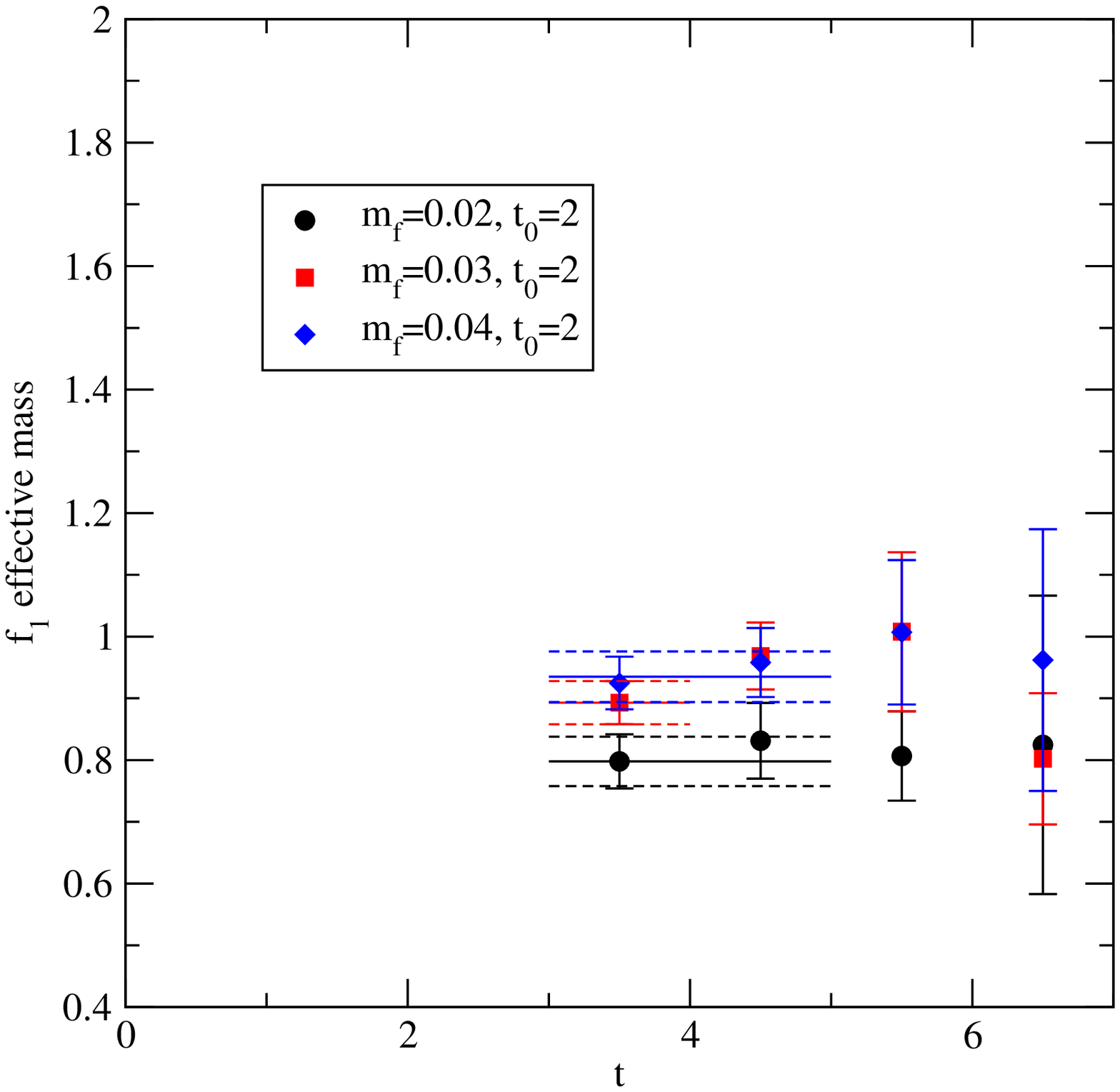}
\end{center}
\caption{
Effective mass of $f_1$ vs $t$ 
using method (A) (left) and method (B) (right).
}
\label{fig:f1-a}
\end{figure}

\begin{figure}[t]
\begin{center}
\includegraphics[angle=-00,scale=0.37,clip=true]{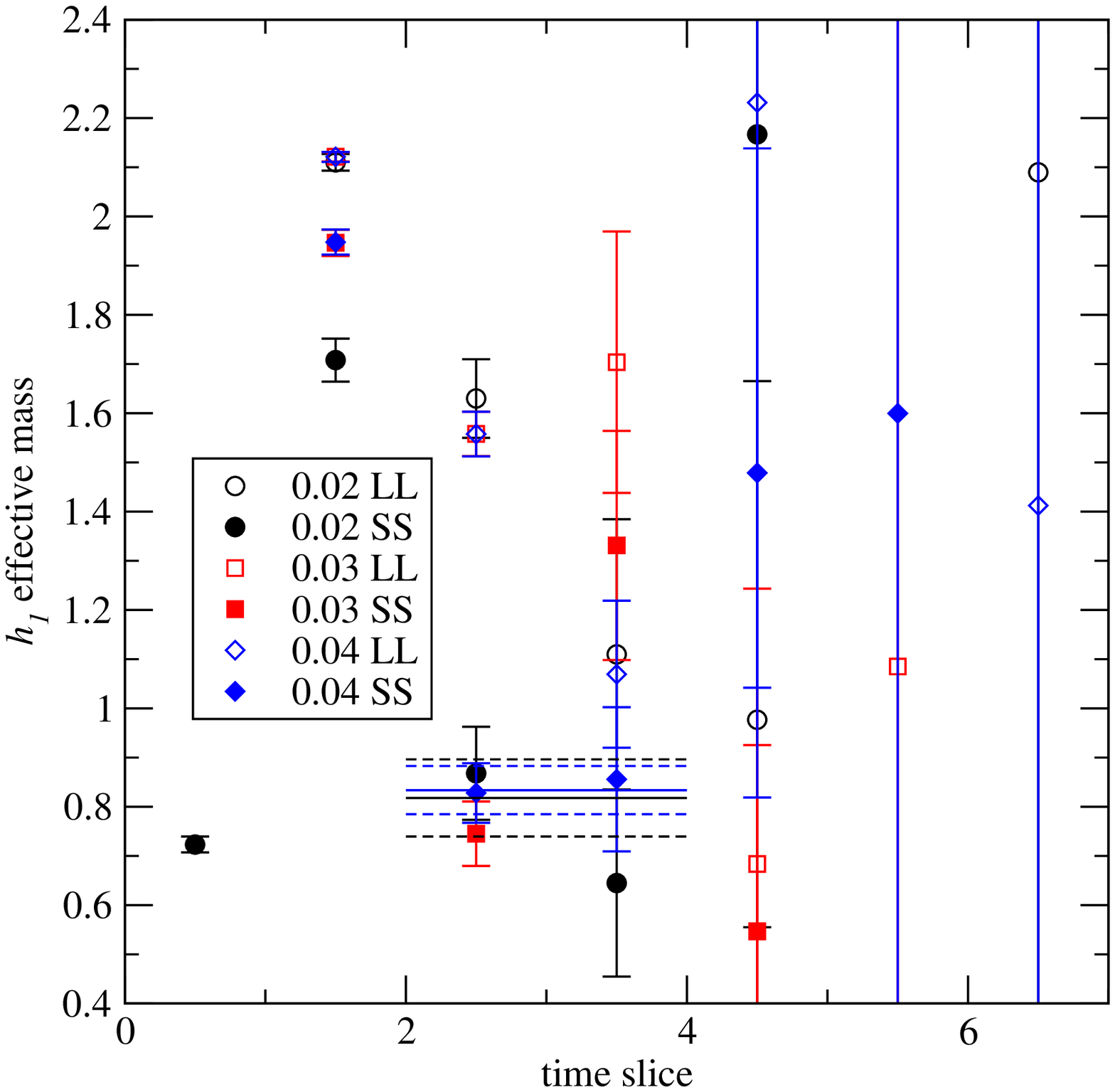}
\includegraphics[angle=-00,scale=0.37,clip=true]{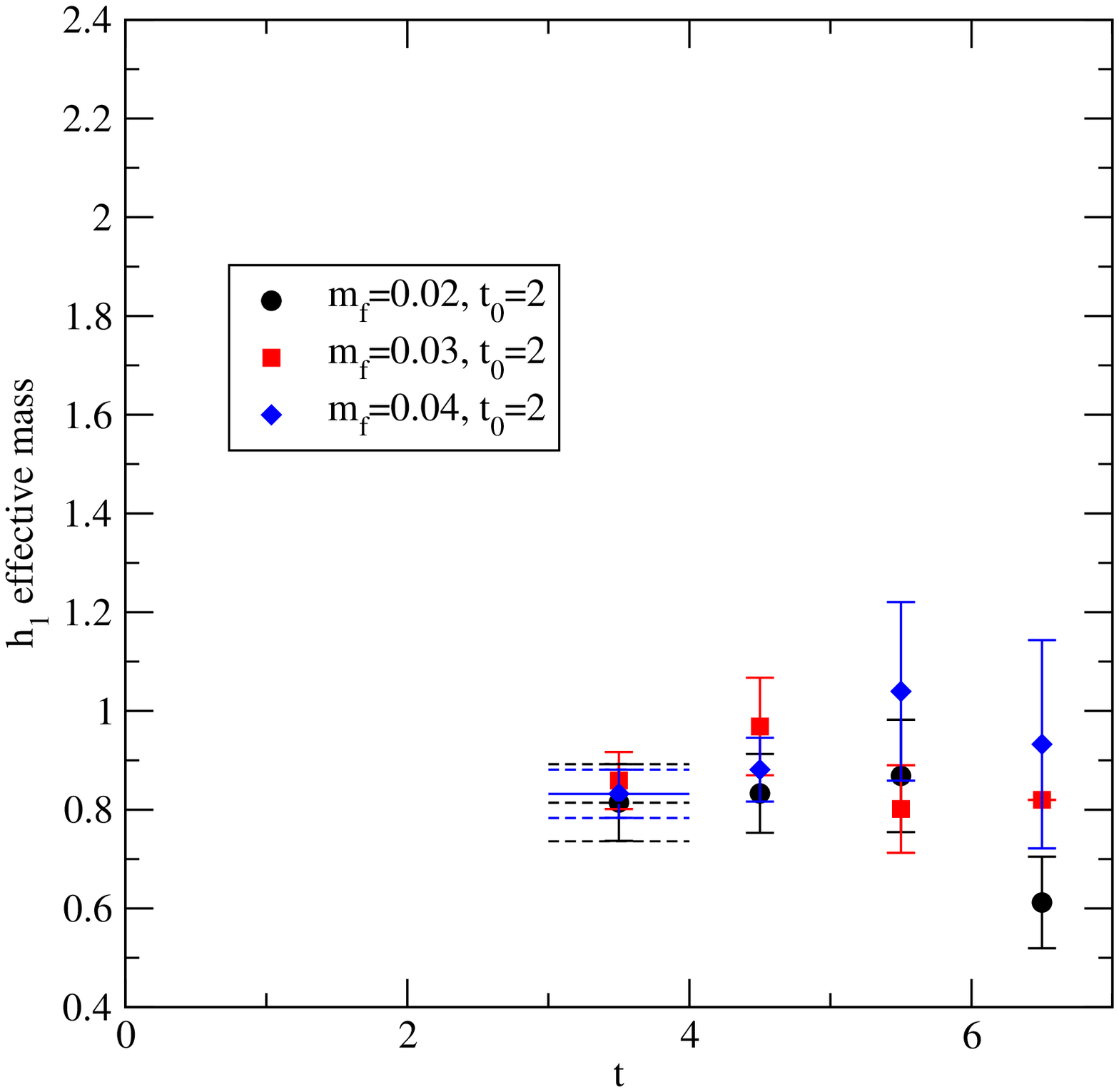}
\end{center}
\caption{
Effective mass of $h_1$  vs $t$ 
using method (A) (left) and method (B) (right).
}
\label{fig:h1-a}
\end{figure}

Figs. \ref{fig:a1-a}, \ref{fig:b1-a}, \ref{fig:f1-a}, and \ref{fig:h1-a} show 
the effective mass obtained using method (A) (left) and method (B) (right), and  Tables \ref{tab:a1-fit}, 
\ref{tab:b1-fit}-\ref {tab:h1-fit} list the results of fits
for $a_1$, $b_1$, $f_1$, and $h_1$, respectively. 
Except for the $h_1$ meson propagator at $m_f=0.03$, the fitting procedure converges.

\clearpage

\begin{table}[t]
\caption{%
$m_{a_1}$.
}
\label{tab:a1-fit}
\begin{center}
\begin{tabular}{cccccc}\hline \hline
$m_{f}$ & $m_{a_1}$ & $t_{0}$ &
$t_{\rm min}$ & $t_{\rm max}$ & method \cr
  \hline
0.02 & 0.816(12) & & 4 & 7 & (A) \cr
 & 0.808(15) & 4 & $t_0+1$ & 9 & (B) \cr
\hline  
0.03 & 0.894(16) & & 5 & 8 & (A) \cr
 & 0.880(11) & 4 &$ t_0+1$ & 9 & (B) \cr
\hline
0.04 & 0.898(11) & & 4 & 8 & (A) \cr
 & 0.895(12) & 4 &$ t_0+1$ & 11 & (B) \cr
\hline 
\end{tabular}
\end{center}
\end{table}

\begin{table}[t]
\caption{%
$m_{b_1}$.
}
\label{tab:b1-fit}
\begin{center}
\begin{tabular}{cccccc}\hline \hline
$m_{f}$ & $m_{b_1}$ & $t_{0}$ &
$t_{\rm min}$ & $t_{\rm max}$ & method \cr
  \hline
0.02 & 0.849(17) & & 4 & 7 & (A) \cr
 & 0.848(21) & 4 & $t_0+1$ & 8 & (B) \cr
\hline  
0.03 & 0.892(14) & & 4 & 8 & (A) \cr
 & 0.898(15) & 4 & $t_0+1$ & 8 & (B) \cr
\hline
0.04 & 0.889(26) & & 6 & 9 & (A) \cr
 & 0.925(13) & 4 & $t_0+1$ & 9 & (B) \cr
\hline 
\end{tabular}
\end{center}
\end{table}

\begin{table}[t]
\caption{%
$m_{f_1}$.
}
\label{tab:f1-fit}
\begin{center}
\begin{tabular}{cccccc}\hline \hline
$m_{f}$ & $m_{f_1}$ & $t_{0}$ &
$t_{\rm min}$ & $t_{\rm max}$ & method \cr
  \hline
0.02 & 0.801(41) & & 2 & 5 & (A) \cr
 & 0.798(40) & 2 & $t_0+1$ & 5 & (B) \cr
\hline  
0.03 & 0.895(34) & & 2 & 4 & (A) \cr
 & 0.893(35) & 2 & $t_0+1$ & 4 & (B) \cr
\hline
0.04 & 0.925(43) & & 2 & 4 & (A) \cr
 & 0.935(41) & 2 & $t_0+1$ & 5 & (B) \cr
\hline 
\end{tabular}
\end{center}
\end{table}

\begin{table}[t]
\caption{%
$m_{h_1}$.
}
\label{tab:h1-fit}
\begin{center}
\begin{tabular}{cccccc}\hline \hline
$m_{f}$ & $m_{h_1}$ & $t_{0}$ &
$t_{\rm min}$ & $t_{\rm max}$ & method \cr
  \hline
0.02 & 0.818(78) & & 2 & 4 & (A) \cr
 & 0.814(78) & 2 & $t_0+1$ & 4 & (B) \cr
\hline  
0.04 & 0.834(49) & & 2 & 4 & (A) \cr
 & 0.832(49) & 2 & $t_0+1$ & 4 & (B) \cr
\hline 
\end{tabular}
\end{center}
\end{table}

\begin{figure}[t]
\begin{center}
\includegraphics[angle=-00,scale=0.40,clip=true]{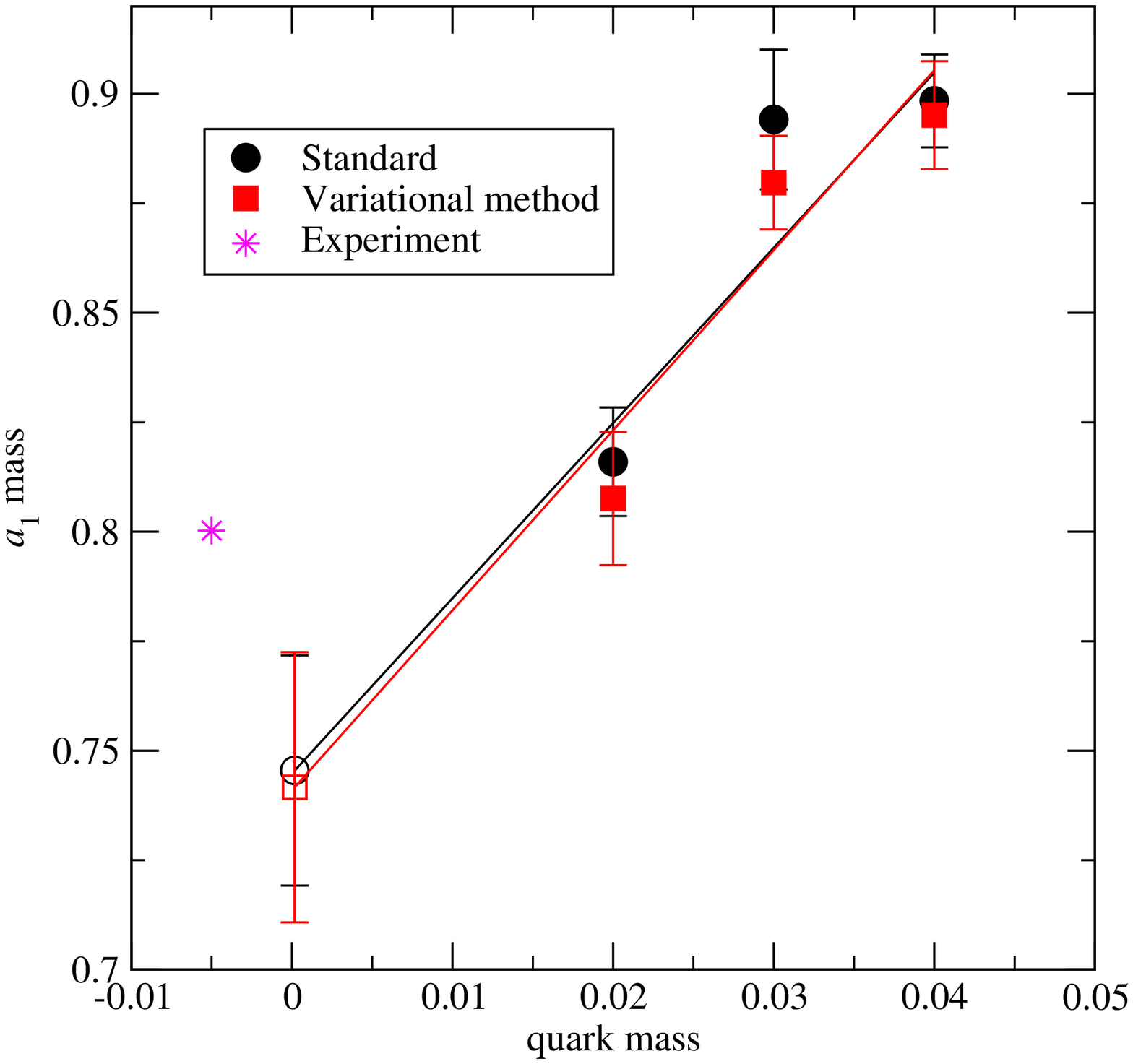}
\end{center}
\caption{
$m_{a_1}$ vs $m_f$.
The asterisk on the left shows the experimental value\cite{Yao:2006px}.
}
\label{fig:a1-chiral}
\end{figure}

\begin{figure}[t]
\begin{center}
\includegraphics[angle=-00,scale=0.40,clip=true]{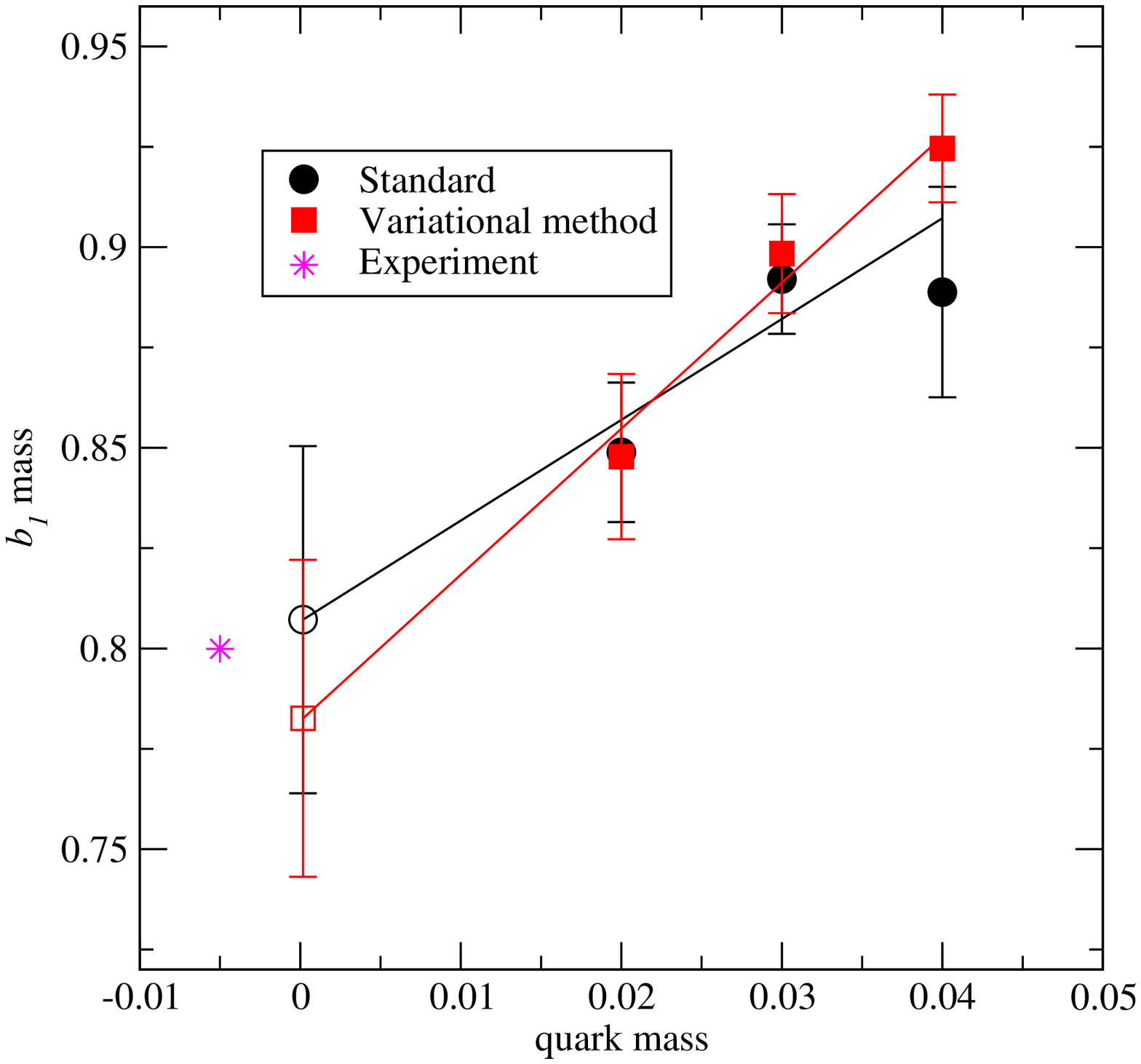}
\end{center}
\caption{
$m_{b_1}$ vs $m_f$.
The asterisk on the left shows the experimental value\cite{Yao:2006px}.
}
\label{fig:b1-chiral}
\end{figure}

\begin{figure}[t]
\begin{center}
\includegraphics[angle=-00,scale=0.40,clip=true]{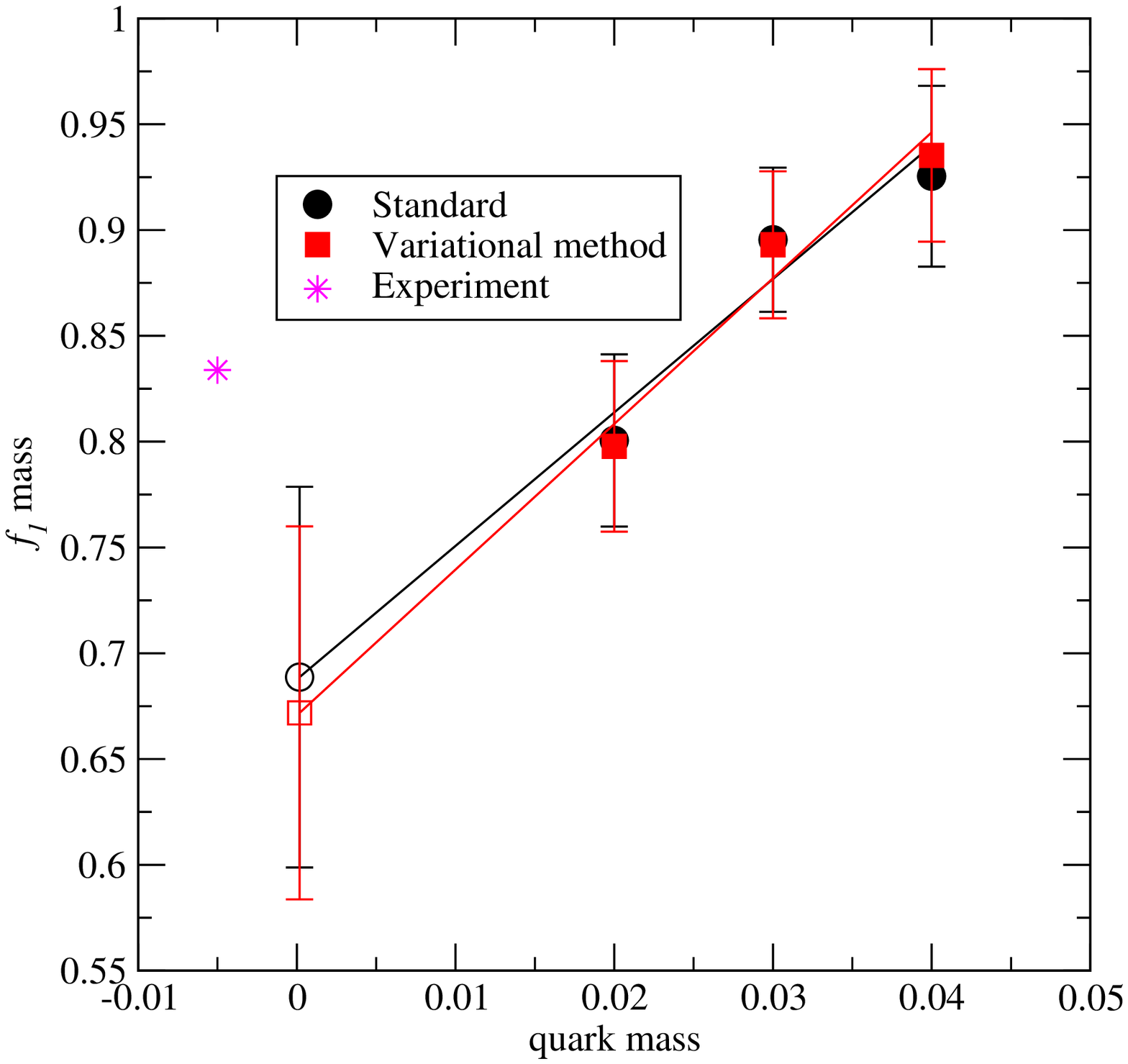}
\end{center}
\caption{
$m_{f_1}$ vs $m_f$.
The asterisk on the left  shows the experimental value\cite{Yao:2006px}.
}
\label{fig:f1-chiral}
\end{figure}

\begin{figure}[t]
\begin{center}
\includegraphics[angle=-00,scale=0.40,clip=true]{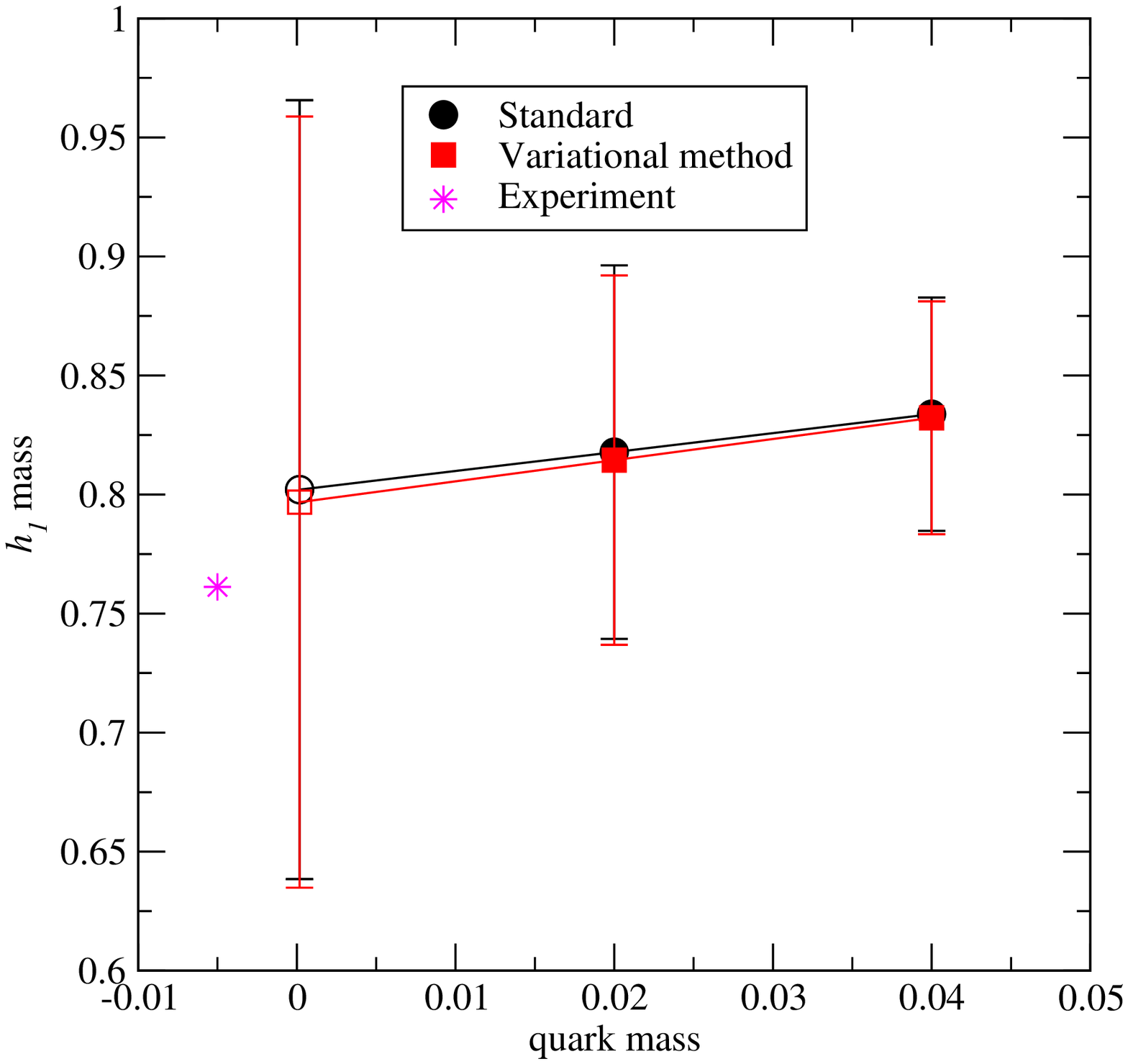}
\end{center}
\caption{
$m_{h_1}$ vs $m_f$.
The asterisk ont he left shows the experimental value\cite{Yao:2006px}.
}
\label{fig:h1-chiral}
\end{figure}

These meson masses are extrapolated linearly to the physical quark mass point,
$m_f=m_{u,d}$, and are shown in Figs. \ref{fig:a1-chiral}-\ref{fig:h1-chiral}. 
The numerical values are summarized
in Tables \ref{tab:a1-chiral}-\ref {tab:h1-chiral} for $a_1$, 
$b_1$, $f_1$, and $h_1$, respectively. 
As the masses are independent of the  method used, within statistical error, 
we choose 
\begin{eqnarray}
&&m_{a_1}^{\text{phys}}=1.140(51) \ \text{GeV}~, \\
&&m_{b_1}^{\text{phys}}=1.203(64) \ \text{GeV}~, \\
&&m_{f_1}^{\text{phys}}=1.033(137) \ \text{GeV}~, \\
&&m_{h_1}^{\text{phys}}=1.225(250) \ \text{GeV}  
\end{eqnarray}
from method (B) as our main values. 

\clearpage

\begin{table}[t]
\caption{%
$m_{a_1}$ at the physical quark mass point $(m_f= m_{u,d})$.
\label{tab:a1-chiral}
}
\begin{center}
\begin{tabular}{cccc}\hline \hline
$m_{a_1}$ & $m_{a_1}^{\text{phys}}$ [MeV] 
& $m_{a_1} r_0$
& method \cr
\hline
0.745(26) & 1,146(45) & 3.19(12) & (A) \cr
0.742(31) & 1,140(51) & 3.17(14) & (B) \cr
\hline 
\end{tabular}
\end{center}
\end{table}

\begin{table}[t]
\caption{%
$m_{b_1}$ at the physical quark mass point 
$(m_f= m_{u,d})$.}
\label{tab:b1-chiral}
\begin{center}
\begin{tabular}{cccc}\hline \hline
$m_{b_1}$ & $m_{b_1}^{\text{phys}}$ [MeV] 
& $m_{b_1} r_0$
& method \cr
\hline
0.807(43) & 1,241(70) & 3.45(19) & (A) \cr
0.783(40) & 1,203(64) & 3.35(17) & (B) \cr
\hline 
\end{tabular}
\end{center}
\end{table}

\begin{table}[t]
\caption{%
$m_{f_1}$ at the physical quark mass point $(m_f= m_{u,d})$.
}
\label{tab:f1-chiral}
\begin{center}
\begin{tabular}{cccc}\hline \hline
$m_{f_1}$ & $m_{f_1}^{\text{phys}}$ [MeV] 
& $m_{f_1} r_0$
& method \cr
\hline
0.689(90) & 1,058(139) & 2.95(39) & (A) \cr
0.672(88) & 1,033(137) & 2.87(38) & (B) \cr
\hline 
\end{tabular}
\end{center}
\end{table}

\begin{table}[t]
\caption{%
$m_{h_1}$ at the physical quark mass point $(m_f= m_{u,d})$.
}
\label{tab:h1-chiral}
\begin{center}
\begin{tabular}{cccc}\hline \hline
$m_{h_1}$ & $m_{h_1}^{\text{phys}}$ [MeV] 
& $m_{h_1} r_0$
& method \cr
\hline
0.802(164) & 1,233(252) & 3.43(70) & (A) \cr
0.797(162) & 1,225(250) & 3.41(69) & (B) \cr
\hline 
\end{tabular}
\end{center}
\end{table}

These numbers may be compared with the experimental results for
$b_1(1235)$, $h_1(1170)$, $a_1(1260)$, and $f_1(1285)$,
the first two of which are in good agreement with the
numerical results. However further investigations 
based on realistic settings 
are clearly needed for more detailed comparisons.

\subsection{Excited meson ($\pi^*$, $\rho^*$) masses}
\begin{figure}[t]
\begin{center} 
\includegraphics[angle=-00,scale=0.37,clip=true]{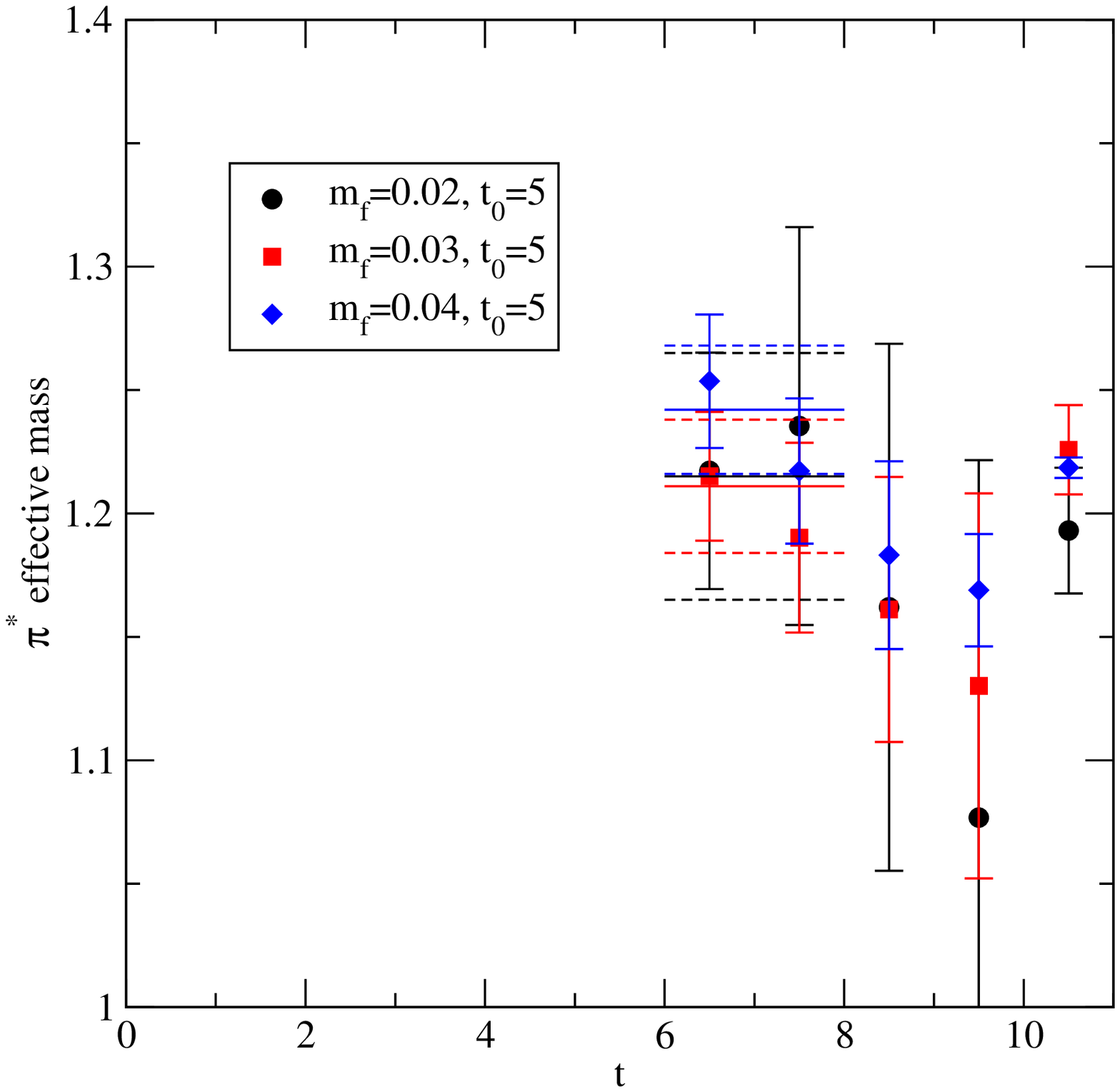}
\includegraphics[angle=-00,scale=0.37,clip=true]{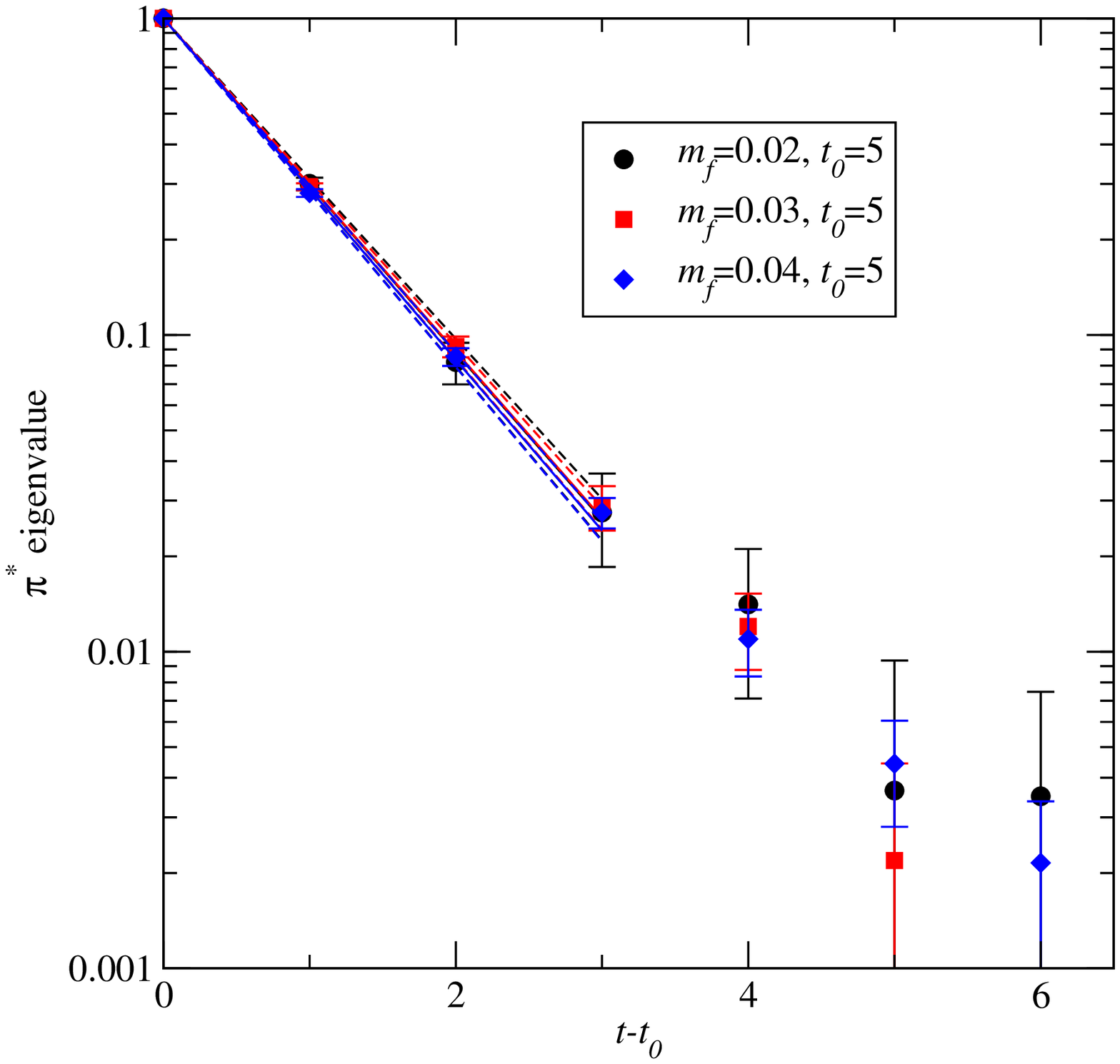}
\end{center}
\caption{
Effective mass of $\pi^*$ and  eigenvalue as functions of $t$ and $t-t_0$.
}
\label{fig:pie-b}
\end{figure}

\begin{figure}[t]
\begin{center}
\includegraphics[angle=-00,scale=0.37,clip=true]{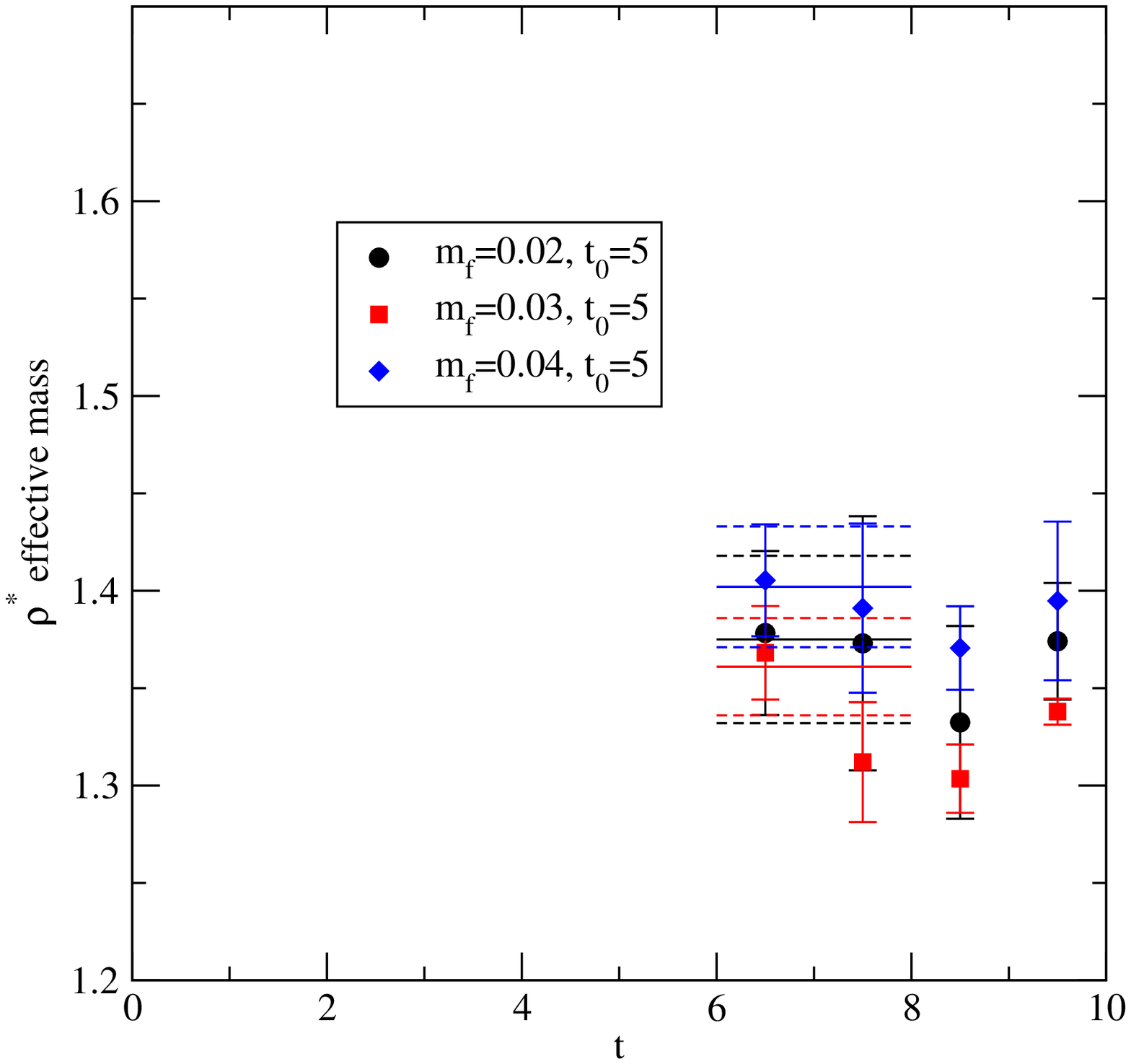}
\includegraphics[angle=-00,scale=0.37,clip=true]{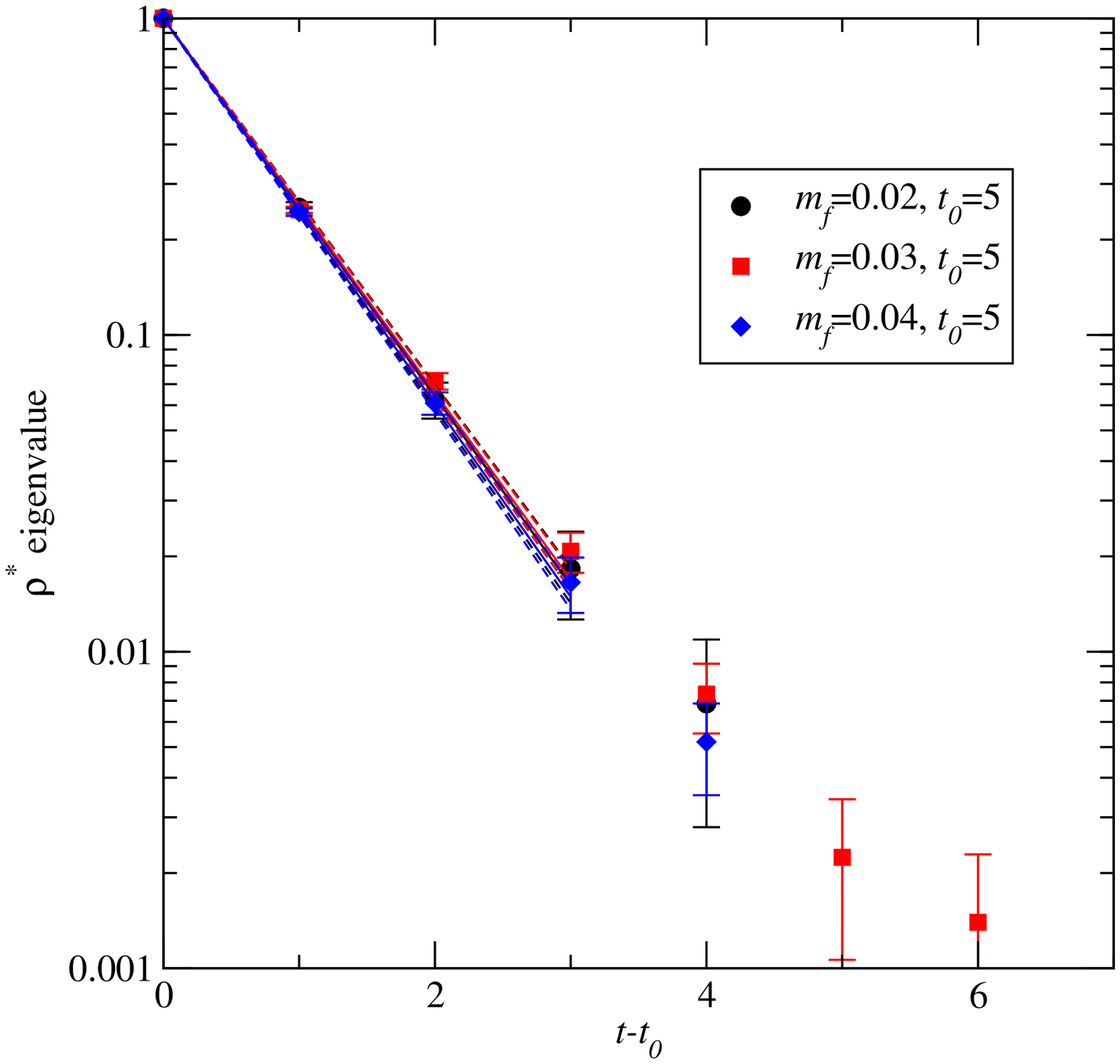}
\end{center}
\caption{
Effective mass of $\rho^*$ and eigenvalue as functions of $t$ and $t-t_0$.
}
\label{fig:rhoe-b}
\end{figure}

In this subsection, the second excited states of the pion and $\rho$ 
meson are discussed. 
Using method (B), we extract the eigenvalue for the second excited state,
$\lambda_{O^*}(t), O=\pi,\rho$ eq. (\ref{eq:variational_diag}), 
which is plotted in the right panels of Fig. \ref{fig:pie-b} 
and \ref{fig:rhoe-b}, respectively. 
Although we only use two different operators for each meson, and
$\lambda_{O^*}(t)$ may have a significant contribution from the higher 
excited state, we fit  $\lambda_{O^*}(t)$ to extract the
temporal exponent, $m_{O^*}$, or the mass of the excited states using
eq.~(\ref{eq:variational_excited_eig}).
The results of the fitting are shown in Tables \ref{tab:pie-fit} and  \ref{tab:rhoe-fit}. 
We checked that the results for $t_0=5$ and $t_0=6$ are consistent with each other.

\begin{table}[t]
\caption{%
$m_{\pi^*}$.}
\label{tab:pie-fit}
\begin{center}
\begin{tabular}{cccccc}\hline \hline
$m_{f}$ & $m_{\pi^*}$ & $t_{0}$ &
$t_{\rm min}$ & $t_{\rm max}$ & method \cr
  \hline
0.02 & 1.215(50) & 5 & $t_0+1$ & 8 & (B) \cr
\hline  
0.03 & 1.211(27) & 5 & $t_0+1$ & 8 & (B) \cr
\hline
0.04 & 1.242(26) & 5 & $t_0+1$ & 8 & (B) \cr
\hline 
\end{tabular}
\end{center}
\end{table}

\begin{table}[t]
\caption{%
$m_{\rho^*}$.}
\label{tab:rhoe-fit}
\begin{center}
\begin{tabular}{cccccc}\hline \hline
$m_{f}$ & $m_{\rho^*}$  & $t_{0}$ 
& $t_{\rm min}$ & $t_{\rm max}$ & method \cr
\hline
0.02 & 1.375(43) & 5 & $t_0+1$ & 8 & (B) \cr
\hline  
0.03 & 1.361(25) & 5 & $t_0+1$ & 8 & (B) \cr
\hline
0.04 & 1.402(31) & 5 & $t_0+1$ & 8 & (B) \cr
\hline 
\end{tabular}
\end{center}
\end{table}

\begin{figure}[ht]
\begin{center}
\includegraphics[angle=-00,scale=0.40,clip=true]{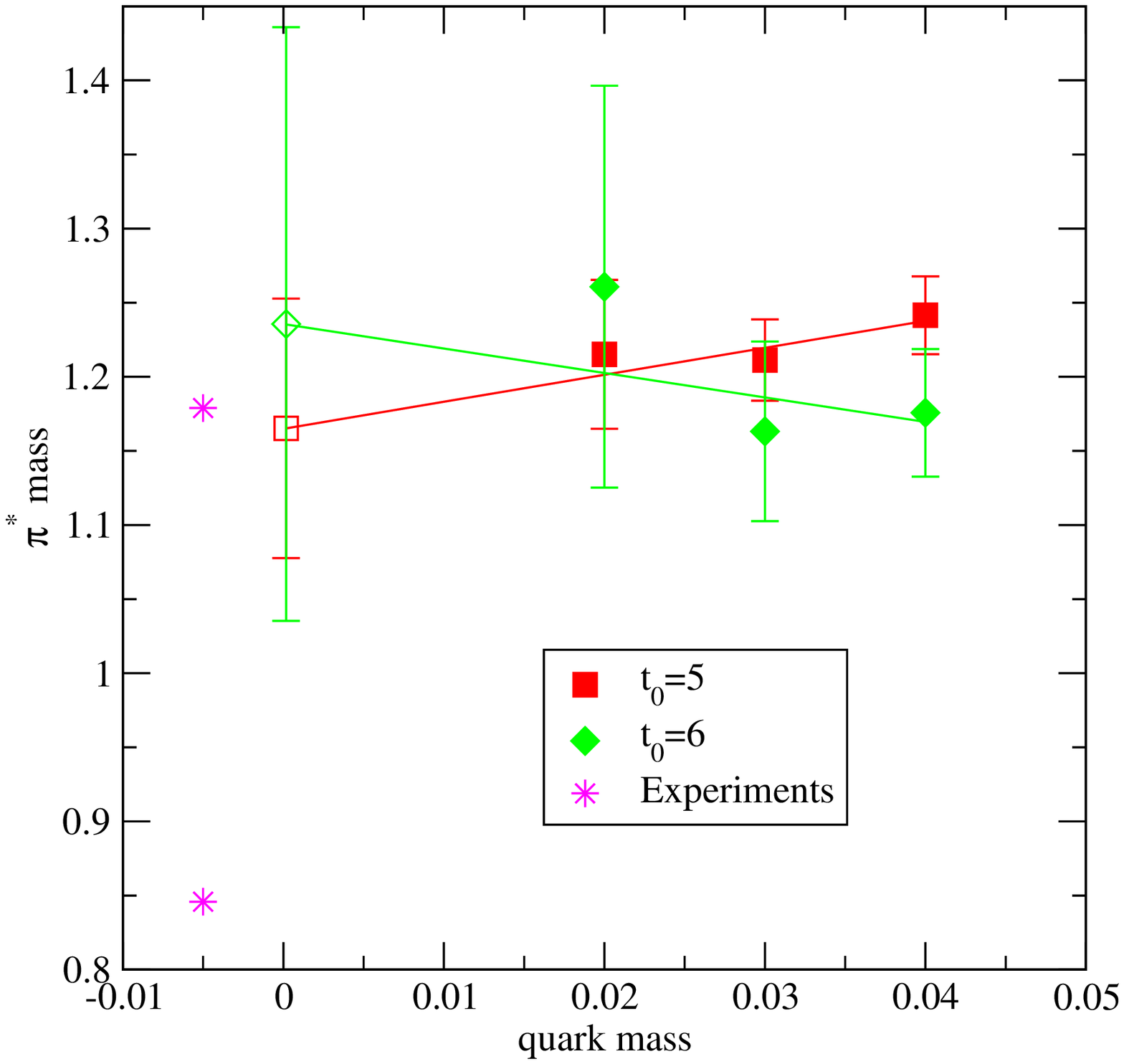}
\end{center}
\caption{
$m_{\pi^*}$ vs. $m_f$.
The left most star symbols show the experimental values\cite{Yao:2006px}
in the real world.
}
\label{fig:pie-chiral}
\end{figure}

\begin{figure}[ht]
\begin{center}
\includegraphics[angle=-00,scale=0.40,clip=true]{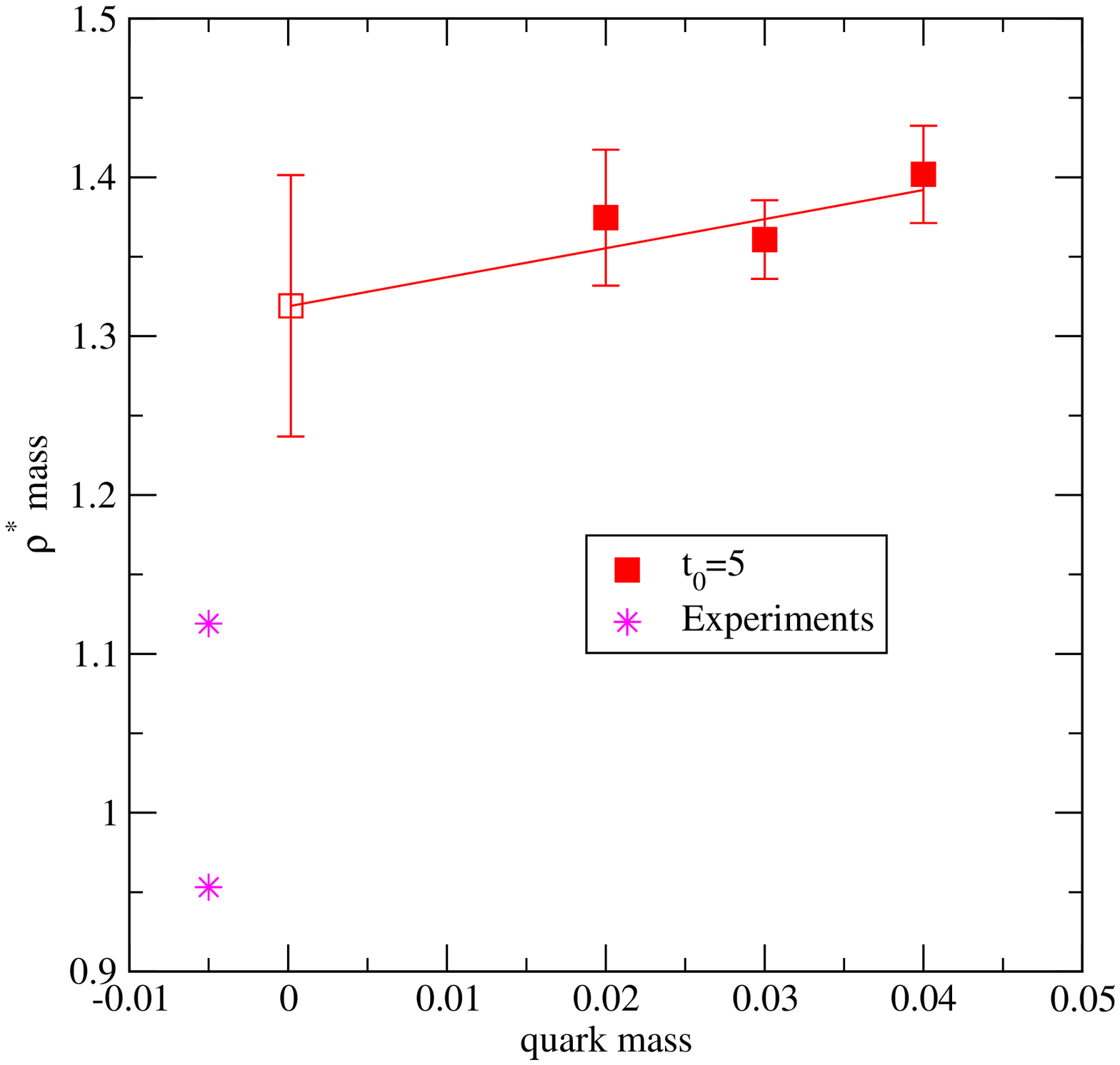}
\end{center}
\caption{
$m_{\rho^*}$ vs. $m_f$
The left most star symbols show the experimental values\cite{Yao:2006px}
in the real world.
}
\label{fig:rhoe-chiral}
\end{figure}

We performed linear extrapolation using eq. (\ref{eq:chiral_lin_formula}) to the
physical quark mass point, and found that
\begin{eqnarray}
&&m_{\pi^*}^{\text{phys}}=1.791(138) \ \text{GeV}~, \\
&&m_{\rho^*}^{\text{phys}}=2.028(131) \ \text{GeV} 
\end{eqnarray}
(see Figs. \ref{fig:pie-chiral} and \ref{fig:rhoe-chiral}, 
Table \ref{tab:pie-chiral} and \ref{tab:rhoe-chiral}).
These states may be interpreted as $\pi(1300) $, and
$\rho(1450) $ or $\rho(1700) $.

\begin{table}[t]
\caption{%
$m_{\pi^*}$ at the physical quark mass point $(m_f= m_{u,d})$
}
\label{tab:pie-chiral}
\begin{center}
\begin{tabular}{cccc}\hline \hline
$m_{\pi^*}$ & $m_{\pi^*}^{\text{phys}}$ [MeV] 
& $m_{\pi^*} r_0$
& method \cr
\hline
1.165(88) & 1,791(138) & 4.98(38) & (B) \cr
\hline 
\end{tabular}
\end{center}
\end{table}

\begin{table}[t]
\caption{%
$m_{\rho^*}$ at the physical quark mass point $(m_f= m_{u,d})$
} 
\label{tab:rhoe-chiral}
\begin{center}
\begin{tabular}{cccc}\hline \hline
$m_{\rho^*}$ & $m_{\rho^*}^{\text{phys}}$ [MeV] 
& $m_{\rho^*} r_0$
& method \cr
\hline
1.319(82) & 2,028(131) & 5.64(36) & (B) \cr
\hline 
\end{tabular}
\end{center}
\end{table}

\subsection{Decay constants}
As the last set of numerical results, we present the leptonic decay constant
in this subsection.
The decay constant of the ground-state pion, $f_\pi$, is determined
using  method (C), and fitting the smeared two-point function to
formula (\ref{eq:method-C_fpi}). 
We also fitted the same two-point functions to the double exponential
formula (\ref{eq:method-b2}) using the values of $m_\pi$ and $m_{\pi^*}$
determined from the variational method (method (B)) to investigate
the decay constant for the second excited state, $f_{\pi^*}$, using
method (D).

Table \ref{tab:fpi-fit} shows the results for each simulated quark mass.
The pion mass and decay constants are consistent with those reported in 
the previous paper\cite{Aoki:2004ht} within statistical error.

\begin{table}[t]
\caption{%
$f_{\pi}$ and $f_{\pi^*}$.
}
\label{tab:fpi-fit}
\begin{center}
\begin{tabular}{cccccccc}\hline \hline
$m_{f}$ & $m_{\pi}$ & $f_\pi$ & $m_{\pi^*}$ & $f_{\pi^*}$ &
$t_{\rm min}$ & $t_{\rm max}$ & method \cr
  \hline
0.02 & 0.2936(13) & 0.09561(40) & --- & --- & 7 & 14 & (C) \cr
 & 0.2934(13)(fixed) & 0.09540(43) & 1.215(50)(fixed) & 0.02244(54) & 4 & 14 & (D) \cr
 & 0.2938(18) $^{\rm a}$ & 0.09494(62) $^{\rm a}$ & --- & --- & 9 & 16 & (C) \cr
\hline  
0.03 & 0.3598(15) & 0.10350(46) & --- & --- & 10 & 16 & (C) \cr
 & 0.3581(10)(fixed) & 0.10370(44) & 1.211(27)(fixed) & 0.03236(65) & 4 & 16 & (D) \cr 
 & 0.3610(18) $^{\rm a}$ & 0.10253(56) $^{\rm a}$ & --- & --- & 9 & 16 & (C) \cr
\hline
0.04 & 0.4098(12) & 0.11002(39) & --- & --- & 8 & 16 & (C) \cr
 & 0.4092(11)(fixed) & 0.10964(40) & 1.242(26)(fixed) & 0.04362(61) & 4 & 16 & (D) \cr
 & 0.4087(16) $^{\rm a}$
& 0.11059(57) $^{\rm a}$ & --- & --- & 9 & 16 & (C) \cr
\hline 
\end{tabular}
\end{center}
\footnotesize{$^{\rm (a)}$ These values are quoted in the previous paper\cite{Aoki:2004ht}.} 
\end{table}

Although the $\pi^*$ decay constant is poorly numerically determined, 
an interesting theoretical prediction can be made. 
AWTI, (\ref{eq:fpi}), for $\pi^*$
describes the equation for its decay constant,
\begin{eqnarray} 
f_{\pi^*} ={2(m_f+\mres) \over m_{\pi^*}^2} \langle 0 | P^a | \pi^* \rangle~.
\end{eqnarray}
If  $m_{\pi^*}$ is not an NG boson, so $m_{\pi^*}$ remains nonzero,
the right-hand side vanishes at the chiral limit, $(m_f\to-\mres)$. 
This prediction was checked on a lattice QCD using Wilson 
fermions\cite{McNeile:2006qy}. and  $f_{\pi^*}$ was 
consistent to be zero at the chiral limit.

\begin{figure}[t]
\begin{center}
\includegraphics[angle=-00,scale=0.40,clip=true]{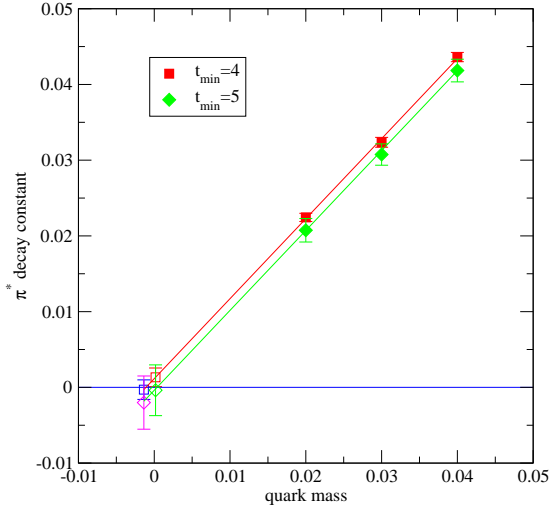}
\end{center}
\caption{
$f_{\pi^*}$ vs $m_f$.
}
\label{fig:pie-decay-chiral}
\end{figure}

Figure \ref{fig:pie-decay-chiral} and Table \ref{tab:fpie-chiral} show the linear 
extrapolation  of $f_{\pi^*}$. At the chiral limit,  the $\pi^*$ decay constant 
is also consistent with the theoretical prediction, {\it i.e.\/} , $f_{\pi^*}\to0$. 

\begin{table}[t]
\caption{%
$f_{\pi^*}$ at the physical quark mass point $(m_f= m_{u,d})$ and the chiral limit 
$(m_f=-\mres)$.
}
\label{tab:fpie-chiral}
\begin{center}
\begin{tabular}{ccccc}\hline \hline
$m_f$ & $f_{\pi^*}$ & $f_{\pi^*}^{\text{phys}}$ [MeV] 
& $f_{\pi^*} r_0$
& method \cr
\hline
$m_{u,d}$ & 0.0013(12) & 20(19) & 0.0057(53) & (D) \cr
$-m_{\text{res}}$ & $-0.0003(13)$ & $-05(20)$ & $-0.0013(57)$ & (D) \cr
\hline 
\end{tabular}
\end{center}
\end{table}

Next we discuss the $\rho$ meson decay constant, $f_{\rho}$. 
The result of the fitting using eq. (\ref{eq:method-C_frho}) is shown 
in Table \ref{tab:frho-fit}. The mass of the $\rho$  extracted by this fitting is
consistent with those obtained from methods (A) and (B) 
within statistical error for all $m_f$. 

\begin{table}[t]
\caption{%
$f_{\rho}/Z_V$.
}
\label{tab:frho-fit}
\begin{center}
\begin{tabular}{cccccc}\hline \hline
$m_{f}$ & $m_{\rho}$ & $f_\rho/Z_V$ &
$t_{\rm min}$ & $t_{\rm max}$ & method \cr
  \hline
0.02 & 0.5730(96) & 0.2011(66) & 9 & 13 & (C) \cr
\hline  
0.03 & 0.6035(64) & 0.2025(50) & 10 & 14 & (C) \cr
\hline
0.04 & 0.6448(51) & 0.2164(37) & 9 & 14 & (C) \cr
\hline 
\end{tabular}
\end{center}
\end{table}

Then the decay constant at the physical quark mass point is obtained as
\begin{eqnarray}
f_\rho^{\text{phys}}=210 (15) \ \text{MeV} 
\end {eqnarray}
by linear extrapolation (see  Table~\ref{tab:frho-chiral}). 
The renormalization factor, $Z_V$, which converts the lattice operator
into the one in the continuum for $\overline{\rm MS}$ at $\mu=2$ GeV 
is necessary to obtain a physical value for the decay constants.
We use $Z_A=0.75734(55)$, which was determined in the previous 
paper\cite{Aoki:2004ht}, and the relation $Z_V=Z_A$, assuming the
good chiral symmetry of the current simulation.

\begin{figure}[ht]
\begin{center}
\includegraphics[angle=-00,scale=0.40,clip=true]{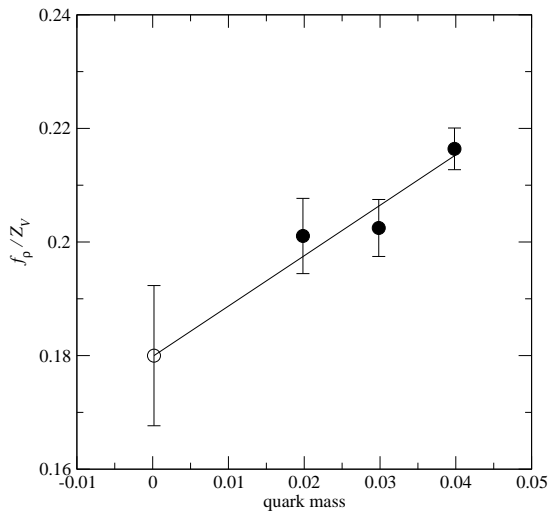}
\end{center}
\caption{
$f_{\rho}/Z_V$ vs $m_f$.
}
\label{fig:rho-decay-chiral}
\end{figure}

\begin{table}[ht]
\caption{%
$f_{\rho}$ at the physical quark mass point $(m_f= m_{u,d})$
}
\label{tab:frho-chiral}
\begin{center}
\begin{tabular}{ccccc}\hline \hline
$f_\rho/Z_V$ & $f_{\rho}$ & $f_{\rho}^{\text{phys}}$ [MeV] 
& $f_{\rho} r_0$
& method \cr
\hline
0.1800(123) & 0.1363(94) & 210(15) & 0.0583(41) & (C) \cr
\hline 
\end{tabular}
\end{center}
\end{table}

\subsection{Systematic uncertainties}

So far we have mainly discussed error due to the limited size of the statistical sample.
Our numerical results were obtained only at one lattice scale,
in one space-time volume,  for three quark masses heavier than the physical values,
and the strange sea quark was neglected.
In this section, various sources of systematic errors are listed and some of their
magnitudes are very roughly estimated to compare our results with those of experiments.

\begin{figure}[t]
\begin{center}
\includegraphics[angle=-00,scale=0.40,clip=true]{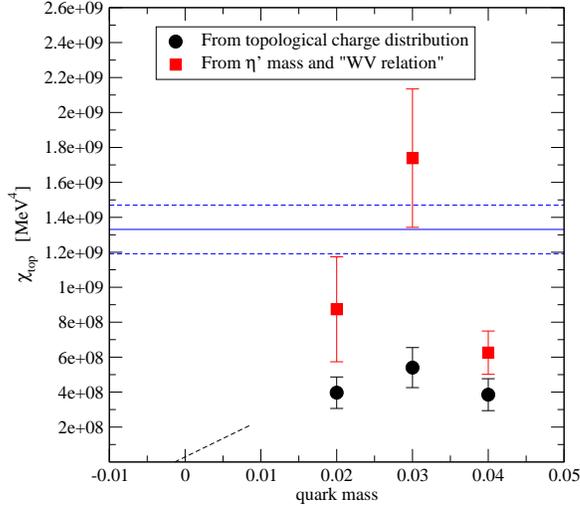}
\end{center}
\caption{
Circles show the measured $\chi_{\text{top}}$ as a function of  $m_f$ \cite{Aoki:2004ht,Berruto:2005hg}
while squares show values calculated from $m_{\eta'}$ and $m_\pi$, as described in the next section.
The horizontal line shows the value obtained from a pure SU(3) YM simulation\cite{DelDebbio:2004ns}.
The dotted line shows the  prediction from chiral perturbation theory.
}
\label{fig:chi-top}
\end{figure}

\begin{figure}[t] 
\begin{center} 
\includegraphics[angle=-00,scale=0.50,clip=true]{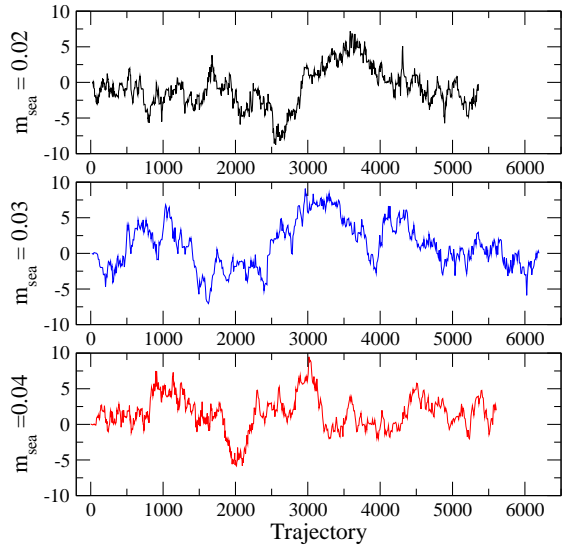}
\end{center}
\caption{
History of the topological charge in the same simulation as that used for Fig.~\ref{fig:chi-top}.
}
\label{fig:top-history}
\end{figure}

\begin{itemize}
\item Approximating the continuous space-time by a discrete lattice results in a
discretization error. Using DWF, the error starts with 
${\cal{O}}(\mres a)+{\cal{O}}(a^2\Lambda_{\text{QCD}}^2)$ . 
The value of $\mres a$ is negligibly small in our simulation compared
with the large statistical error involved except in the case of the pion.
Our results are closer to their continuum values than
those obtained using a Wilson-type fermion on similar lattice scale.

For quenched DWF QCD, the physical values of $f_\pi$, $f_K$, and $f_K/f_\pi$ shift 
by $\sim$5\%, 3\%, and 2\%, respectively, when the lattice scale changes from 
$a^{-1}=2$ GeV to continuum limit\cite{Aoki:2005ga}, 
which are equal or less than current statistical error. 
\item Because of the limited number of quark mass points calculated in our simulation,
we restricted ourselves to using the simplest linear chiral extrapolations 
(\ref{eq:chiral_lin_formula}) and that obtained from the AWTI (\ref{eq:chiral_sqrt_formula}).
A more appropriate extrapolation based on a larger number of quark mass points
is the chiral fitting formula from the (partially quenched) chiral perturbation theory.
While the mass of $\eta'$, which is investigated as the main topic in this work, 
shows little dependence on quark mass, a more precise chiral extrapolation to the physical quark mass point using lighter quark masses is needed to obtain more reliable results.
\item Although our assumption, that the ground state is a one-particle state is certainly wrong for some quantum numbers, some of the decay channel in nature 
are prohibited in simulations using degenerate up and down quarks with heavier mass in a relatively small spatial box (2 fm)$^3$ without a strange quark.
More sophisticated investigations such as calculating the scattering amplitudes between multiparticles are needed to verify  our spectrum results for the decaying meson.
\item Without results obtained from a larger volume, it is difficult to estimate the 
finite-volume effect, although it might be smaller than that for baryons. 
\item Strange sea quark effect: 
The number of quark flavors that play dynamical roles in the $\eta'$  meson
may be very important as seen in the WV relation , $m_{\eta'}^2\propto N_f$.
By increasing $N_f$ from 2 to 3 by including strange quark, the WV prediction
for $m_{\eta'}$ becomes $\sim$ 20\% larger. 
A strange quark  is, however, heavier than up/down quarks, and 
the mass of $\eta'$ in  the $N_f=2+1$ QCD is likely to be in between
the results of $N_f=2$ and 3. 
\item Topological charge distribution and its effects to $\eta'$ meson: In our simulation, 
we deliberately used a special gauge action, DBW2, for  good chiral symmetry. However,  the autocorrelation time of the topological charge in the simulation becomes longer. 

The samples taken in our simulation may not be sufficiently long 
for the reliable estimation of the autocorrelation time for $Q_\text{top}$. 
The growth of the binned-jackknife error for $\vev{Q_\text{top}}$ 
with increasing bin size was monitored, and we estimated 
the autocorrelation time of roughly $\sim$ 300 trajectories 
for the $m_f=0.02$ ensemble and $\sim$ 200 trajectories 
for $m_f=0.03, 0.04$.
Because of the less frequent tunneling between different topological sectors, the charge distribution sampled in our simulation may be statistically skewed. 
In fact, $\vev{Q_\text{top}}=-0.7(7), 1.4(6),$ and 1.8(4) for $m_f=0.02, 0.03,$ and 0.04, respectively.
Note that the central value for $m_f=0.04$ is more than four standard deviations  away from zero.
It is conceivable that this poor sample of the topological sectors causes  significant systematic errors in $\eta'$ spectrum, particularly for the $m_f=0.04$ ensemble.

Figure \ref{fig:chi-top} shows the topological susceptibility, $\chi_\text{top}$ in (\ref{eq:Qtop}) as a function of quark mass.
The fact that the susceptibility for all three masses is constant within 
two standard deviations implies  that the simulation points are  far from the lighter-quark-mass region, where the susceptibility may vanish as a linear function of quark mass. It is also possible the tunneling between different topological sectors does not occur sufficiently frequently, as shown in Figure \ref{fig:top-history}; thus,  the estimation for the susceptibility has a larger systematic error.
\end{itemize}


Of course, more reliable estimation of the magnitude of these systematic errors may be
carried out by future simulations on a finer and larger lattice using
lighter quark masses with the strange sea quark effect, 
and with a larger statistical sample size.

\section{Summary and discussions} 

We have measured light meson propagators in all channel (flavor nonsinglet/singlet 
pseudoscalar, vector, scalar, pseudovector, and tensor meson
$\pi$,  $\rho$, $a_0$, $a_1$, $b_1$; $\eta'$, $\omega$, $f_0$, $f_1$, $h_1$)
and estimated the ground state meson masses and some of
leptonic decay constants, as well as, the excited state mass, 
in two flavors of domain wall QCD.

The size of the statistical sample used in the calculation 
is increased by five to ten times higher than that
reported previously \cite{Aoki:2004ht}.
By  applying  the gauge-invariant Wuppertal smearing because of 
quark operators for their better overlap  with the ground-state, 
the statistical error of the pion and $\rho$ masses is reduced by approximately 50\% and  
the reduction for $\eta'$ is more than 100\%, {\it i.e.\/}, we were only able to 
obtain the nonzero signal by using smeared field. To extract 
values for the meson mass and decay constant by fitting the propagators,
we use two methods, the standard and  variational methods.
The results of these methods are consistent with each other, which indicates that
excited-state contamination of the ground-state is controlled by the smearing.

The systematic uncertainties discussed in the previous section are
difficult to estimate; thus, we only quote results 
with statistical errors.
Our results linearly extrapolated to the physical quark mass point are
\begin{eqnarray*}
&&a_{m_\rho}^{-1}=1.537(26) \ {\text {GeV} }~,\\
&&r_0=0.5491(93)\ {\text{fm}}
\end{eqnarray*}
for quantities directly related to the lattice scale, 
\begin{eqnarray*}
&&f_{\rho}=210(15) \ {\text {MeV} }~,\\
&&f_{\pi^*}=20(19) \ {\text {MeV } }
\end{eqnarray*}
for decay constants, and
\begin{eqnarray*}
&&m_{a_{0}}=1.111(81) \ {\text {GeV} }~,\\
&&m_{\eta'}=819(127) \ {\text {MeV} }~,\\
&&m_{\omega}=790(194) \ {\text {MeV} }~,\\
&&m_{\pi^*}=1.791(138) \ {\text {GeV} }~,\\
&&m_{\rho^*}=2.028(131) \ {\text {GeV} }~, \\
&&m_{a_1}=1.140(51) \ {\text {GeV} }~, \\ 
&&m_{b_1}=1.203(64) \ {\text {GeV} }~, \\
&&m_{f_1}=1.033(137) \ {\text {GeV} }~, \\
&&m_{h_1}=1.225(250) \ {\text {GeV} }~,
\end{eqnarray*}
for the mass spectrum.
The lattice scale is set from $m_\rho=775.49$ MeV. 

\begin{figure}[t] 
\begin{center} 
\includegraphics[angle=-00,scale=0.50,clip=true]{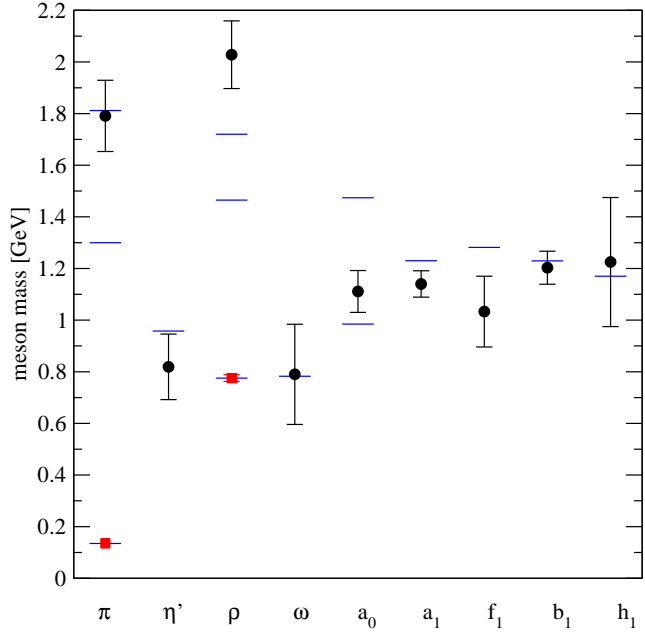}
\end{center}
\caption{
Comparison of simulation results with experimental values\cite{Yao:2006px} in
the real world. Horizontal bars show the experimental values and
filled circles show the simulation results. The error bars indicate
statistical errors only. The squares show the quantities used to set the
lattice spacing and the physical quark mass point.
}
\label{fig:meson-1}
\end{figure}

In Fig.~\ref{fig:meson-1}, the  meson masses obtained 
in this work are compared with the experimental values\cite{Yao:2006px}.
Horizontal bars show the experimental values and
filled circles show the simulation results. The error bars indicate
statistical errors only. 

The decay constant of the excited pseudoscalar meson turns out to be 
consistent with zero at the chiral limit as expected:
\begin{equation} 
f_{\pi^*} = -05(20) \ {\text {MeV }}~.
\end{equation}
The flavor singlet scalar meson, ${f_0}$, was too noisy
to obtain its mass in our data.

In this paper we chose the simplest noise method, complex ${\boldsymbol Z}_2$,  for evaluating
the quark loop amplitudes. More elaborate and/or sophisticated methods 
\cite{Foley:2005ac,Duncan:2001ta,DeGrand:2004qw,Michael:1998sg,Collins:2007mh} 
may improve the statistical accuracy of the calculation.

The recalculated pion mass is consistent with the previous result, 
but the results for $\rho$ and $a_0$ meson masses are significantly different. 
The central value of $m_\rho$ is 10\% larger, {\it i.e.\/}, $a^{-1}_{m_\rho}$ is 10\% smaller, and the error bar is reduced by 50\%  compared with the previous 
results \cite{Aoki:2004ht}.
Both the central value and the error bar of $m_{a_0}$ are 25\% smaller than those in previous results\cite{Prelovsek:2004jp}. 

We confirm that the flavor singlet pseudoscalar meson,  $\eta'$, 
is not an NG boson, and $m_{\eta'}$ is not likely to be zero at the chiral limit, 
which is consistent with the standard understanding of the axial anomaly.
Assuming the WV relation (\ref{eq:witten-veneziano}) is exact at $N_c=3$,
one can calculate the mass gap, $m_0^2$, and topological susceptibility, 
$\chi_\text{top}$, from our values of $m_{\eta'}$ and $m_\pi$,
\begin{equation}
\chi_\text{top}(\text{WV}) = {f^2_\pi\over 2 N_f} m_0^2,~~
m_0^2 = m_\eta'^2-m_\pi^2 pp
\end{equation}
The value of $\chi_\text{top}(\text{WV})$ for $N_f=2$ are plotted in Fig.~\ref{fig:chi-top}
as squares. The horizontal line is  $\chi_\text{top}$ obtained from 
a pure $SU(3)$ YM simulation\cite{DelDebbio:2004ns}. For $m_f=0.02$ and  0.03,
$\chi_\text{top}(\text{WV})$ is consistent with the quenched value, while
$m_f=0.04$ point $\chi_\text{top}(\text{WV})$ undershoots the line significantly.
By linearly extrapolating to the chiral limit, we obtained 
$m_0^2=(808 (129){\text{ MeV}})^2$ and 
$\chi_{\text{top}}(\text{WV})= (193 (15){\text{ MeV}})^4$, which 
is consistent with the quenched value\cite{DelDebbio:2004ns}
$(191 (5){\text{ MeV}})^4$. 
The agreement, which may imply only small $1/N_c$ correction, 
is interesting and  deserves further investigation in future.

These results are susceptible to various systematic errors.
First, we have only two flavors of dynamical quarks.
The omission of the strange quark and antiquark pairs in vacuum,
whose mass is comparable to the dynamical scale of the QCD,  
may skew our results significantly.
The limited number of quark masses, three unitary points, restricted
us to examining only the simplest function for the quark mass
dependence of the physical results. Thus, the chiral extrapolation has 
a systematic error due to the omission of curvature resulting from the chiral logarithms
and higher order terms although many of our results show little dependence on
quark mass.
The ensemble was generated only on
a $16^3\times 32$ lattice with periodic boundary condition in the space directions;
thus, all the meson  spectrum is affected by the  ``mirror'' images located
 $\sim 2$ fm away from the original image in each of the three spatial directions.
The effects may be as large as $\sim$ 10\% for the lightest 
quark mass points. The lattice discretization error in this study is small,
${\cal{O}}(\mres a)+{\cal{O}}(a^2\Lambda_{\text{QCD}}^2)\sim {\cal O}(1{\rm\%})$.
The previous careful
studies\cite{Aoki:2004ht} on the scaling violation show a $\sim$ 5\% level shift
for $a\sim 0.1$ fm lattices.
The omission of the isospin violation due to the differences in quark mass
and electric charge is likely to be negligible compared with other sources of errors,
but this issue can also be studied nonperturbatively using
lattice\cite{Blum:2007cy}.
Despite the significant statistical error and 
the various remaining systematic uncertainties, this study should
serve a benchmark calculation for the statistical features of difficult
physical quantities, disconnected diagrams,  and computational feasibility tests.
The results for the mass of $\eta'$ were close 
to the experimental value,
indicates that further improvements can be made particularly calculation 
using an $N_f=2+1$ DWF ensemble\cite{Antonio:2006px,Allton:2007hx}.

\section*{Acknowledgements}

We thank RIKEN, Brookhaven National Laboratory, and the U.S. Department of 
Energy for providing the facilities essential for the completion of this work. 
We are grateful  to  members of the RBC collaboration, especially to
T.~Blum, N.~Christ, C.~Dawson, R.~Mawhinney,  K.~Orginos, and A.~Soni for their
various contributions in the early stages of this work and  their continuous 
encouragement.
The QCDOC supercomputer at the RIKEN-BNL Research Center (RBRC) 
was used for the numerical calculations in this work.
K.H. thanks RBRC for its hospitality while this work 
was partly performed. 
We are grateful to the authors and maintainers of the CPS\cite{CPS}, which was used in this work.
This work is supported in part by the Grants-in-Aid for 
Scientific Research from the Ministry of Education, Culture, Sports, Science and Technology (No. 17750050).

\end{document}